\documentclass[11pt,a4paper,openright]{article}
\usepackage[english]{babel} 
\usepackage[latin1]{inputenc}
\usepackage{amsmath,amssymb,amsthm,amsfonts,amscd,authblk,dsfont,color,cancel,diagbox,enumerate,epsfig,eucal,fancyhdr,latexsym,lmodern,mathrsfs,mathtools,multirow,musicography,phaistos,skull,simplewick,subcaption,tikz-cd}
\usepackage[normalem]{ulem}
\usepackage{graphicx}
\usepackage{a4wide}
\usepackage{microtype}
\usepackage{bbm}
\usepackage{slashed}
\usepackage{times}

\usepackage[
final=true,                    
plainpages=false,              
pdfpagelabels=true,            
pdfencoding=auto,              
unicode=true,                  
hypertexnames=true,            
naturalnames=true              
]{hyperref}
\usepackage[all,cmtip]{xy}
\usepackage[autostyle, english = american]{csquotes}
\MakeOuterQuote{"}

\usepackage[multiuser]{fixme}
\FXRegisterAuthor{nd}{annd}{Nico} 
\FXRegisterAuthor{ab}{anab}{Albi} 
\FXRegisterAuthor{cd}{ancd}{Cla}

\newcommand{\llangle}{\langle\!\langle}
\newcommand{\rrangle}{\rangle\!\rangle}

\definecolor{darkred}{rgb}{0.8,0.1,0.1}
\hypersetup{
     colorlinks=true,         
     linkcolor=darkred,
     citecolor=blue,
}

\usepackage{comment}
\usepackage[margin=2.2cm]{geometry}
\usepackage[all,cmtip]{xy}

\usepackage{marvosym}

\theoremstyle{plain}
\newtheorem{theo}{Theorem}[section]

\newtheorem{graphrules}{Graph Rules}

\theoremstyle{definition}

\newtheorem{ex}[theo]{Example}

\definecolor{NiColor}{RGB}{77,77,255}
\definecolor{NiColoRed}{RGB}{255,77,77}
\definecolor{NiCitation}{RGB}{77,255,77}

\definecolor{AlbiColor}{rgb}{0.8, 0.25, 0.33}

\newtheoremstyle{TheoremStyle}
{3pt}
{3pt}
{}
{}
{\bf}
{:}
{.5em}
{}
\newtheoremstyle{ExampleAndRemarkStyle}
{3pt}
{3pt}
{}
{}
{\bf}
{:}
{.5em}
{}
\newtheoremstyle{ProofStyle}
{3pt}
{3pt}
{}
{}
{\bf}
{:}
{.5em}
{}

\theoremstyle{TheoremStyle}
\newtheorem{theorem}{Theorem}

\newtheorem{proposition}[theorem]{Proposition}
\newtheorem{lemma}[theorem]{Lemma}
\newtheorem{Definition}[theorem]{Definition}
\theoremstyle{ExampleAndRemarkStyle}
\newtheorem{remark}[theorem]{Remark}
\newtheorem{Example}[theorem]{Example} 
\theoremstyle{ProofStyle}

\usetikzlibrary{shapes,snakes} 
\usetikzlibrary{arrows} 
\newcommand{\midarrow}{
	\begin{tikzpicture}[every node/.style={sloped,allow upside down}]
		\draw[->] (0,0) -- +(.1,0);
	\end{tikzpicture}
}

\newcommand{\xvertex}[1]{
	\begin{tikzpicture}[every node/.style={sloped,allow upside down}]
		\draw (0,0) -- (0,0.5);
		\filldraw (0,0) circle (1.5pt) node[right] {#1};
	\end{tikzpicture}
}
\newcommand{\tildexvertex}[1]{
	\begin{tikzpicture}[every node/.style={sloped,allow upside down}]
		\draw[snake] (0,0) -- (0,0.6);
		\filldraw (0,0) circle (1.5pt) node[right] {#1};
	\end{tikzpicture}
}
\newcommand{\xivertex}[1]{
	\begin{tikzpicture}[every node/.style={sloped,allow upside down}]
		\draw (0,0) circle (1.5pt);
	\end{tikzpicture}
}

\title{%
An algebraic correspondence between stochastic differential equations and the Martin-Siggia-Rose formalism
}

\author{%
Alberto Bonicelli$^{1,a}$, Claudio Dappiaggi$^{1,b}$, Nicol\`o Drago$^{2,c}$ 
\vspace{4mm}\\
{\small $^1$ Dipartimento di Fisica, Università  di Pavia and INFN and INdAM, Sezione di Pavia, Via Bassi 6, I-27100 Pavia, Italy.}\vspace{2mm}\\
{\small $^2$ Dipartimento di Matematica, Università  di Trento and INFN-TIFPA and INdAM, Via Sommarive 14, I-38123 Povo, Italy}
\vspace{4mm}\\
 {\footnotesize  ~$^a$ \href{mailto:alberto.bonicelli01@universitadipavia.it }{alberto.bonicelli01@universitadipavia.it}~,~$^b$ \href{mailto:claudio.dappiaggi@unipv.it}{claudio.dappiaggi@unipv.it}~,~$^c$ 
 \href{mailto:nicolo.drago@unige.it}{nicolo.drago@unige.it}~}
 }

\date{\today}

\begin{document}

\maketitle

\begin{abstract}
	In the realm of complex systems, dynamics is often modeled in terms of a non-linear, stochastic, ordinary differential equation (SDE) with either an additive or a multiplicative Gaussian white noise. In addition to a well-established collection of results proving existence and uniqueness of the solutions, it is of particular relevance the explicit computation of expectation values and correlation functions, since they encode the key physical information of the system under investigation. A pragmatically efficient way to dig out these quantities consists of the Martin-Siggia-Rose (MSR) formalism which establishes a correspondence between a large class of SDEs and suitably constructed field theories formulated by means of a path integral approach. Despite the effectiveness of this duality, there is no corresponding, mathematically rigorous proof of such correspondence. We address this issue using techniques proper of the algebraic approach to quantum field theories which is known to provide a valuable framework to discuss rigorously the path integral formulation of field theories as well as the solution theory both of ordinary and of partial, stochastic differential equations. In particular, working in this framework, we establish rigorously, albeit at the level of perturbation theory, a correspondence between correlation functions and expectation values computed either in the SDE or in the MSR formalism.
\end{abstract}
\paragraph*{Keywords:}
Algebraic Quantum Field Theory,
Martin-Siggia-Rose Formalism,
Stochastic Differential Equations


\section{Introduction}
\label{Sec: introduction}

Complex systems are a rather important and thriving realm with manifold applications in different areas often trespassing the traditional borders of modern physics and statistical mechanics, becoming a valuable framework to describe phenomena ranging from biology, to neuroscience or even to finance, see {\it e.g.} \cite{Arnold-2013, Gardiner-1985}.
Among the many reasons for such versatility, first and foremost one must take into account that in the formulation of many models, encompassing the behaviour of many degrees of freedom leads naturally to considering that either or both the underlying key parameters of the model and the main variables of interest display a random behaviour. 

One of the main reasons for the success of such line of thought can be ascribed to the existence of a very well-developed, computationally efficient, underlying mathematical framework which combines tools and techniques from very different areas such as probability theory, functional analysis and mathematical physics. In particular, depending on the specific case of interest, the main object of investigation is a random variable whose dynamics is codified either by a {\bf stochastic differential equation (SDE)} or by a stochastic partial differential equation (SPDE).

In this work, we shall focus our attention on a large class of SDEs of the form
\begin{align}\label{Eq: SDE - general form}
	\dot{x}_\eta(t)=\alpha(x_\eta(t),t)+\beta(x_\eta(t),t)\sqrt{\sigma}\eta(t)\,,
\end{align}
where $\sigma>0$, while $x_\eta$ is a $\mathcal{D}'(\mathbb{R})$-valued random variable where $\mathcal{D}'(\mathbb{R})$ denotes the space of distributions, defined as the the topological dual of the set of compactly supported functions $\mathcal{D}(\mathbb{R})$.
In Equation \eqref{Eq: SDE - general form}, $\alpha(x(t),t)$ and $\beta(x(t),t)$ are polynomial functions in $x(t)$ with coefficients smoothly dependent on $t$. In addition $\eta$, a white noise, is a stationary Gaussian stochastic process with vanishing mean and covariance whose integral kernel reads $\mathbb{E}_\eta[\eta(t)\eta(t')]=\delta(t-t')$, $\mathbb{E}_\eta$ standing for the expectation value.

To set the nomenclature, we shall refer to $\alpha$ (\textit{resp.} $\beta$) as the \textbf{additive term} (\textit{resp.} \textbf{multiplicative term}) of the SDE, the latter being non-trivial if and only if   $\beta_1:=\partial_x\beta\neq0$. From the analytic viewpoint the solution theory of Equation \eqref{Eq: SDE - general form} is well-understood and fully developed, see {\it e.g.} \cite{Oksendal} for a review of the key existence and uniqueness results. At the same time, in the development and validation of concrete models, one is mainly interested in finding an explicit form of the expectation value both of the solutions of Equation \eqref{Eq: SDE - general form} and of the associated correlation functions.  

In the physics literature this problem is often addressed dealing with Equation \eqref{Eq: SDE - general form} by means of a completely different framework which goes under the name of \textbf{Martin-Siggia-Rose (MSR) formalism} \cite{Martin-Siggia-Rose-73}, also known as the \textbf{Janssen-De Dominicis-Peliti path integral formalism} on account of \cite{De Dominicis-Peliti.1978, Jannsen-1976}. 
It is a very efficient and fruitful approach, see {\it e.g.} \cite{Aron-Barci-Cugliandolo-Gonzales-Arenas-Lozano-2016,Buice-Chow-2010,Hertz-Roudi-Sollich-2017,Tauber-Howards-Vollmayr-2005,Zinn-Justin-1989} which can also be suitably modified to study SPDEs  \cite{Wiese-1998}. The main rationale behind this scheme lies in associating to any possibly non-linear SDE of the form of Equation \eqref{Eq: SDE - general form} a corresponding field theory which allows to evaluate the correlation functions of the solutions of Equation \eqref{Eq: SDE - additive case} in terms of suitable counterparts which, in turn, are computed by means of a path-integral formulation.

In short, in the MSR formalism, one introduces an auxiliary purely imaginary field $\tilde{x}(t)$ and the sought correlation functions are computed by means of the identity
\begin{align}\label{Eq: correlation and formal path integral}
	\mathbb{E}_\eta[x_\eta(t_1)\cdots x_\eta(t_k)]
	=\int\limits_{\mathcal{C}}\mathrm{D}x\mathrm{D}\tilde{x}\,e^{-S[x,\tilde{x}]}
	x(t_1)\cdots x(t_k)\,,
\end{align}
where $\mathcal{C}$ is a suitable (and often implicitly defined) space of configurations and where the action $S[x,\tilde{x}]$ reads
\begin{equation}\label{Eq: Action}
	S[x,\tilde{x}]
	=\int\limits_{\mathbb{R}}\big[\tilde{x}(t)\left(\dot{x}(t)-\alpha(x(t),t)-x_0\delta_{t_0}(t)\right)
	-\frac{\sigma}{2}\beta^2(x(t),t)\tilde{x}^2(t)\big]\mathrm{d}t,
\end{equation}	
where $\sigma>0$, $\alpha,\beta$ are as per Equation \eqref{Eq: SDE - general form} while the term $x_0\delta_{t_0}$ is implementing the initial condition $x_\eta(t_0)=x_0$. We shall not derive Equation \eqref{Eq: correlation and formal path integral} and we recommend to an interested reader to refer either to \cite{Buice-Chow-2010} or to \cite[App. D]{Aron-Barci-Cugliandolo-Gonzales-Arenas-Lozano-2016}. Yet we strongly emphasize that, to the best of our knowledge, all derivations of the connections between SDEs and a path-integral formulation are based on several formal manipulations and computations and Equation \eqref{Eq: correlation and formal path integral} itself is not based on a firm mathematical ground. At the same time, as mentioned before, the MSR formalism is often applied since, at a numerical level, an analysis based on path-integrals can be computationally very efficient. 

Therefore the main goal of this work is to circumvent these deficiencies showing that one can give a rigorous meaning to Equation \eqref{Eq: correlation and formal path integral}, though using techniques proper of perturbation theory whenever non-linear terms are present in Equation \eqref{Eq: SDE - general form}.

In order to achieve our goal, we shall consider a third framework, known as {\em algebraic quantum field theory (AQFT)}, see {\it e.g.} \cite{Brunetti:2015vmh}. Originally it has been devised in order to establish a rigorous formulation of any quantum field theory, disentangling the role and the structural properties of the algebra of physical observables from the choice of a quantum state. Among the many successes which can be ascribed to AQFT, to our purposes, most notable is the formulation of a mathematically satisfactory, covariant theory of renormalization using techniques proper of microlocal analysis \cite{Brunetti-Fredenhagen-2000,Hollands-wald-2001}.
Starting from these premises, slowly but steadily, it has become manifest that the algebraic approach is a valuable tool both to give a mathematically precise formulation of theories where pathological singularities occur and to provide an efficient framework to account for any underlying renormalization freedom at an intrinsic level. This statement holds particularly true both when studying SPDEs or the MSR formalism, as we outline succinctly in the following.

\vskip .2cm

\noindent{\bf S(P)DEs --}
Recently, in \cite{Dappiaggi-Drago-Rinaldi-Zambotti-2022}, see also \cite{Bonicelli-Dappiaggi-Rinaldi-2021}, it has been developed a novel method to construct algorithmically at a perturbative level the solutions and the associated correlation functions for a large class of non-linear SPDEs with an additive Gaussian white noise, using techniques inspired by the algebraic approach to quantum field theory. In detail we have considered equations of the form
\begin{align}\label{Eq: Generic equation}
	E\widehat{\Phi}=\xi+F[\widehat\Phi],
\end{align}
where $\widehat{\Phi}$ is a $\mathcal{D}'(M)$-valued random variable, $M$ being an arbitrary Riemannian manifold. In addition $E$ is a linear operator either of elliptic or of parabolic type, while $F\colon\mathbb{R}\to\mathbb{R}$ is a, possibly non-linear, polynomial function. In a few words, following \cite{Brunetti:2009qc} and being inspired by the so-called functional formalism \cite{Carfora-Dappiaggi-Drago-Rinaldi-2020,Dappiaggi-Drago-Rinaldi-2020, Fredenhagen:2014lda}, we consider a  specific class of distributions with values in polynomial functionals over $C^\infty(M)$. The main ingredients are two distinguished elements
\begin{align*}
	\langle\Phi(\varphi),f\rangle=\int_M \varphi(x) f(x)\mu(x)\,,
	\qquad
	\langle\boldsymbol{1}(\varphi),f\rangle=\int_M f(x)\mu(x)\,,
\end{align*}
where $\mu$ is a strictly positive density over $M$, $\varphi\in C^\infty(M)$ while $f\in C^\infty_0(M)$. These two functionals are employed as generators of a commutative algebra $\mathcal{A}$ whose composition is the pointwise product. At this stage, $\mathcal{A}$ does not contain any information on the underlying stochastic process.

These are encoded in the algebra by deforming its product 
so to codify the information on the covariance of the underlying white noise.
In this framework, taking an expectation value with respect to the underlying stochastic process is translated in the evaluation of the relevant functionals at the configuration $\varphi=0$.

If we switch the attention back to Equation \eqref{Eq: SDE - general form}, one can realize that it is possible to adapt the algebraic framework also to this case. More precisely, we consider an integral version of Equation \eqref{Eq: SDE - general form},
\begin{multline}\label{Eq: SDE fixed point}
	x_\eta(t):=x_0
	+\int_{-\infty}^t\chi(s)\left[\alpha(x_\eta(s),s)+\beta(x_\eta(s),s)\sqrt{\sigma}\eta(s)\right]\mathrm{d}s
	\\
	=x_0
	+[G\ast\chi(\alpha(x_\eta)+\beta(x_\eta)\sqrt{\sigma}\eta)](t)\,,
\end{multline}
where here the role of the linear operator $E$ is played by $L=\frac{\mathrm{d}}{\mathrm{d}t}$ while $G=\vartheta$ is the Heaviside function.
Observe that we have inserted a cut-off $\chi\in\mathcal{D}(\mathbb{R})$ in order to make the convolution $\ast$ well-defined.
This entails that, we are modifying the right-hand side of Equation \eqref{Eq: SDE - general form} adding an overall multiplicative factor $\chi(t)$.
Throughout the paper we shall comment on its significance and on the possibility of or on the obstructions in removing it.
Moreover, notice that Equation \eqref{Eq: SDE fixed point} prescribes the initial datum $x_0$ at $t_0=-\infty$.
Fully in the spirit of \cite{Dappiaggi-Drago-Rinaldi-Zambotti-2022}, we consider a polynomial algebra, $\mathcal{P}_\xi$ generated by the functionals
\begin{align*}
	\langle\mathbb{I}(\xi),f\rangle
	=\int\limits_{\mathbb{R}}f(t)dt\,,
	\quad
	\langle G(\xi),f\rangle=\int\limits_{\mathbb{R}}[G\ast\chi\xi](t)f(t)dt\,,
	\quad\xi\in C^{\infty}(\mathbb{R}), f\in C_0^{\infty}(\mathbb{R})\,.
\end{align*}
Much in the spirit of \cite{Dappiaggi-Drago-Rinaldi-Zambotti-2022}, we can deform the pointwise algebra product introducing the map
\begin{align*}
	\Gamma_{\delta/2}\colon\mathcal{P}_\xi\to\mathcal{P}_\xi\,,
	\qquad
	\Gamma_{\delta/2}:=\exp[\Upsilon_{\delta/2}]
	\qquad
	\Upsilon_{\delta/2}(F):=\frac{1}{2}\int\limits_{\mathbb{R}^2}\delta(s-s')
	\frac{\delta^2 F}{\delta\xi(s)\delta\xi(s')}\mathrm{d}s\mathrm{d}s'\,.
\end{align*}
Similarly to the SPDE scenario, this deformation codifies in the algebra of functionals the information on the covariance of the underlying Gaussian stochastic process, while the expectation value amounts to evaluating each functional at the configuration $\xi=0$. In other words, barring further technical aspects which are discussed in the main body of this work, we shall end up with 
\begin{align}
	\label{Eq: correlation and Gamma delta map}
	\mathbb{E}_\eta[x_\eta(t_1)\cdots x_\eta(t_k)]
	&=\Gamma_{\delta/2}[
	x_\xi(t_1)\cdots
	x_\xi(t_k)]|_{\xi=0}\,.
\end{align}
To wit, for the particular case of $\alpha=0$, $\beta=1$ and $x_0=0$ one finds, at first order in perturbation theory,
\begin{align*}
	x_\xi(t)=\int\limits_{\mathbb{R}} G(t,s)\sqrt{\sigma}\chi(s)\xi(s)\mathrm{d}s
	\quad\Rightarrow\quad
	\Upsilon_{\delta/2}[x_\xi(t_1)x_\xi(t_2)]|_{\xi=0}
	=\sigma^2\int\limits_{\mathbb{R}} G(t_1,s)G(t_2,s)\chi(s)^2\mathrm{d}s\,,
\end{align*}
which matches with the corresponding expression for $\mathbb{E}_\eta[x_\eta(t_1)x_\eta(t_2)]$.

Notice in particular that it will turn out that the right-hand side of Equation \eqref{Eq: correlation and Gamma delta map} is meaningful only if the solutions $x_\eta$ of Equation \eqref{Eq: SDE - general form} are constructed at a perturbative level and in the sense of Stratonovich ---\textit{cf.} Remark \ref{Rmk: SDE formalism deals with Stratonovich only}. However, this last feature is not a restriction as one may always hop from an It\^o-like interpretation to an equivalent Stratonovich counterpart by a suitable modification of the underlying SDE, in particular of the functions $\alpha$ and $\beta$, see \textit{e.g.} \cite[\S 3]{Oksendal}.

\vskip .2cm

\noindent{\bf MSR --} In this formalism the main technical challenge is to give meaning to the right-hand side of Equation \eqref{Eq: correlation and formal path integral}. Similar hurdles appear in many instances in several quantum field theoretical models and one satisfactory way to overcome them is to make use of the perturbative approach to quantum field theory (pAQFT), see {\it e.g.} \cite{Brunetti:2009qc,Rejzner:2016hdj}.
Without entering into many technical details at this stage, we highlight that the underlying rationale consists of introducing a suitable tensor algebra of polynomial functionals $\mathcal{P}_{x\tilde{x}}$ which is in turn generated by a class of local functionals, see Definition \ref{Def: xtildex polynomial functional} and \ref{Def: xtildex local functionals},
\begin{equation}\label{Eq: MSR Functionals Intro}
	F:C^\infty(\mathbb{R})\times C^\infty(\mathbb{R}^2)\to\mathcal{D}^\prime(\mathbb{R})\quad (x,\tilde{x})\mapsto \langle F(x,\tilde{x}),f\rangle=\int\limits_{\mathbb{R}}f(t)p(x(t),\tilde{x}(t))dt,
\end{equation}
where $f\in\mathcal{D}(\mathbb{R})$ while $p$ is a polynomial function in the variables $x$ and $\tilde{x}$ with coefficients lying in $C^\infty(\mathbb{R})$. 

Similarly to the preceding scenario, the correlation functions on the right hand side of Equation \eqref{Eq: correlation and formal path integral} can be constructed at a perturbative level by encoding their information in a deformation of the underlying algebra structure. In other words we introduce a linear map $\Gamma_{G}:=\exp[\Upsilon_G]\colon\mathcal{P}_{x\tilde{x}}\to\mathcal{P}_{x\tilde{x}}$ where the action of $\Upsilon_G$ on the functionals as in Equation \eqref{Eq: MSR Functionals Intro} reads
\begin{align*}
	\Upsilon_G(F):=\int\limits_{\mathbb{R}^2} G(t,t')\frac{\delta^2 F}{\delta x(t)\delta \tilde{x}(t')}\mathrm{d}t\mathrm{d}t'\,,
\end{align*}
where $G(t,t^\prime)=\vartheta(t-t^\prime)$, $\vartheta$ being the Heaviside function. By means of $\Gamma_{G}$ we can identify an associative and commutative algebra $\mathcal{P}_G:=\Gamma_G(\mathcal{P}_{x\tilde{x}})$ whose product is
\begin{align*}
	F_1\cdot_G F_2:=\Gamma_G[\Gamma_G^{-1}(F_1)\Gamma_G^{-1}(F_2)]
	\qquad
	\forall F_1,F_2\in\mathcal{P}_G\,.
\end{align*}
Following the main rationale of pAQFT \cite{Brunetti:2009qc}, the key advantage of working with $\mathcal{P}_G$ lies in the fact that one can define
\begin{align}\notag
	\int\limits_{\mathcal{C}}\mathrm{D}x\mathrm{D}\tilde{x}\,e^{-S_0[x,\tilde{x}]}
	x(t_1)\cdots x(t_k)\tilde{x}(s_1)\cdots\tilde{x}(s_\ell)
	&:=\Gamma_{G}\left[x(t_1)\cdots x(t_n)\tilde{x}(s_1)\cdots\tilde{x}(s_\ell)\right]|_{\substack{x=x_0\\\tilde{x}=0}}
	\\
	\label{Eq: formal path integral and Gamma G0 product}
	&=[x(t_1)
	\cdots_G x(t_n)
	\cdot_G\tilde{x}(s_1)
	\cdots_G\tilde{x}(s_\ell)]|_{\substack{x=x_0\\\tilde{x}=0}}\,.
\end{align}
where $S_0[x,\tilde{x}]$ is the free action, obtained from Equation \eqref{Eq: Action} setting $\alpha=\beta=0$.

\vskip .2cm

\noindent{\bf SDE-MSR Correspondence --} Having given meaning independently to the analysis of a non-linear SDE and to the path-integral approach by means of the algebraic framework, we can tackle the question whether one can make precise the SDE-MSR correspondence and prove it. The first hurdle can be overcome thanks to procedures outlined before. As a matter of fact, we shall translate the well-posedness of the SDE-MSR correspondence into the question whether the following identity holds true, where the left hand side is defined via Equation \eqref{Eq: correlation and Gamma delta map}, while the right hand one by means of Equation \eqref{Eq: formal path integral and Gamma G0 product}:
\begin{align}\notag
	\mathbb{E}_\eta[x_\eta(t_1)\cdots x_\eta(t_k)]
	&\stackrel{?}{=}\Gamma_{G}[
	e^{\langle\chi,\tilde{x}\alpha\rangle
		+\frac{\sigma}{2}\langle\chi^2,\tilde{x}^2\beta^2\rangle}
	x(t_1)\cdots x(t_k)]\bigg|_{\substack{x=x_0\\\tilde{x}=0}}
	\\
	\label{Eq: correlation and G0 product}
	&=[e_{\cdot_G}^{\langle\chi,\tilde{x}\alpha\rangle
		+\frac{\sigma}{2}\langle\chi^2, \tilde{x}^2\beta^2\rangle}
	\cdot_G x(t_1)\cdots_G x(t_k)]\bigg|_{\substack{x=x_0\\\tilde{x}=0}}\,.
\end{align}
Here $\alpha$ and $\beta$ are as per Equation \eqref{Eq: SDE - general form}, $x_0\in\mathbb{R}$ is an initial condition, while $\chi\in\mathcal{D}(\mathbb{R})$ is a cut-off function. In addition $\langle\chi,\tilde{x}\alpha\rangle:=\int\limits_{\mathbb{R}}dt\,\chi(t)\tilde{x}(t)\alpha(x(t),t)$ and a similar definition applies to $\langle\chi^2,\tilde{x}^2\beta^2\rangle$. Observe that, since $\Gamma_{G}$ is only defined on a tensor algebra generated by polynomial functionals, the right-hand side of Equation \eqref{Eq: correlation and G0 product} is meaningful only in the realm of perturbation theory, where the role of the expansion parameter is played by the cut-off $\chi$. In the main body of this work we shall address the question raised in Equation \eqref{Eq: correlation and G0 product} proving that the equality holds true. At this point it is not worth to enter in the interpretative aspects of the identity under investigation or in a discussion of the technical content of the proof. We limit ourselves to reporting the main result of this work:

\begin{theorem}\label{Thm: main results}
	Let $\alpha,\beta$ be polynomial functions in the variable $x$ with coefficients smoothly depending on $t$ and let $x_\xi(t)\in\mathcal{P}_\xi[[\chi]]$ be a formal series in powers of $\chi$ with coefficients in $\mathcal{P}_\xi$ which is a solution to the SDE \eqref{Eq: SDE - general form}, obtained as a perturbative expansion in the cut-off $\chi$ of Equation \eqref{Eq: SDE fixed point}.
	Letting $\Gamma_G$ be as per Definition \ref{Def: Gamma G map}, Equation \eqref{Eq: correlation and G0 product} holds true if either $\alpha=0$ or $\beta=1$. Furthermore, for all $F\in\mathcal{P}_{x}$ it descends
	\begin{align}\label{Eq: SDE-MSR correspondence - delta and G product for local functionals}
		\Gamma_{\delta/2}[F(x_\xi)]|_{\xi=0}
		=\left[e_{\cdot_G}^{\langle\chi,\tilde{x}(\alpha+\frac{1}{2}\chi\sigma\beta\beta_1)\rangle	+\frac{\sigma}{2}\langle\chi^2,\tilde{x}^2\beta^2\rangle}
		\cdot_G F\right]\bigg|_{\substack{x=x_0\\\tilde{x}=0}}\,,
	\end{align}
	where $F(x_\xi)\in\mathcal{P}_\xi[[\chi]]$.
\end{theorem}

Notice that Equation \eqref{Eq: SDE-MSR correspondence - delta and G product for local functionals} is nothing but the counterpart of Equation \eqref{Eq: correlation and G0 product} for a generic polynomial functional $F\in\mathcal{P}_x$. Theorem \ref{Thm: main results} links the MSR approach and the AQFT one at the level of SDEs. This is an important step in the comparison of the two methods and it can be generalized in several directions. To begin with, one may extend the proof of Theorem \ref{Thm: main results} to the case of arbitrary $\alpha,\beta$. This extension is mainly technical and we do not foresee any specific obstruction rather than dealing with a more complicated class of graphs than those in the proofs of Theorems \ref{Thm: SDE-MSR correspondence - additive case}-\ref{Thm: SDE-MSR correspondence - multiplicative case}.

A second avenue of research consists of considering SPDEs rather than SDEs. This would provide a more robust connection between the MSR approach and the one developed in \cite{Dappiaggi-Drago-Rinaldi-Zambotti-2022}.
From a technical point of view, this would require to replace $\frac{\mathrm{d}}{\mathrm{d}t}$ on $\mathbb{R}$ with a heat-like operator like $\frac{\partial}{\partial t}-\Delta_x$ on $\mathbb{R}\times M$, where $\Delta_x$ denotes the Laplacian on a given Riemannian manifold $M$.
The basic definitions introduced in this work would remain the same but renormalization would now play a relevant role in the definition of the product $\cdot_G$.
The latter would require to cope with the fact that powers of $G$ are not well-defined.
The development of the "algebraic" MSR approach within this setting would allow to introduce a notion of "local and covariant renormalization" in the context of SPDE.
The latter concept has been introduced in \cite{Brunetti_Fredenhagen_Verch_2003} and successfully applied to QFT on curved spacetimes to reduce the ambiguity in the definition of renormalized quantities, \textit{e.g.}, the stress-energy tensor \cite{Hollands-wald-2001,Hollands-Wald-2005}.
In the SPDE framework we expect that a similar construction could provide new insights on the nature of the renormalization constants appearing in more standard approaches.
Eventually, it would be extremely interesting to move from the perturbative construction used in this paper to a non-perturbative one.
This can potentially be achieved by considering a path-integral approach to deal with the formal MSR integral in Equation \eqref{Eq: correlation and formal path integral}.

The paper is structured as follows: In Section \ref{Sec: algebraic approach to Martin-Sigga-Rose formalism} we outline the algebraic approach to the MSR formalism. In particular the definition of the map $\Gamma_G$ and its properties are discussed, together with some general results on the underlying expectation values -- see  Definition \ref{Def: MSR expectation value}.
Similarly in Section \ref{Sec: algebraic approach to SDE} we give a succinct introduction to the algebraic approach to stochastic differential equations adapting to the case in hand the framework in \cite{Dappiaggi-Drago-Rinaldi-Zambotti-2022}.
Section \ref{Sec: proof of the SDE-MSR correspondence} is the core of this work and therein we discuss the issue of giving a proof of Theorem \ref{Thm: main results}. We divide the analysis in two cases. In the first one, at the heart of  Section \ref{Subsec: additive SDE: beta=1} we focus on Equation \eqref{Eq: SDE - general form} with an additive term. In the second one, discussed in Section \ref{Subsec: multiplicative SDE: alpha=0}, we prove Theorem \ref{Thm: main results} for the multiplicative scenario. In both instances the proof is based on a suitable graph expansion for both sides of Equation \eqref{Eq: SDE-MSR correspondence - delta and G product for local functionals}, although the procedure is much more involved in the second rather than in the first scenario.
Appendix \ref{Sec: appendix with proofs} recollects the proof of technical results.

\section{Algebraic approach to the Martin-Siggia-Rose formalism}
\label{Sec: algebraic approach to Martin-Sigga-Rose formalism}

In this section we outline an algebraic formulation of the Martin-Siggia-Rose (MSR) formalism \cite{Martin-Siggia-Rose-73} following an approach akin to the one used in \cite{Carfora-Dappiaggi-Drago-Rinaldi-2020,Dappiaggi-Drago-Rinaldi-2020,Keller-2009,Klehfoth-Wald-2022,Wald-1979}. 

To fix the notation, given $u\in\mathcal{D}'(\mathbb{R})$ and $f\in\mathcal{D}(\mathbb{R})$, we shall denote by $\langle u,f\rangle:=u(f)$ the dual pairing between distributions and test functions. At the same time, we define the {\em universal tensor algebra over $\mathcal{D}'(\mathbb{R})$} ({\em resp.} $\mathcal{D}(\mathbb{R})$) as 
\begin{align}\label{Eq: tensor algebra}
	\mathcal{T}'(\mathbb{R})
	:=\mathcal{T}(\mathcal{D}'(\mathbb{R}))
	:=\mathbb{C}\oplus\bigoplus_{n\geq 1}\mathcal{D}'(\mathbb{R})^{\otimes n}\quad\textrm{and}\quad \mathcal{T}(\mathbb{R})
	:=\mathcal{T}(\mathcal{D}(\mathbb{R}))
	:=\mathbb{C}\oplus\bigoplus_{n\geq 1}\mathcal{D}(\mathbb{R})^{\otimes n}\,,
\end{align}
If $T\in\mathcal{T}'(\mathbb{R})$ and $f\in\mathcal{T}(\mathbb{R})$ we set $\langle T,f\rangle:=\sum_{n\geq 0}\langle T_n,f_n\rangle$. The sum is always convergent since $T\equiv(T_n)_{n\in\mathbb{N}}\in\mathcal{T}'(\mathbb{R})$ if and only if $\exists \bar{n}_T\in\mathbb{N}$ such that $T_n=0$ for all $n\geq\bar{n}_T$. The same constraint applies to $f\in\mathcal{T}(\mathcal{D}(\mathbb{R}))$. In the following we introduce the key building block of our construction,  

\begin{Definition}\label{Def: xtildex polynomial functional}
	We call $\operatorname{Pol}_{x\tilde{x}}$ the vector space of all \textbf{polynomial} $\boldsymbol{x\tilde{x}}$\textbf{-functionals}. These are maps
	\begin{align*}
		F\colon C^\infty(\mathbb{R})^2\to\mathcal{T}^\prime(\mathbb{R})\quad (x,\tilde{x})\mapsto F(x,\tilde{x}),
	\end{align*}
	for which there exist $n,\tilde{n}\in\mathbb{N}\cup\{0\}$ such that, for all $x,y,\tilde{x},\tilde{y}\in C^\infty(\mathbb{R})$ the maps $\lambda,\tilde{\lambda}\mapsto F(x+\lambda y,\tilde{x}+\tilde{\lambda}\tilde{y})$ are polynomials of degree $n$ in $\lambda$ and $\tilde{n}$ in $\tilde{\lambda}$. In addition we endow $\operatorname{Pol}_{x\tilde{x}}$ with the unital, associative algebra structure such that $\forall F_1,F_2\in\operatorname{Pol}_{x\tilde{x}}$,
	\begin{align*}
		F_1F_2(x,\tilde{x})
		:=F_1(x,\tilde{x})\otimes F_2(x,\tilde{x})\in\operatorname{Pol}_{x\tilde{x}}\,.
	\end{align*}
\end{Definition}


\begin{remark}
	Observe that Definition \ref{Def: xtildex polynomial functional} is equivalent to the notion of polynomial functional which is often used in the  literature on algebraic quantum field theory, see {\it e.g.} \cite{Dappiaggi-Drago-Rinaldi-Zambotti-2022}. As a matter of fact, if $F\in\operatorname{Pol}_{x\tilde{x}}$
	\begin{align*}
		\frac{\delta^{k+\tilde{k}}F}{\delta x^k\delta\tilde{x}^{\tilde{k}}}=0
		\qquad\forall k> n\,,\,\tilde{k}>\tilde{n}\,,
	\end{align*}
	where $\dfrac{\delta F}{\delta x},\dfrac{\delta F}{\delta \tilde{x}}\colon C^\infty(\mathbb{R})^2\to\mathcal{T}'(\mathbb{R})$ are the functional derivatives
	\begin{multline}\label{Eq: x tildex functional derivative kernel}
		\left\langle\frac{\delta F}{\delta x}(x,\tilde{x})[y],f\right\rangle
		:=\frac{\partial}{\partial\lambda}
		\langle F(x+\lambda y,\tilde{x}),f\rangle\bigg|_{\lambda=0}
		=\left\langle\int\limits_{\mathbb{R}}\frac{\delta F}{\delta x(t)} y(t)\mathrm{d}t, f\right\rangle
		\\
		\left\langle\frac{\delta F}{\delta \tilde{x}}(x,\tilde{x})[\tilde{y}],f\right\rangle
		:=\frac{\partial}{\partial\tilde{\lambda}}
		\langle F(x,\tilde{x}+\tilde{\lambda}\tilde{y}),f\rangle\bigg|_{\tilde{\lambda}=0}
		=\left\langle \int\limits_{\mathbb{R}} \frac{\delta F}{\delta \tilde{x}(t)}\tilde{y}(t)\mathrm{d}t,f\right\rangle\,,
	\end{multline}
	where $x,\tilde{x},y\in C^\infty(\mathbb{R})$, $f\in\mathcal{D}(\mathbb{R})$ while the last term in both equations should be understood as a formal definition.
\end{remark}

\noindent Henceforth we focus on the following notable subclass of polynomial $x\tilde{x}$-functionals.
\begin{Definition}\label{Def: xtildex local functionals}
	We denote by $\mathcal{P}_{x\tilde{x}}^{\textsc{loc}}\subset\operatorname{Pol}_{x\tilde{x}}$ the subspace of polynomial {\bf local} $x\tilde{x}$-functionals.
	More precisely $F\in\mathcal{P}_{x\tilde{x}}^{\textsc{loc}}$ if, for all $x,\tilde{x}\in C^\infty(\mathbb{R})$, $F(x,\tilde{x})\in\mathcal{D}'(\mathbb{R})$ and in particular
	\begin{align}\label{Eq: local functionals condition}
		\langle F(x,\tilde{x}),f\rangle
		=\int_\mathbb{R}f(t)p(x(t),\tilde{x}(t))\mathrm{d}t
		\qquad\forall f\in\mathcal{D}(\mathbb{R})\,,\,
		\forall x,\tilde{x}\in C^\infty(\mathbb{R})\,,
	\end{align}
	where $p$ is a fixed polynomial in the variables $x,\tilde{x}$, with coefficients lying in $C^\infty(\mathbb{R})$. At the same time $\mathcal{P}_{x\tilde{x}}^{\textsc{loc}}$ generates a subalgebra of $\operatorname{Pol}_{x\tilde{x}}$ denoted by $\mathcal{P}_{x\tilde{x}}$, whose elements are referred to as {\bf multi-local} polynomial functionals.
\end{Definition}	

For later convenience, notice that a generic element $F\in\mathcal{P}_{x\tilde{x}}$ is a finite linear combination of monomials $x(t_1)\cdots x(t_k)\tilde{x}(s_1)\cdots\tilde{x}(s_\ell)$ with the possibility that some among the base points $t_1,\dots t_k$ and $s_1,\dots s_\ell$ coincide. All coefficients are smooth functions in the underlying variables.

\begin{remark}\label{Rem: tildex-independent-functionals}
	In the following we shall be interested also in the collection of local polynomial functionals which are $\tilde{x}$-independent. We denote them by $\mathcal{P}_x^{\textsc{loc}}\subset\mathcal{P}_{x\tilde{x}}^{\textsc{loc}}$ and, at the same time, $\mathcal{P}_x$ stands for the subalgebra of $\mathcal{P}_{x\tilde{x}}$ generated by $\mathcal{P}^{\textsc{loc}}_x$.
\end{remark}

\begin{remark}\label{Rem: sum_of_polynomials}
	Since in the following we shall consider extensively multilocal polynomial $x\tilde{x}$-functionals $F\in\mathcal{P}_{x\tilde{x}}$, it is convenient to consider the associated integral kernels. Focusing for simplicity and without loss of generality on the generators of $\mathcal{P}_{x\tilde{x}}$, we shall write them as $x(t_1)\cdots x(t_n)\tilde{x}(s_1)\cdots\tilde{x}(s_k)$.
	In addition we will be interested in considering their sum, for which we shall employ a notation which we outline here by means of an exhaustive example. Consider $F_i:C^\infty(\mathbb{R}^2)\to\mathcal{D}^\prime(\mathbb{R})^{\otimes 2}\subset\mathcal{T}^\prime(\mathbb{R})$, $i=1,2$, such that
	$$F_1(x,\tilde{x})=(x\tilde{x})\otimes 1,\quad F_2=1\otimes\tilde{x}^2,$$
	where $1$ denotes the distribution generated by the unit, constant function. In view of the underlying vector space structure we can consider $F_1+F_2:C^\infty(\mathbb{R}^2)\to\mathcal{D}^\prime(\mathbb{R})^{\otimes 2}\subset\mathcal{T}^\prime(\mathbb{R})$. With a slight abuse of notation we denote the associated integral kernel as $x(t_1)\tilde{x}(t_1)+\tilde{x}^2(t_2)$.
\end{remark}

\begin{Example}\label{Exam: interacting vertex}
	Among the local, polynomial $x\tilde{x}$-functionals as per Definition \ref{Def: xtildex local functionals}, a notable example in the analysis of the MSR approach can be constructed as follows. Let $\chi\in\mathcal{D}(\mathbb{R})$, $\sigma\in\mathbb{R}$ while $\alpha\equiv\alpha(x(t),t)$ and $\beta\equiv\beta(x(t),t)$ are two polynomial functions in $x(t)$ with smooth time-dependent coefficients and $\vartheta_0\in[0,1]$.
	We call \textbf{interacting vertex} the element $V_{\alpha,\beta,\vartheta_0}\in\mathcal{P}_{x\tilde{x}}^{\textsc{loc}}$, whose integral kernel $V_{\alpha,\beta,\vartheta_0}(x,\tilde{x})$ reads
	\begin{align}\label{Eq: MSR interacting potential - vartheta0 generic case}
		V_{\alpha,\beta,\vartheta_0}(x,\tilde{x})
		=\tilde{x}[\chi\alpha
		+\sigma\vartheta_0\chi^2\beta\beta_1]
		+\frac{\sigma}{2}\chi^2\tilde{x}^2\beta^2\,,
	\end{align}
	where $\beta_1:=\partial_x\beta$.
	The nomenclature will become clear once we discuss the path-integral formulation of the MSR approach.
	Observe that $V_{\alpha,\beta,\vartheta_0}$ is the functional appearing in the right-hand side of Equation \eqref{Eq: SDE-MSR correspondence - delta and G product for local functionals}.
	Moreover, on account of the cut-off function $\chi(t)$, $\forall x,\tilde{x}\in C^\infty(\mathbb{R})$, $V_{\alpha,\beta,\vartheta_0}(x,\tilde{x})\in\mathcal{E}^\prime(\mathbb{R})$ and therefore it is meaningful to consider
	\begin{align}\label{Eq: MSR interacting potential - smeared version}
		\langle V_{\alpha,\beta,\vartheta_0}\rangle\colon
		C^\infty(\mathbb{R})^2\to\mathbb{C}\,,
		\qquad
		\langle V_{\alpha,\beta,\vartheta_0}\rangle (x,\tilde{x})
		:=\langle V_{\alpha,\beta,\vartheta_0}(x,\tilde{x}),1\rangle\,.
	\end{align}
\end{Example}

It is convenient to depict the generators of $\mathcal{P}_{x\tilde{x}}$ by means of graph endowed with a set of rules, which are also covering the possibility of taking linear combinations of monomials and of integrating some of their arguments. 

\begin{graphrules}\label{Graph: Rules 1}
	Let $f(t_1,\ldots,t_k)$ be a polynomial in $x(t_1),\ldots x(t_k)$, $\tilde{x}(t_1)\ldots,\tilde{x}(t_k)$. We associate to it a graph according to the following rules:
	\begin{enumerate}
		\item
		
		The evaluation points $t_1,\ldots, t_k$ of $f(t_1,\ldots,t_k)$ are represented by labeled vertices:
		\begin{tikzpicture}[every node/.style={sloped,allow upside down}]
			\filldraw (0,0) circle (1.5pt) node[left] {$t_1$};
		\end{tikzpicture}
		$\ldots$
		\begin{tikzpicture}[every node/.style={sloped,allow upside down}]
			\filldraw (0,0) circle (1.5pt) node[left] {$t_k$};
		\end{tikzpicture}
		
		\item
		$f(t)=x(t)$ is represented by a "straight external leg"
		from the vertex
		\begin{tikzpicture}[every node/.style={sloped,allow upside down}]
			\filldraw (0,0) circle (1.5pt) node[left] {$t$};
		\end{tikzpicture}\,:
		\begin{tikzpicture}[every node/.style={sloped,allow upside down}]
			\filldraw (0,0) circle (1.5pt) node[left] {$t$};
			\draw (0,0) -- (0.5,0);
		\end{tikzpicture}
		
		\item
		$f(t)=\tilde{x}(t)$ is represented by a "snaky external leg"
		from the vertex
		\begin{tikzpicture}[every node/.style={sloped,allow upside down}]
			\filldraw (0,0) circle (1.5pt) node[left] {$t$};
		\end{tikzpicture}\;:
		\begin{tikzpicture}[every node/.style={sloped,allow upside down}]
			\filldraw (0,0) circle (1.5pt) node[left] {$t$};
			\draw (0,0) edge[snake] (1,0);
		\end{tikzpicture}
		
		\item
		$f(t)g(t')$ is represented by the disjoint union of the graphs associated to $f(t)$ and $g(t')$.
		If $t=t'$ the graphs are joined at the vertex $t$.
		
		\item
		$f(t)+g(t)$ is represented by the formal sum of the graphs associated to $f$ and $g$, see also Remark \ref{Rem: sum_of_polynomials}.
		
		\item
		If $f(t_1,\ldots,t_k)=\int g(t_1,\ldots,t_{k+1})\mathrm{d}t_{k+1}$, the graph of $f$ is obtained from the graph of $g$ by removing the label from the vertex $t_{k+1}$.
	\end{enumerate}
\end{graphrules}

\begin{Example}\label{Exam: graph}
	Following slavishly the above rules we can depict $x(t_1)\tilde{x}(t_1)+\tilde{x}(t_2)^2+x(t_3)^3+x(t_4)\tilde{x}(t_5)$ as follows
	\begin{center}
		\begin{tikzpicture}[every node/.style={sloped,allow upside down}]
			\filldraw (0,0) circle (1.5pt) node[left] {$t_1$};
			\draw[snake] (0,0) -- (0.5,0.5);
			\draw (0,0) -- (-0.5,0.5);
		\end{tikzpicture}
		$+$
		\begin{tikzpicture}[every node/.style={sloped,allow upside down}]
			\filldraw (0,0) circle (1.5pt) node[left] {$t_2$};
			\draw[snake] (0,0) -- (0.5,0.5);
			\draw[snake] (0,0) -- (-0.5,0.5);
		\end{tikzpicture}
		$+$
		\begin{tikzpicture}[every node/.style={sloped,allow upside down}]
			\filldraw (0,0) circle (1.5pt) node[left] {$t_3$};
			\draw (0,0) -- (0.5,0.5);
			\draw (0,0) -- (0,0.5);
			\draw (0,0) -- (0.5,0);
		\end{tikzpicture}
		$+$
		\begin{tikzpicture}[every node/.style={sloped,allow upside down}]
			\filldraw (0,0) circle (1.5pt) node[left] {$t_4$};
			\filldraw (0.5,0) circle (1.5pt) node[right] {$t_5$};
			\draw (0,0) -- (0,0.5);
			\draw[snake] (0.5,0) -- (0.5,0.6);
		\end{tikzpicture}
	\end{center}
\end{Example}

\noindent To conclude the section we introduce the map $\Gamma_{G}$ already mentioned in Section \ref{Sec: introduction}. We recall that, in our construction, $\Gamma_G$ is the main ingredient necessary to give via the right-hand side of Equation \eqref{Eq: correlation and G0 product} a rigorous definition of the expectation value of polynomial functionals $F\in\mathcal{P}_{x\tilde{x}}$. These are otherwise computed only via formal path integral methods. The following definitions are much inspired by the analysis in \cite{Bonicelli-Dappiaggi-Rinaldi-2021}, see also \cite{Carfora-Dappiaggi-Drago-Rinaldi-2020,Dappiaggi-Drago-Rinaldi-Zambotti-2022}.

\begin{Definition}\label{Def: Gamma G map}
	We denote by $\Gamma_G\colon\mathcal{P}_{x\tilde{x}}\to\operatorname{Pol}_{x\tilde{x}}$ the linear map defined by
	\begin{align*}
		\Gamma_G:=\exp[\Upsilon_G]
		\qquad
		\Upsilon_G(F):=
		\int_{\mathbb{R}^2} G(t,t')\frac{\delta^2 F}{\delta x(t)\delta \tilde{x}(t')}\mathrm{d}t\mathrm{d}t'
		\,,
	\end{align*}
	where $G(t,t'):=\vartheta(t-t')$ is the Heaviside function and where we adopted the convention $G(t,t)=\vartheta_0:=0$ while $\exp[\Upsilon_G]$ denotes the exponential of $\Upsilon_G$ in the operator sense.
\end{Definition}
At its core, the maps $\Upsilon_G$, $\Gamma_G$ implement a "Wick rule" for the fields $x,\tilde{x}$.
In particular we have
\begin{align*}
	\Upsilon_G(x(t_1)x(t_2))
	=0\,,
	\qquad
	\Upsilon_G(\tilde{x}(t_1)\tilde{x}(t_2))
	=0\,,
	\qquad
	\Upsilon_G(x(t_1)\tilde{x}(t_2))
	=G(t_1,t_2)\,.
\end{align*}

\begin{remark}
	Notice that the integral defining $\Upsilon_G(F)$ converges because for all $f\in\mathcal{D}(\mathbb{R}^n)$ we have that
	\begin{align*}
		\frac{\delta^2}{\delta x(t)\delta\tilde{x}(t')}\langle F(x,\tilde{x}),f\rangle\,,
	\end{align*}
	vanishes if $(t,t')\notin (-\ell,\ell)^2$, where $\ell>0$ is such that $\operatorname{supp}(f)\subset(-\ell,\ell)^n$.
\end{remark}

	\noindent In the following proposition we adapt to the case in hand a standard result concerning algebra deformations induced by maps which are structurally akin to $\Gamma_{G}$ as in Definition \ref{Def: Gamma G map}, see {\it e.g.} \cite{Bonicelli-Dappiaggi-Rinaldi-2021,Dappiaggi-Drago-Rinaldi-2020,Dappiaggi-Drago-Rinaldi-Zambotti-2022,Keller-2009}.
	
	\begin{proposition}\label{Prop: G product}
		Let $\mathcal{P}_{G}:=\Gamma_{G}(\mathcal{P}_{x\tilde{x}})$.
		Then $\mathcal{P}_{G}$ is an associative algebra when equipped with the $\cdot_G$ product such that, for all $F_1,F_2\in\mathcal{P}_{x\tilde{x}}$
		\begin{align}\label{Eq: G-product}
			F_1\cdot_G F_2
			:=\Gamma_{G}\left[
			\Gamma_{G}^{-1}(F_1)\Gamma_{G}^{-1}(F_2)
			\right]\,.
		\end{align}
	\end{proposition}
	
	\begin{remark}\label{Rem: on theta0}
		If $F\in\mathcal{P}_{x\tilde{x}}^{\textsc{loc}}$, $\dfrac{\delta^2 F}{\delta x(t)\delta\tilde{x}(t')}$ is a distribution supported only on $t=t'$.
		This implies that $\Upsilon_G(F)$ involves formally the product of the Heaviside function $\vartheta(t-t')$ with the Dirac delta distribution $\delta(t-t')$, \textit{e.g.}
		\begin{align*}
			\Upsilon_G(x(t)\tilde{x}(t'))=x(t)\tilde{x}(t')+\vartheta(t-t^\prime)\delta(t-t^\prime)\,.
		\end{align*}
		The product $\vartheta(t-t^\prime)\delta(t-t^\prime)$, which we denoted in Definition \ref{Def: Gamma G map} with a slight abuse of notation as $G(t,t)$, is ill-defined since different regularizations $\vartheta_\varepsilon\in C^\infty(\mathbb{R})$ of $\vartheta$ -- \textit{i.e.} such that $\vartheta_\varepsilon\underset{\varepsilon\to 0}{\longrightarrow}\vartheta$ in $\mathcal{D}'(\mathbb{R})$ -- lead to different limiting values $\overline{\vartheta}:=\lim_{\varepsilon\to 0}\vartheta_\varepsilon\delta$.
		In Definition \ref{Def: Gamma G map} we are implicitly considering an arbitrary but fixed choice for $\overline{\vartheta}$.
		Different values for $\overline{\vartheta}$ are always admissible leading in turn to different maps $\Gamma_{G,\overline{\vartheta}}$.
		In the following we adopt the convention, $\overline{\vartheta}=0$, which simplifies the graphical expansion of the action of $\Gamma_{G}$, \textit{cf.} the graph rules \ref{Graph: Rules 2}.
		In addition, observe that, in view of Proposition \ref{Prop: G product}, different choices of $\overline{\vartheta}$ lead to isomorphic algebras. 
	\end{remark}

	
	\begin{graphrules}\label{Graph: Rules 2}
		The map $\Gamma_{G}$ is completely determined by its action on monomials, which can be represented in terms of a graphical expansion as follows:
		\begin{enumerate}
			\item
			Consider the graph associated with $x(t_1)\cdots x(t_n)\tilde{x}(s_1)\cdots\tilde{x}(s_k)$ as per the graph rules \ref{Graph: Rules 1};
			
			\item
			To compute the graphical counterpart of
			$\Upsilon_G[x(t_1)\cdots x(t_n)\tilde{x}(s_1)\cdots\tilde{x}(s_k)]$ replace a pair made by a straight external line
			\begin{tikzpicture}[every node/.style={sloped,allow upside down}]
				\filldraw (0,0) circle (1.5pt) node[left] {$t$};
				\draw (0,0) -- (0.5,0);
			\end{tikzpicture}
			and a snaky external line
			\begin{tikzpicture}[every node/.style={sloped,allow upside down}]
				\filldraw (0,0) circle (1.5pt) node[left] {$s$};
				\draw[snake] (0,0) -- (1,0);
			\end{tikzpicture}
			with a directed edge
			\begin{tikzpicture}[every node/.style={sloped,allow upside down}]
				\filldraw (0,0) circle (1.5pt) node[left] {$t$};
				\filldraw (0.5,0) circle (1.5pt) node[right] {$s$};
				\draw (0,0) -- node {\midarrow} (0.5,0);
			\end{tikzpicture}
			with source $t$ and target $s$.
			This has to be repeated for all pairs of vertices, unless $t=s$.
			For example
			\begin{center}
				$\Upsilon_G[x(s_1)x(s_2)\tilde{x}(s_3)]=$
				\begin{tikzpicture}[every node/.style={sloped,allow upside down}]
					\filldraw (0,0) circle (1.5pt) node[left] {$s_1$};
					\draw (0,0) -- (0,0.5);
					\filldraw (1,0) circle (1.5pt) node[left] {$s_2$};
					\draw (1,0) -- node {\midarrow} (2,0);
					\filldraw (2,0) circle (1.5pt) node[right] {$s_3$};
				\end{tikzpicture}
				$+$
				\begin{tikzpicture}[every node/.style={sloped,allow upside down}]
					\filldraw (0,0) circle (1.5pt) node[left] {$s_1$};
					\draw (0,0) -- node {\midarrow} (1,0);
					\filldraw (1,0) circle (1.5pt) node[right] {$s_3$};
					\filldraw (2,0) circle (1.5pt) node[right] {$s_2$};
					\draw (2,0) -- (2,0.5);
				\end{tikzpicture}
			\end{center}
			
			\item
			To convert a given graph in the expansion to a closed expression:
			\begin{enumerate}
				\item 
				Write a factor $G(t,s)$ for each directed edge
				\begin{tikzpicture}[every node/.style={sloped,allow upside down}]
					\filldraw (0,0) circle (1.5pt) node[left] {$t$};
					\filldraw (0.5,0) circle (1.5pt) node[right] {$s$};
					\draw (0,0) -- node {\midarrow} (0.5,0);
				\end{tikzpicture}
				with source $t$ and target $s$, with the convention that $G(t,s)=0$ if $t=s$;
				\item
				Write a factor $x(t)$ for every straight external line
				\begin{tikzpicture}[every node/.style={sloped,allow upside down}]
					\filldraw (0,0) circle (1.5pt) node[left] {$t$};
					\draw (0,0) -- (0.5,0);
				\end{tikzpicture}
				associated with the vertex $t$.
				Similarly a factor $\tilde{x}(s)$ is assigned to each snaky external line
				\begin{tikzpicture}[every node/.style={sloped,allow upside down}]
					\filldraw (0,0) circle (1.5pt) node[left] {$s$};
					\draw[snake] (0,0) -- (1,0);
				\end{tikzpicture}
				linked with the vertex $s$.
			\end{enumerate}
		\end{enumerate}
	\end{graphrules}
	\begin{Example}\label{Ex: GammaG grphs}
		To make the reader more comfortable with the notation introduced above we propose examples showing explicit computations of the action of $\Gamma_{G}$ as well as their graphical representation.
		\begin{center}
			$\Gamma_{G}(x(t)\tilde{x}(s))=x(t)\tilde{x}(s)+G(t,s)=$
			\xvertex{$t$}\tildexvertex{$s$}
			$+$
			\begin{tikzpicture}[every node/.style={sloped,allow upside down}]
				\filldraw (0,0) circle (1.5pt) node[left] {$t$};
				\draw (0,0) -- node {\midarrow} (1,0);
				\filldraw (1,0) circle (1.5pt) node[right] {$s$};
			\end{tikzpicture}
			\\
			$\Gamma_{G}(x(t)\tilde{x}(t))
			=x(t)\tilde{x}(t)+0=$
			\begin{tikzpicture}[every node/.style={sloped,allow upside down}]
				\filldraw (0,0) circle (1.5pt) node[left] {$t$};
				\draw (0,0)  edge (0,0.5);
				\draw[snake] (0,0) -- (1,0);
			\end{tikzpicture}
			\\
			\bigskip
			$\Gamma_{G}(x(t)\tilde{x}(s)x(s)^2)
			=x(t)\tilde{x}(s)x(s)^2
			+G(t,s)x(s)^2$
			$=$
			\begin{tikzpicture}[every node/.style={sloped,allow upside down}]
				\filldraw (0,0) circle (1.5pt) node[right] {$t$};
				\draw (0,0) -- (0,0.5);
				\filldraw (1,0) circle (1.5pt) node[right] {$s$};
				\draw (1,0) -- (1,0.5);
				\draw (1,0) -- (1+0.5,0.5);
				\draw[snake] (1,0) -- (1-0.5,0.5);
			\end{tikzpicture}
			$+$
			\begin{tikzpicture}[every node/.style={sloped,allow upside down}]
				\filldraw (0,0) circle (1.5pt) node[left] {$t$};
				\draw (0,0) -- node {\midarrow} (1,0);
				\filldraw (1,0) circle (1.5pt) node[right] {$s$};
				\draw (1,0) -- (1+0.5,0.5);
				\draw (1,0) -- (1,0.5);
			\end{tikzpicture}
			$\Gamma_{G}(x(t)^2\tilde{x}(s)^2)=x(t)^2\tilde{x}(s)^2
			+4G(t,s)x(t)\tilde{x}(s)
			+2G(t,s)^2$
			$=$
			\begin{tikzpicture}[every node/.style={sloped,allow upside down}]
				\filldraw (0,0) circle (1.5pt) node[left] {$t$};
				\draw (0,0) -- (-0.5,0.5);
				\draw (0,0) -- (0.5,0.5);
				\filldraw (1.5,0) circle (1.5pt) node[left] {$s$};
				\draw[snake] (1.5,0) -- (1,0.5);
				\draw[snake] (1.5,0) -- (2,0.5);
			\end{tikzpicture}
			\\
			$+4$
			\begin{tikzpicture}[every node/.style={sloped,allow upside down}]
				\filldraw (0,0) circle (1.5pt) node[left] {$t$};
				\draw (0,0) -- (0,0.5);
				\draw (0,0) -- node {\midarrow} (1,0);
				\filldraw (1,0) circle (1.5pt) node[right] {$s$};
				\draw[snake] (1,0) -- (1,0.5);
			\end{tikzpicture}
			$+2$
			\begin{tikzpicture}[every node/.style={sloped,allow upside down}]
				\filldraw (0,0) circle (1.5pt) node[left] {$t$};
				\draw (0,0) edge [bend left] node {\midarrow} (1,0);
				\draw (0,0) edge [bend right] node {\midarrow} (1,0);
				\filldraw (1,0) circle (1.5pt) node[right] {$s$};
			\end{tikzpicture}
		\end{center}
		
		Notice that, for all $F\in\mathcal{P}_{x\tilde{x}}$, the graph corresponding to $\Gamma_{G}(F)$ is a directed graph with neither closed directed paths nor closed loops.
		Indeed, closed directed paths like
		
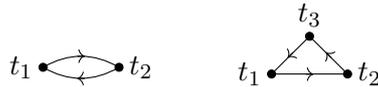
\begin{center}
			\begin{tikzpicture}[every node/.style={sloped,allow upside down}]
				\filldraw (0,0) circle (1.5pt) node[left] {$t_1$};
				\filldraw (1,0) circle (1.5pt) node[right] {$t_2$};
				\draw (0,0)  edge [bend left]  node {\midarrow} (1,0);
				\draw (1,0)  edge [bend left] node {\midarrow} (0,0);
			\end{tikzpicture}
			\qquad
			\begin{tikzpicture}[every node/.style={sloped,allow upside down}]
				\filldraw (0,0) circle (1.5pt) node[left] {$t_1$};
				\filldraw (1,0) circle (1.5pt) node[right] {$t_2$};
				\filldraw (0.5,0.5) circle (1.5pt) node[above] {$t_3$};
				\draw (0,0)  edge  node {\midarrow} (1,0);
				\draw (1,0)  edge  node {\midarrow} (0.5,0.5);
				\draw (0.5,0.5)  edge  node {\midarrow} (0,0);
			\end{tikzpicture}
		\end{center}
		vanish on account of the retardation properties of $G$, since \textit{e.g.} $\vartheta(t_1-t_2)\vartheta(t_2-t_3)\vartheta(t_3-t_1)=0$.
		Similarly, closed loops are excluded because they account for factors proportional to $\vartheta_0=0$.
	\end{Example}
	
	\subsection{Expectation values in the algebraic MSR formalism}
	\label{Subsec: expectation values in the algebraic MSR formalism}
	
	In this section  we employ the map $\Gamma_{G}$ introduced in Definition \ref{Def: Gamma G map} to give meaning to the right hand side of Equation \eqref{Eq: formal path integral and Gamma G0 product}. We recall that the aim is to compute in the algebraic framework both the expectation values and the correlations of the underlying field configurations in the MSR formalism.
	
	\begin{Definition}\label{Def: MSR expectation value}
		Let $V_{\alpha,\beta,\vartheta_0}\in\mathcal{P}_{x\tilde{x}}^{\textsc{loc}}$ be as per Equation \eqref{Eq: MSR interacting potential - vartheta0 generic case}.
		We define $\llangle\;\rrangle_{\alpha,\beta,\vartheta_0}\colon\mathcal{P}_x\to\mathcal{T}'(\mathbb{R})[[\chi]]$ as
		\begin{align}\label{Eq: MSR expectation value}
			\llangle F\rrangle_{\alpha,\beta,\vartheta_0}
			:=\Gamma_{G}[e^{\langle V_{\alpha,\beta,\vartheta_0}\rangle}F]
			\bigg|_{\substack{x=x_0\\\tilde{x}=0}}
			=[e_{\cdot_G}^{\langle V_{\alpha,\beta,\vartheta_0}\rangle}\cdot_GF]
			\bigg|_{\substack{x=x_0\\\tilde{x}=0}}\,,
		\end{align}
		where $\langle V_{\alpha,\beta,\vartheta_0}\rangle:=\langle V_{\alpha,\beta,\vartheta_0}(x,\tilde{x}),1\rangle$.
		Here in the second equality we used Proposition \ref{Prop: G product} together with the identity $\Gamma_{G}(F)=F$ for all $F\in\mathcal{P}_x$. The space $\mathcal{T}'(\mathbb{R})[[\chi]]$ consists of formal power series in the cut-off $\chi$ with coefficients lying in $\mathcal{T}'(\mathbb{R})$.
	\end{Definition}
	
	\noindent Observe that, if $f\in\mathcal{D}(\mathbb{R})^{\otimes k}$, $k\in\mathbb{N}$, $\langle \llangle F\rrangle_{\alpha,\beta,\vartheta_0},f\rangle$ reads
	\begin{align*}
		\langle \llangle F\rrangle_{\alpha,\beta,\vartheta_0},f\rangle
		:=\sum_{n\geq 0}\frac{1}{n!}
		\langle[(V_{\alpha,\beta,\vartheta_0})_{\cdot_G}^n\cdot_G F]\bigg|_{\substack{x=0\\\tilde{x}=0}},
		1^{\otimes n}\otimes f\rangle
		=:\sum_{n\geq 0}\frac{1}{n!}
		\langle[\langle V_{\alpha,\beta,\vartheta_0}\rangle_{\cdot_G}^n\cdot_G F]\bigg|_{\substack{x=0\\\tilde{x}=0}},
		f\rangle\,.
	\end{align*}
	To wit, for the particular case $\alpha=0$, $\beta=1$ and $x_0=0$ we may compute, at first order in perturbation theory,
	\begin{align*}
		\llangle x(t_1)x(t_2)\rrangle_{0,1,\vartheta_0}
		=\frac{\sigma^2}{2}\Gamma_G[\langle\chi^2\tilde{x}^2\rangle x(t_1)x(t_2)]\Big|_{\substack{x=0\\\tilde{x}=0}}
		=\sigma^2\int\limits_{\mathbb{R}} G(t,s)^2\chi(s)^2\mathrm{d}s\,,
	\end{align*}
	which matches with the corresponding expression for $\mathbb{E}_\eta[x_\eta(t_1)x_\eta(t_2)]$.

	\begin{remark}\label{Rmk: vartheta0-generalized interacting vertex; multiplicative ambiguity in MSR}
		At this stage it is worth commenting the interpretation of the parameter $\vartheta_0$ appearing in Definition \ref{Def: MSR expectation value}.
		This is related to the existing ambiguity in solving Equation \eqref{Eq: SDE - general form} when applying the MSR formalism whenever $\beta$ is such that $\beta_1\neq 0$.
		As a matter of fact, it is a standard feature of stochastic differential equations, see {\it e.g.} \cite{Gardiner-1985}, that values of $\vartheta_0\in[0,1]$ correspond to different interpretations of the solution theory of the underlying SDE, $\vartheta_0=0$ being associated to the It\^o convention while $\vartheta_0=1/2$ to the Stratonovich one \cite{Ito-1951,Stratonovich-1966}.
		Within the MSR formalism such ambiguity can be accounted for by considering a $\vartheta_0$-dependent interacting vertex $V_{\alpha,\beta,\vartheta_0}$ as per Equation \eqref{Eq: MSR interacting potential - vartheta0 generic case}.
		The expectation value of a functional $F\in\mathcal{P}_x$ corresponds to $\llangle F\rrangle_{\alpha,\beta,\vartheta_0}$.
	\end{remark}
	
	\begin{remark}
		Equation \eqref{Eq: MSR expectation value} is similar in spirit to the Talay-Tubaro representation \cite{Talay_Tubaro_1990}.
		The latter is based on the observation that, for a suitable $\varphi\in C^\infty(\mathbb{R})$, it holds $\mathbb{E}[\varphi(x_\eta(t))]=e^{-t\mathcal{L}}\varphi(x_0)$, where $\mathcal{L}:=\alpha(x,t)\partial_x+\frac{1}{2}\sigma\beta^2(x,t)\partial_x^2$  is the generator of Equation \eqref{Eq: SDE - general form}.
		Indeed, Equation \eqref{Eq: MSR expectation value} can be seen as a perturbative expansion of the Talay-Tubaro representation for $F=\varphi$ in the It\^o prescription $\vartheta_0=0$. Yet, observe that our formulation extends to arbitrary integration prescriptions. 
	\end{remark}

	\noindent Our next goal is to determine a graphical expansion of $\llangle F\rrangle_{\alpha,\beta,\vartheta_0}$ and we start by discussing a preliminary result whose proof can be found in Appendix \ref{Sec: appendix with proofs}.
	\begin{proposition}\label{Prop: vanishing expectation if tildex is everywhere}
		Let $n\in\mathbb{N}$ and let $F_1,\ldots,F_n\in\mathcal{P}_{x\tilde{x}}^{\textsc{loc}}$ be local $x\tilde{x}$-functionals as per Definition \ref{Def: xtildex local functionals}. 
		Then it holds that
		\begin{align}
			F_j(x,0)=0\;
			\forall j\in{1,\ldots,n}
			\quad\Rightarrow\quad
			\Gamma_G(F_1\cdots F_n)\big|_{\tilde{x}=0}
			=0\,.
		\end{align}
	\end{proposition}
	\noindent Letting $F\in\mathcal{P}_x$ and considering the corresponding expectation value
	\begin{align}\label{Eq: series expansion of MSR expectation value - general case}
		\llangle F\rrangle_{\alpha,\beta,\vartheta_0}
		=\sum_{n\geq 0}\frac{1}{n!}
		[\langle V_{\alpha,\beta,\vartheta_0}\rangle_{\cdot_G}^n\cdot_G F]
		\bigg|_{\substack{x=x_0\\\tilde{x}=0}}\, ,
	\end{align}
	it is convenient to adopt a graph representation of the right hand side. In the following we set the corresponding rules.
	
	\begin{graphrules}\label{Graph: Rules 3}
		
		Taking into account the graph rules \ref{Graph: Rules 1} and \ref{Graph: Rules 2},  we set
		\begin{itemize}
			\item \begin{tikzpicture}
				\node[rectangle,draw,fill=white] at (0,0) {$\alpha$};
			\end{tikzpicture} is the contribution to the graph from $\chi\alpha$,
			\item 	\begin{tikzpicture}
				\node[rectangle split, rectangle split horizontal, rectangle split parts=2,draw,fill=white] at (2,0) {\nodepart{one} $\beta$ \nodepart{two} $\beta$};
			\end{tikzpicture} is that from $\sigma\chi^2\beta^2$,
			\item \begin{tikzpicture}
				\node[rectangle split, rectangle split horizontal, rectangle split parts=2,draw,fill=white] at (2,0) {\nodepart{one} $\beta$ \nodepart{two} $\beta_1$};
			\end{tikzpicture} is that from $\sigma\chi\beta\beta_1$.
		\end{itemize}
		Hence the graph expansion of $\langle V_{\alpha,\beta,\vartheta_0}\rangle$ reads
		\begin{center}
			\begin{tikzpicture}[every node/.style={sloped,allow upside down}]
				\draw (-1,0) node[left] {$\langle V_{\alpha,\beta,\vartheta_0}\rangle=$};
				\draw[snake] (0,0) -- (0,1);
				\node[rectangle,draw,fill=white] at (0,0) {$\alpha$};
				\draw (1,0) node {$+\frac{1}{2}$};
				\draw[snake] (2,0) -- (1.5,1);
				\draw[snake] (2,0) -- (2.5,1);
				\node[rectangle split, rectangle split horizontal, rectangle split parts=2,draw,fill=white] at (2,0) {\nodepart{one} $\beta$ \nodepart{two} $\beta$};
				\draw (3,0) node {$+\vartheta_0$};
				\draw[snake] (4.25,0) -- (4.25,1);
				\node[rectangle split, rectangle split horizontal, rectangle split parts=2,draw,fill=white] at (4,0) {\nodepart{one} $\beta$ \nodepart{two} $\beta_1$};
			\end{tikzpicture}
		\end{center}
	\end{graphrules}
	
	Observe that, when computing $[\langle V_{\alpha,\beta,\vartheta_0}\rangle_{\cdot_G}^n\cdot_G F]
	\bigg|_{\substack{x=x_0\\\tilde{x}=0}}$, the resulting graphical computation requires that all interacting vertices $\langle V_{\alpha,\beta,\vartheta_0}\rangle$ are connected with the graph associated with $F$.
	In particular, since evaluation at $\tilde{x}=0$ is considered, all snaky external lines of $\langle V_{\alpha,\beta,\vartheta_0}\rangle$ have to be converted to directed edges using the graph rules  \ref{Graph: Rules 2}.
	This entails that, if we denote by $\mathcal{G}_{\alpha,\beta}^{(n,F)}$ the set of graphs obtained by considering the graph expansion of
	\begin{align*}
		F\cdot_G
		\underbrace{\langle V_{\alpha,\beta,\vartheta_0}\rangle\cdot_G\ldots\cdot_G
			\langle V_{\alpha,\beta,\vartheta_0}\rangle}_{n}\bigg|_{\substack{x=x_0\\\tilde{x}=0}}\,,
	\end{align*}
	then the expansion of the expectation value $\llangle F\rrangle_{\alpha,\beta,\vartheta_0}$ reads
	\begin{align}\label{Eq: graph series expansion of MSR expectation value - general case}
		\llangle F\rrangle_{\alpha,\beta,\vartheta_0}
		=\sum_{n\geq 0}\sum_{\gamma\in\mathcal{G}_{\alpha,\beta}^{(n,F)}}\frac{1}{|\operatorname{Aut}_{\mathcal{G}_{\alpha,\beta}^{(n,F)}}(\gamma)|}\gamma\,.
	\end{align}
	In this identity $|\operatorname{Aut}_{\mathcal{G}_{\alpha,\beta}^{(n,F)}}(\gamma)|$ is the cardinality of the automorphism group of $\gamma$. An \textbf{automorphism of} $\boldsymbol{\gamma}\in\mathcal{G}_{\alpha,\beta}^{(n,F)}$ is an arbitrary permutation of the $n$ interacting vertices $\langle V_{\alpha,\beta,\vartheta_0}\rangle$ which preserves the number and orientation of the directed edges.
	The connection between Equations \eqref{Eq: series expansion of MSR expectation value - general case} and \eqref{Eq: graph series expansion of MSR expectation value - general case} is realized by observing that any $\gamma\in\mathcal{G}_{\alpha,\beta}^{(n,F)}$ appears exactly $\big(n!/|\operatorname{Aut}_{\mathcal{G}_{\alpha,\beta}^{(n,F)}}(\gamma)|\big)$-times in Equation \eqref{Eq: series expansion of MSR expectation value - general case}.
	More precisely, let us consider $\alpha=0$, $\beta=1$, $x_0=0$ and
	\begin{align*}
		x(t_1)x(t_2)\cdot_G\langle V_{0,1,\vartheta_0}\rangle\,.
	\end{align*}
	In this case the resulting contribution is obtained by considering the graphs
	\begin{center}
		\begin{tikzpicture}[every node/.style={sloped,allow upside down}]
			\filldraw (1.5,1) circle (1pt) node[left] {$t_1$};
			\filldraw (2.5,1) circle (1pt) node[right] {$t_2$};
			\draw (1.5,1) edge node {\midarrow} (2,0);
			\draw (2.5,1) edge node {\midarrow} (2,0);
			\node[rectangle split, rectangle split horizontal, rectangle split parts=2,draw,fill=white] at (2,0) {\nodepart{one} $\beta$ \nodepart{two} $\beta$};
		\end{tikzpicture}
		\qquad
		\begin{tikzpicture}[every node/.style={sloped,allow upside down}]
			\filldraw (1.5,1) circle (1pt) node[left] {$t_2$};
			\filldraw (2.5,1) circle (1pt) node[right] {$t_1$};
			\draw (1.5,1) edge node {\midarrow} (2,0);
			\draw (2.5,1) edge node {\midarrow} (2,0);
			\node[rectangle split, rectangle split horizontal, rectangle split parts=2,draw,fill=white] at (2,0) {\nodepart{one} $\beta$ \nodepart{two} $\beta$};
		\end{tikzpicture}
	\end{center}
	Since there is a single interacting vertex the automorphism group is trivial.
	It is also worth observing that the analytic contributions of these graphs coincide.
	We will give a finer description of these graphs in Section \ref{Sec: proof of the SDE-MSR correspondence}.
	
	\begin{Example}\label{Ex: effective graph expansion with linear term}
		The graph expansion discussed in the graph rules \ref{Graph: Rules 3} has the following simple yet effective application.
		Assume that
		\begin{align*}
			\alpha(x(t),t)=\alpha_1(t) x(t)+\alpha'(x(t),t)\,,
		\end{align*}
		where $\alpha_1\in C^\infty(\mathbb{R})$.
		Within this setting we would like to prove that the expectation value $\llangle F\rrangle_{\alpha,\beta,\vartheta_0}$ of any $F\in\mathcal{P}_x$ can be computed as
		\begin{align}\label{Eq: background independence}
			\llangle F\rrangle_{\alpha,\beta,\vartheta_0}
			=\llangle F\rrangle^1_{\alpha',\beta,\vartheta_0}
			:=\Gamma_{G_1}(e^{V_{\vartheta_0}'}F)\big|_{\tilde{x}=0}
			\qquad
			V_{\vartheta_0}'(x,\tilde{x})
			:=\tilde{x}[\chi\alpha'
			+\sigma\vartheta_0\chi^2\beta\beta_1]
			+\frac{\sigma^2}{2}\chi^2\tilde{x}^2\beta^2\,,
		\end{align}
		where $\Gamma_{G_1}$ is defined as per Definition \ref{Def: Gamma G map} with $G(t,t')=\vartheta(t-t')$ replaced by 
		\begin{align*}
			G_1(t,t')
			=e^{\int_{t'}^t\chi(s)\alpha_1(s)\mathrm{d}s}G(t,t')\,,
		\end{align*}
		the fundamental solution of $\frac{\mathrm{d}}{\mathrm{d}t}-\chi(t)\alpha_1(t)$ ---once again with the convention $G_1(t,t)=0$ \textit{cf.} Remark \ref{Rem: on theta0}.
		Notice that the Equation \eqref{Eq: background independence} involves a partial resummation of the perturbative series because $G_1$ depends on $\chi$.
		This is an instance of a feature which is common in quantum field theory and which is referred to as background independence or principle of perturbative agreement, see {\it e.g.} \cite{Drago:2015ewa,Hollands-Wald-2005,Tehrani-Zahn-2020,Zahn-2015}.
		
		The graph expansion of $\llangle\;\rrangle_{\alpha,\beta,\vartheta_0}$ allows to establish a rather direct argument justifying Equation \eqref{Eq: background independence}.
		Indeed we observe that the graph expansion for $\tilde{x}\alpha$ can be written as
		\begin{center}		
			\begin{tikzpicture}[every node/.style={sloped,allow upside down}]
				\draw[snake] (0,0) -- (0,1);
				\node[rectangle,draw,fill=white] at (0,0) {$\alpha$};
				\draw (0.5,0) node[right] {$=$};
				\filldraw (1.5,0) circle (1.5pt) node[below] {$\alpha_1$};
				\draw[snake] (1.5,0) -- (1.5,1);
				\draw (1.5,0) -- (2,0);
				\draw[snake] (3,0) -- (3,1);
				\draw (2,0) node[right] {$+$};
				\node[rectangle,draw,fill=white] at (3,0) {$\alpha'$};
			\end{tikzpicture}
		\end{center}
		Within the graph expansion of $\llangle F\rrangle_{\alpha,\beta,\vartheta_0}$, $F\in\mathcal{P}_x$, one may recollect the graphs which contain the contribution
		\begin{tikzpicture}[every node/.style={sloped,allow upside down}]
			\filldraw (0,0) circle (1.5pt) node[above] {$\alpha_1$};
			\draw[snake] (0,0) -- (-1,0);
			\draw (0,0) -- (0.5,0);
		\end{tikzpicture}
		in such a way that the directed edges
		\begin{tikzpicture}[every node/.style={sloped,allow upside down}]
			\draw (0,0) edge node {\midarrow} (0.5,0);
		\end{tikzpicture}
		associated with a contribution $G$ are replaced by an "effective" directed edge
		\begin{tikzpicture}[every node/.style={sloped,allow upside down}]
			\draw[thick,densely dotted] (0,0) edge node {\midarrow} (0.5,0);
		\end{tikzpicture}
		associated with $G_1$.
		The proof proceeds as follows.
		Let $\gamma_0\in\bigcup_{n\in\mathbb{N}}\mathcal{G}_{\alpha,\beta}^{(n,F)}$ be an arbitrary graph contributing to $\llangle F\rrangle_{\alpha,\beta,\vartheta_0}$.
		Let us fix an edge of $\gamma_0$, here depicted as
		\begin{center}		
			\begin{tikzpicture}[every node/.style={sloped,allow upside down}]
				\draw (0,0) edge node {\midarrow} (1,0);
				\filldraw (0,0) circle (1.5pt) node[below] {$t$};
				\filldraw (1,0) circle (1.5pt) node[below] {$t'$};
				\draw[dashed] (0,0) -- (-0.5,0);
				\draw[dashed] (1,0) -- (1.5,0);
			\end{tikzpicture}
		\end{center}
		Within the sum over $\bigcup_{n\in\mathbb{N}}\mathcal{G}_{\alpha,\beta}^{(n,F)}$ we may now select a graph $\gamma_1$ which is identical to $\gamma_0$ except for the directed edge drawn above, which is replaced by
		\begin{center}		
			\begin{tikzpicture}[every node/.style={sloped,allow upside down}]
				\draw[dashed] (0,0) -- (-0.5,0);
				\filldraw (0,0) circle (1.5pt) node[below] {$t$};
				\draw (0,0) edge node {\midarrow} (1,0);
				\filldraw (1,0) circle (1.5pt) node[below] {$\alpha_1$};
				\draw (1,0) edge node {\midarrow} (2,0);
				\filldraw (2,0) circle (1.5pt) node[below] {$t'$};
				\draw[dashed] (2,0) -- (2.5,0);
			\end{tikzpicture}
		\end{center}
		Thus, $\gamma_1$ differs from $\gamma_0$ only by the insertion of a contribution of the form
		\begin{tikzpicture}[every node/.style={sloped,allow upside down}]
			\filldraw (0,0) circle (1.5pt) node[above] {$\alpha_1$};
			\draw[snake] (0,0) -- (-1,0);
			\draw (0,0) -- (0.5,0);
		\end{tikzpicture}:
		Notice that this raises the perturbation order in $\chi$.
		Following a similar procedure we consider a sequence of graphs $\{\gamma_k\}_k\subset\bigcup_{n\in\mathbb{N}}\mathcal{G}_{\alpha,\beta}^{(n,F)}$, each of which is equal to $\gamma_0$ except for the insertion of $k$ contributions
		\begin{tikzpicture}[every node/.style={sloped,allow upside down}]
			\filldraw (0,0) circle (1.5pt) node[above] {$\alpha_1$};
			\draw[snake] (0,0) -- (-1,0);
			\draw (0,0) -- (0.5,0);
		\end{tikzpicture}
		between the considered vertices.
		The sum $\sum_n\gamma_n$ is recollected as a unique graph $\gamma^1_0$ where the directed edge
		\begin{tikzpicture}[every node/.style={sloped,allow upside down}]
			\draw (0,0) edge node {\midarrow} (0.5,0);
			\filldraw (0,0) circle (1.5pt) node[left] {$t$};
			\filldraw (0.5,0) circle (1.5pt) node[right] {$t'$};
		\end{tikzpicture}
		has been replaced by the "effective" edge
		\begin{center}
			\begin{tikzpicture}[every node/.style={sloped,allow upside down}]
				\filldraw (-2,0) circle (1.5pt) node[below] {$t$};
				\draw[thick,densely dotted] (-2,0) edge node {\midarrow} (-1,0);
				\filldraw (-1,0) circle (1.5pt) node[below] {$t'$};
				\filldraw (-0.5,0) node {$=$};
				\draw (0,0) edge node {\midarrow} (1,0);
				\filldraw (0,0) circle (1.5pt) node[below] {$t$};
				\filldraw (1,0) circle (1.5pt) node[below] {$t'$};
				\filldraw (1,0) node[right] {$+$};
			\end{tikzpicture}
			\begin{tikzpicture}[every node/.style={sloped,allow upside down}]
				\filldraw (0,0) circle (1.5pt) node[below] {$t$};
				\draw (0,0) edge node {\midarrow} (1,0);
				\draw (1,0) edge node {\midarrow} (2,0);
				\filldraw (1,0) circle (1.5pt) node[below] {$\alpha_1$};
				\filldraw (2,0) circle (1.5pt) node[below] {$t'$};
				\filldraw (2,0) node[right] {$+$};
			\end{tikzpicture}
			\begin{tikzpicture}[every node/.style={sloped,allow upside down}]
				\filldraw (0,0) circle (1.5pt) node[below] {$t$};
				\draw (0,0) edge node {\midarrow} (1,0);
				\filldraw (1,0) circle (1.5pt) node[below] {$\alpha_1$};
				\draw (1,0) edge node {\midarrow} (2,0);
				\filldraw (2,0) circle (1.5pt) node[below] {$\alpha_1$};
				\draw (2,0) edge node {\midarrow} (3,0);
				\filldraw (3,0) circle (1.5pt) node[below] {$t'$};
				\filldraw (3,0) node[right] {$+\ldots+$};
			\end{tikzpicture}
			\begin{tikzpicture}[every node/.style={sloped,allow upside down}]
				\filldraw (0,0) circle (1.5pt) node[below] {$t$};
				\draw (0,0) edge node {\midarrow} (1,0);
				\filldraw (1,0) circle (1.5pt) node[below] {$\alpha_1$};
				\draw[dotted] (1,0) edge (2,0);
				\draw (1,0) edge node {\midarrow} (2,0);
				\draw[dotted] (2,0) edge (3,0);
				\filldraw (2.5,0) node[above] {$n$};
				\draw (3,0) edge node {\midarrow} (4,0);
				\draw (4,0) edge node {\midarrow} (5,0);
				\filldraw (4,0) circle (1.5pt) node[below] {$\alpha_1$};
				\filldraw (5,0) circle (1.5pt) node[below] {$t'$};
				\filldraw (5,0) node[right] {$+\ldots$};
			\end{tikzpicture}
		\end{center}
		The analytic contribution associated with
		\begin{tikzpicture}[every node/.style={sloped,allow upside down}]
			\draw[thick,densely dotted] (0,0) edge node {\midarrow} (0.5,0);
		\end{tikzpicture}
		can be computed explicitly, namely
		\begin{multline*}
			G(t,t')\Bigl(1
			+\sum_{n=1}^\infty
			\int_{t\geq s_1\geq \ldots\geq s_n\geq t'} \chi(s_1)\alpha_1(s_1)\cdots\chi(s_n)\alpha_1(s_n)\mathrm{d}s_1\cdots\mathrm{d}s_n\,\Bigr)
			\\=G(t,t')\sum_{n=0}^\infty\frac{1}{n!}
			\bigg[\int_{t'}^t\chi(s)\alpha_1(s)\mathrm{d}s\bigg]^n
			=e^{\int_{t'}^t\chi(s)\alpha_1(s)\mathrm{d}s}G(t,t')\,,
		\end{multline*}
		which coincides with the fundamental solution $G_1(t,t')$ of $\dfrac{\mathrm{d}}{\mathrm{d}t}-\chi(t)\alpha_1(t)$.
		
		This shows how to change a single directed edge
		\begin{tikzpicture}[every node/.style={sloped,allow upside down}]
			\draw (0,0) edge node {\midarrow} (0.5,0);
		\end{tikzpicture}
		of a given graph $\gamma_0$
		by an "effective" directed edge
		\begin{tikzpicture}[every node/.style={sloped,allow upside down}]
			\draw[thick, densely dotted] (0,0) edge node {\midarrow} (0.5,0);
		\end{tikzpicture}.
		Proceeding analogously for all edges and then for all graphs $\gamma\in\bigcup_{n\in\mathbb{N}}\mathcal{G}_{\alpha,\beta}^{(n,F)}$ Equation \eqref{Eq: background independence} descends.
	\end{Example}

	We conclude this section by proving a proposition of independent interest, discussing a factorization property of the functional $V_{\alpha,\beta,\vartheta_0}$ introduced in Equation \eqref{Eq: MSR interacting potential - vartheta0 generic case}.
	The proof can be found in Appendix \ref{Sec: appendix with proofs}.
	
	\begin{proposition}\label{Prop: alphabeta expectation value useful identity}
		For all $F\in\mathcal{P}_x$ it holds
		\begin{align}\label{Eq: alphabeta expectation value useful identity}
			\llangle F\rrangle_{\alpha,\beta,\vartheta_0}
			=\Gamma_G\left[T_{\frac{\sigma}{2}Q_{\chi^2\beta^2}}\left(
			Fe^{\langle\chi,\tilde{x}\alpha\rangle+\langle\chi^2,\sigma\vartheta_0\tilde{x}\beta\beta_1\rangle}
			\right)\right]\bigg|_{\tilde{x}=0}\,,
		\end{align}
		where
		\begin{align}
			\label{Eq: TQ-map}
			T_{\frac{\sigma}{2}Q_{\chi^2\beta^2}}
			:=\exp_T\left[
			\Upsilon_{\frac{\sigma}{2} Q_{\chi^2\beta^2}}\right]\,,
			&\qquad
			\Upsilon_{\frac{\sigma}{2}Q_{\chi^2\beta^2}}(F)
			:=\int\limits_{\mathbb{R}^2}\frac{\sigma}{2} Q_{\chi^2\beta^2}(t,t')\frac{\delta^2 F}{\delta x(t)\delta x(t')}\mathrm{d}t\mathrm{d}t'
			\\
			\label{Eq: Qbeta-operator}
			Q_{\chi^2\beta^2}(t,t';x)
			&:=\int\limits_{\mathbb{R}} G(t,s)\chi^2(s)\beta^2(x(s),s)G(t',s)\mathrm{d}s\,.
		\end{align}
		Here $\exp_T$ denotes the \textbf{time-ordered exponential} defined by
		\begin{align*}
			\exp_T\left[
			\Upsilon_{\frac{\sigma}{2} Q_{\chi^2\beta^2}}\right]
			&:=\sum_{n=0}^{\infty}\frac{1}{n!}\textrm{T-}\int\limits_{\mathbb{R}^{2n}}\mathrm{d}t_1\ldots\mathrm{d}t_{2n}\prod_{j=1}^n\frac{\sigma}{2}Q_{\chi^2\beta^2}(t_j,t_{j+n})
			\frac{\delta^2}{\delta x(t_j)\delta x(t_{j+n})}
			\\&:=\sum_{n=0}^\infty\int_{\tau_1\geq\ldots\geq\tau_n}\mathrm{d}t_1\ldots\mathrm{d}t_{2n}\prod_{j=1}^n\frac{\sigma}{2}Q_{\chi^2\beta^2}(t_j,t_{j+n})
			\frac{\delta^2}{\delta x(t_j)\delta x(t_{j+n})}\,,
		\end{align*}
		where $\tau_j=\min\{t_j,t_{j+n}\}$.
	\end{proposition}

	\begin{remark}\label{Rmk: alphabeta expectation value useful identity - particular cases}
		It is worth observing that two distinguished instances of Equation \eqref{Eq: alphabeta expectation value useful identity} are obtained by setting either $\alpha=0$ or $\beta=1$. In this last case, 
		\begin{align*}
			\llangle F\rrangle_{\alpha,1,\vartheta_0}
			=T_{\frac{\sigma}{2}Q_{\chi^2}}\left[\Gamma_G(Fe^{\langle\chi,\tilde{x}\alpha\rangle})\big|_{\tilde{x}=0}\right]
			=\Gamma_{\frac{\sigma}{2}Q_{\chi^2}}\left[\Gamma_G(Fe^{\langle\chi,\tilde{x}\alpha\rangle})\big|_{\tilde{x}=0}\right]\,,
		\end{align*}
		where we used that $Q_{\chi^2}$ is $x$-independent and, therefore, $T_{\frac{\sigma}{2}Q_{\chi^2}}=\exp[\Upsilon_{\frac{\sigma}{2}Q_{\chi^2}}]=\Gamma_{\frac{\sigma}{2}Q_{\chi^2}}$. Observe that this last map has already appeared in the literature, {\it e.g.} in  \cite{Bonicelli-Dappiaggi-Rinaldi-2021,Dappiaggi-Drago-Rinaldi-Zambotti-2022} as a tool to analyze stochastic partial differential equations with an additive Gaussian noise.
		As we shall discuss in  Section \ref{Subsec: additive SDE: beta=1}, in these works $T_{\frac{\sigma}{2}Q_{\chi^2}}$ is a key ingredient in the evaluation of the correlations between $x$-configurations, whereas $\Gamma_G(Fe^{\langle\chi,\tilde{x}\alpha\rangle})\big|_{\tilde{x}=0}$ plays a role in the analysis of the perturbative series due to the non-linearity encoded by $\alpha$ in Equation \eqref{Eq: SDE - general form}.
		
		\vskip .2cm		
		
		\noindent Suppose instead that $\alpha=0$. In this case
		\begin{align*}
			\llangle F\rrangle_{0,\beta,\vartheta_0}
			=\Gamma_{G}\left[
			T_{\frac{\sigma}{2}Q_{\chi^2\beta^2}}(
			Fe^{\langle\chi^2,\sigma\vartheta_0\tilde{x}\beta\beta_1\rangle})
			\right]\bigg|_{\tilde{x}=0}\,.
		\end{align*}
		Whenever $\vartheta_0=0$ it descends that the action of $\Gamma_G$ becomes trivial and thus
		\begin{align*}
			\llangle F\rrangle_{0,\beta,0}
			=T_{\frac{\sigma}{2}Q_{\chi^2\beta^2}}(F)\, .
		\end{align*}
		Hence the $\tilde{x}$-configuration produces no effect and there exists a non-trivial correlation between the $x$-configurations:
		\begin{align*}
			\llangle x(t)x(t')\rrangle_{0,\beta,0}=\sigma Q_{\chi^2\beta^2}(t,t')\,.
		\end{align*}
	\end{remark}
	
\begin{ex}
	A relevant application of 
	Example \ref{Ex: effective graph expansion with linear term} and Remark \ref{Rmk: alphabeta expectation value useful identity - particular cases} is obtained by considering $\alpha(x(t),t)=\alpha_1x(t)$, $\alpha_1\in\mathbb{R}$, and $\beta=1$.
	In this case the value $\vartheta_0$ does not play any role in the computation of $\llangle\;\rrangle_{\alpha,1,\vartheta_0}$.
	Moreover, Example \ref{Ex: effective graph expansion with linear term} and Remark \ref{Rmk: alphabeta expectation value useful identity - particular cases} lead to
	\begin{align*}
		\llangle F\rrangle_{\alpha,1,\vartheta_0}
		=\Gamma_{G_1}(F)
		=\Gamma_{\frac{\sigma}{2}Q_{\chi^2}}(F)
		\qquad \forall F\in\mathcal{P}_x\,,
	\end{align*}
	where
	\begin{align*}
		G_1(t,t')
		=e^{\alpha_1\int_{t'}^t\chi(s)\mathrm{d}s}\vartheta(t,t')\,,
		\qquad
		Q_{\chi^2}(t,t')
		=\int G_1(t,s)\chi^2(s)G_1(t',s)\mathrm{d}s\,,
	\end{align*}
	If $\alpha_1<0$ we may consider the limit $\chi\to 1$ of the last expression, \textit{e.g.} first by considering $\chi\in\mathcal{D}(\mathbb{R})$ such that $\chi(0)=1$, then replacing $\chi$ by $\chi_n(t):=\chi(t/n)$ and eventually taking the limit $n\to\infty$ of the resulting expectation values.
	Overall one obtains
	\begin{align*}
		\lim_{\chi \to 1}Q_{\chi^2}(t,t')
		=\int^{\min\{t,t'\}}_{-\infty}e^{\alpha_1(t+t'-2s)}
		\mathrm{d}s
		=\frac{e^{\alpha_1|t-t'|}}{2\alpha_1}\,.
	\end{align*}
	The ensuing expectation values $\llangle F\rrangle_{\alpha,1,\vartheta_0}$ obtained in this limit coincide with those associated with the stationary, unconditioned Ornstein-Uhlenbeck process.
\end{ex}

	\begin{remark}\label{Rem: on the cut-off}
		Through this whole paper we have assumed that $\chi\in\mathcal{D}(\mathbb{R})$.
		At a technical level, this is mainly due to the need to make sense of the interacting vertex $V_{\alpha,\beta,\vartheta_0}\in\mathcal{P}_{x\tilde{x}}$, \textit{cf.} Equation \eqref{Eq: MSR interacting potential - vartheta0 generic case}.
		The issue of removing $\chi$ is known as the adiabatic limit problem and, currently, solutions are known only for rather specific cases.	In the algebraic framework one can consider an alternative, weaker, notion of adiabatic limit known as \emph{algebraic adiabatic limit} \cite{Brunetti-Fredenhagen-2000,Drago-Gerard-2017,Duch-2018,Duch-2021,Lindner-Fredenhagen-2014}.
		It especially useful when one considers functionals supported in a specific finite time interval $I\subseteq\mathbb{R}$.
		Therein, the algebraic adiabatic limit shows that the particular form of the cut-off does not matter, as long as it abides by mild requirements.
		For completeness we will briefly describe this approach in the present setting.
		The initial observation is that when dealing with Equation \eqref{Eq: SDE - general form} it would be natural to assume $\chi\in C^\infty(\mathbb{R})$ and of the form
		\begin{align}\label{Eq: cut-off support properties}
			\chi(t)=
			\begin{dcases}
				1\qquad t\geq t_1
				\\
				0\qquad t\leq t_0
			\end{dcases}\,,
		\end{align}
		for $t_0,t_1\in\mathbb{R}$.
		In an idealized, limiting case, one would even like to set $\chi(t):=\vartheta(t-t_0)$, so that the solution $x_\xi$ defines a solution to Equation \eqref{Eq: SDE - general form} with $x_\xi(t_0)=x_0$.
		In the MSR approach we cannot allow for $\chi$ as per Equation \eqref{Eq: cut-off support properties}, since $V_{\alpha,\beta,\vartheta_0}\in\mathcal{P}_{x\tilde{x}}$ would be ill-defined.
		However, since we are mainly interested in the expectation values $\llangle F\rrangle_{\alpha,\beta,\vartheta_0}$ we may circumvent this hurdle by using the retardation properties of $G$.
		For definiteness let assume that $\operatorname{supp}(\chi)\subset(t_0,t_1)$:
		In this case, if $F\in\mathcal{P}_x$ is such that
		\begin{align*}
			\operatorname{supp}\bigg(\frac{\delta F(x)}{\delta x(t)}\bigg)\subset (-\infty,t_0)
			\qquad
			\forall x\in C^\infty(\mathbb{R})\,, 
		\end{align*}
		then
		\begin{align*}
			\Upsilon_G(e^{V_{\alpha,\beta,\vartheta_0}}F)
			=\int\limits_{\mathbb{R}^2} G(t,s)\frac{\delta F}{\delta x(t)}
			\frac{\delta V_{\alpha,\beta,\vartheta_0}}{\delta\tilde{x}(s)}
			e^{V_{\alpha,\beta,\vartheta_0}}\mathrm{d}t\mathrm{d}s
			=0\,.
		\end{align*}
		In other words $\llangle F\rrangle_{\alpha,\beta,\vartheta_0}=F(x_0,0)$ and, therefore, for all $F\in\mathcal{P}_x$ with
		\begin{align*}
			\operatorname{supp}\bigg(\frac{\delta F}{\delta x(t)}\bigg)\subset (-\infty,a)\,,
		\end{align*}
		the value $\llangle F\rrangle_{\alpha,\beta,\vartheta_0}$ is not affected by changes of $\chi$ within the interval $(a,+\infty)$.
		Let now $I=(t_0,t_1)\subset\mathbb{R}$ be an open interval and let $\mathcal{P}_G(I)$ be the $\cdot_G$-subalgebra of $\mathcal{P}_G$ of all $x\tilde{x}$-functionals $F\in\mathcal{P}_G$ such that
		\begin{align*}
			\operatorname{supp}\bigg(\frac{\delta F(x)}{\delta x(t)}\bigg)
			\subset I
			\qquad
			\forall x\in C^\infty(\mathbb{R})\,.
		\end{align*}
		Thereon we can consider the functional $\llangle\;\rrangle_{\alpha,\beta,\vartheta_0}^\chi$ defined as per Equation \eqref{Eq: MSR expectation value} where $\chi\in\mathcal{D}(\mathbb{R})$ is such that $\chi|_I=1$.
		On account of the previous line of reasoning $\llangle\;\rrangle_{\alpha,\beta,\vartheta_0}^\chi$ does not change if we modify $\chi$ within $[t_1,+\infty)$.
		Therefore, when evaluating $\llangle F\rrangle_{\alpha,\beta,\vartheta_0}^\chi$, $F\in\mathcal{P}_G(I)$, we may consider $\chi$ abiding to Equation \eqref{Eq: cut-off support properties}.
		This procedure has "removed algebraically" the cut-off $\chi$, at the price of focusing on a particular class of functionals supported in a finite time interval.
		\cite{Brunetti-Fredenhagen-2000,Drago-Gerard-2017,Duch-2018,Duch-2021,Lindner-Fredenhagen-2014}.
	\end{remark}
	
	\section{Algebraic approach to SDE}\label{Sec: algebraic approach to SDE}
	
	In this section we outline the algebraic approach to the analysis of stochastic differential equations (SDE) as in Equation \eqref{Eq: SDE - general form}. The following discussion can be seen as an adaptation to the case in hand of the approach developed in \cite{Dappiaggi-Drago-Rinaldi-Zambotti-2022}.
	Similarly to the procedure followed in Section \ref{Sec: algebraic approach to Martin-Sigga-Rose formalism}, the main rationale calls for introducing $\mathcal{P}_\xi$, an algebra of distribution-valued polynomial functionals over smooth configurations. The randomness due to the Gaussian white noise present in Equation \eqref{Eq: SDE - general form} is subsequently encoded algebraically by means of a suitable deformation of the algebra structure, which is reminiscent of Equation \eqref{Def: Gamma G map}.
	
	
	\begin{Definition}\label{Def: Pxi-functionals}
		We call $\operatorname{Pol}_\xi$, the vector space of all \textbf{polynomial $\xi$-functionals}. Taking into account Equation \eqref{Eq: tensor algebra}, these are maps 
		\begin{align*}
			F\colon C^\infty(\mathbb{R})\to\mathcal{T}'(\mathbb{R}),\quad
			\xi\mapsto F(\xi),
		\end{align*}
		such that, for all $\xi,\xi'\in C^\infty(\mathbb{R})$ there exists $n\in\mathbb{N}\cup\{0\}$ such that the maps $\lambda\mapsto F(\xi+\lambda \xi')$ are polynomial of degree $n$ in $\lambda$. In addition we endow $\operatorname{Pol}_\xi$ with the natural, unital associative algebra structure induced by the tensor product on $\mathcal{T}^\prime(\mathbb{R})$. 
	\end{Definition}
	
	In order to give meaning within this formalism to $x_\xi$ as per Equation \eqref{Eq: SDE fixed point}, it is convenient to individuate a distinguished subalgebra of $\operatorname{Pol}_\xi$. To fix the notation, observe that, in the following, given $\mathcal{O}$ any subset of the space of functionals $K:C^\infty(\mathbb{R})\to\mathcal{D}^\prime(\mathbb{R})$, with $C^\infty[\mathcal{O}]$ we denote the smallest, polynomial, $C^\infty(\mathbb{R})$-ring containing $\mathcal{O}$. 
	
	
	\begin{Definition}\label{Def: Pxi smallest algebra}
		We denote by $\mathcal{P}_\xi$ the smallest algebra generated by the following requirements:
		\begin{enumerate}[(a)]
			\item\label{Item: smooth functions are in Pxi}
			$C^\infty(\mathbb{R})\subset\mathcal{P}_\xi$, where $g\in C^\infty(\mathbb{R})$ is regarded as a constant $\xi$-functional.
			
			\item\label{Item: Gxi is in Pxi}
			For all $x_0\in C^\infty(\mathbb{R})$ the distribution-valued functional
			\begin{align*}
				\xi\mapsto G\ast\chi x_0\xi\in C^\infty(\mathbb{R})\subset \mathcal{D}'(\mathbb{R})\,,
			\end{align*}
			belongs to $\mathcal{P}_\xi$.
			
			\item\label{Item: alpha-beta closedness in Pxi}
			Let $F\in\mathcal{P}_\xi$ be such that $F(\xi)\in\mathcal{D}'(\mathbb{R})$ and in particular let us assume that there exists a polynomial $p$ in the variable $\xi$ with coefficients in $C^\infty(\mathbb{R})$ such that
			\begin{align*}
				\langle F(\xi),f\rangle
				=\int_{\mathbb{R}}p(\xi(t))f(t)\mathrm{d}t
				\qquad\forall f\in \mathcal{D}(\mathbb{R})\,.
			\end{align*}
			Then the following functionals still belong to $\mathcal{P}_\xi$:
			\begin{multline*}
				\langle G\ast\chi\alpha(F)(\xi),f\rangle
				:=\int_{\mathbb{R}}[G\ast\chi\alpha(p(\xi))](t)f(t)\mathrm{d}t
				\\
				\langle G\ast\chi\xi\beta(F)(\xi),f\rangle
				:=\int_{\mathbb{R}}[G\ast\chi\xi\beta(p(\xi))](t)f(t)\mathrm{d}t
				\qquad\forall f\in\mathcal{D}(\mathbb{R})\,.
			\end{multline*}
		\end{enumerate}
	\end{Definition}
	
	Properties \ref{Item: smooth functions are in Pxi}-\ref{Item: Gxi is in Pxi}-\ref{Item: alpha-beta closedness in Pxi} are the minimal requirements to ensure the well-definiteness of the $\xi$-functional $x_\xi$ defined by Equation \eqref{Eq: SDE fixed point}.
	The latter can now be regarded as an element lying in the algebra $\mathcal{P}_\xi[[\chi]]$ of formal power series in the cut-off $\chi$ with coefficients in $\mathcal{P}_\xi$.
	In order to encode the probabilistic features due to the Gaussian white noise in Equation \eqref{Eq: SDE - general form}, we adopt the same procedure as in \cite{Dappiaggi-Drago-Rinaldi-Zambotti-2022} and as in  Definition \ref{Def: Gamma G map} within the MSR framework:
	
	\begin{Definition}\label{Def: Gamma delta map}
		We denote by $\Gamma_{\delta/2}\colon\mathcal{P}_\xi\to\operatorname{Pol}_\xi$ the linear map defined by
		\begin{gather}\label{Eq: delta map}
			\Gamma_{\delta/2}:=\exp[\Upsilon_{\delta/2}]
			\qquad
			\Upsilon_{\delta/2}(F):=\frac{1}{2}\int\delta(t,t')\frac{\delta^2F}{\delta \xi(t)\delta \xi(t')}\mathrm{d}t\mathrm{d}t'\,.
		\end{gather}
	\end{Definition}
	
	\noindent Although it appears to play no role in our construction, we report for completeness that, similarly to Proposition \ref{Prop: G product}, one can employ the map $\Gamma_{\delta/2}$ to deform the algebra structure of $\mathcal{P}_\xi$. The proof is omitted since, mutatis mutandis, it is identical to the one in \cite[Corol. 32]{Dappiaggi-Drago-Rinaldi-Zambotti-2022}:
	
	\begin{proposition}\label{Prop: Algebraic Structure SDE}
		Let $\Gamma_{\delta/2}$ be as per Definition \ref{Def: Gamma delta map}. Then the vector space  $\mathcal{P}_\delta:=\Gamma_{\delta/2}[\mathcal{P}_\xi]$ is a unital, commutative, associative $\mathbb{C}$-algebra if endowed with the product 
		\begin{align}\label{Eq: delta product}
			F_1\cdot_{\delta/2} F_2
			:=\Gamma_{\delta/2}[\Gamma_{\delta/2}^{-1}(F_1)\Gamma_{\delta/2}^{-1}(F_2)]
			\qquad\forall F_1,F_2\in\mathcal{P}_\delta\,.
		\end{align}
	\end{proposition}
	
	Observe that Definition \ref{Def: Gamma delta map} allows us to recover the correlation functions of the solutions of Equation \eqref{Eq: SDE - general form} as 
	\begin{align*}
		\mathbb{E}_\eta[x_\eta(t_1)\cdots x_\eta(t_k)]
		=\Gamma_{\delta/2}[x_\xi(t_1)\cdots x_\xi(t_k)]|_{\xi=0}\in\mathbb{C}[[\chi]]\,,
	\end{align*}
	where $x_\xi\in\mathcal{P}_\xi[[\chi]]$ is defined as per Equation \eqref{Eq: SDE fixed point}.
	As a matter of fact, given any functional $F\in\operatorname{Pol}_\xi$, one can extend the above correspondence defining
	\begin{align}\label{Eq: expectation values in the algebraic SDE approach}
		\mathbb{E}_\eta[F(x_\eta)]
		:=\Gamma_{\delta/2}[F(x_\xi)]|_{\xi=0}\,,
	\end{align}
	where $\xi\mapsto F(x_\xi)\in\mathcal{P}_\xi[[\chi]]$.
	
	\begin{remark}\label{Rmk: SDE formalism deals with Stratonovich only}
		It is worth stressing that the algebraic approach described by Equations \eqref{Eq: correlation and Gamma delta map}-\eqref{Eq: expectation values in the algebraic SDE approach} allows to compute at a level of perturbation theory the expectation values $\mathbb{E}_\eta[F(x_\eta)]$ when Equation \eqref{Eq: SDE - general form} is solved in the Stratonovich sense \cite{Stratonovich-1966}.
		This is in sharp contrast to the algebraic MSR formalism developed in Section \ref{Sec: algebraic approach to Martin-Sigga-Rose formalism}, which allows to compute the correlation functions for the solution $x_\xi$ considered either in the It\^o or in the Stratonovich sense, see in particular Remark \ref{Rmk: vartheta0-generalized interacting vertex; multiplicative ambiguity in MSR}.
		
		As a matter of fact, the reason forcing us to consider only solutions à la Stratonovich in Equation \eqref{Eq: correlation and Gamma delta map} is related to the use of $\mathcal{P}_\xi[[\chi]]$, as we argue in the following.
		Indeed to draw this conclusion we can consider, without loss of generality, the case $\alpha=0$ and $\beta(x(t),t)=x(t)$: At second order in perturbation theory, Equation \eqref{Eq: SDE fixed point} reads
		\begin{align*}
			x_\xi=x_0
			+x_0G\ast\chi\xi
			+x_0G\ast[\chi\xi G\ast\chi\xi]+\mathcal{O}(\chi^3)\,.
		\end{align*}
		Focusing in particular on the term $\Upsilon_{\delta/2}(G\ast[\chi\xi G\ast\chi\xi])$ we find
		\begin{align*}
			\Upsilon_{\delta/2}(G\ast[\chi\xi G\ast\chi\xi])
			&=\Upsilon_{\delta/2}\bigg[\;
			\int\limits_{\mathbb{R}^2} G(t,s_1)\chi(s_1)\xi(s_1) G(s_1,s_2)\chi(s_2)\xi(s_2)\mathrm{d}s_1\mathrm{d}s_2
			\bigg]
			\\
			&=\int\limits_{\mathbb{R}} G(t,s)\chi(s)^2G(s,s)\mathrm{d}s\,,
		\end{align*}
		where the ambiguity in defining $G(s,s)$ has already been discussed in Remark \ref{Rem: on theta0}.
		At the same time the identity
		\begin{gather*}
			\int\limits_{\mathbb{R}^2} G(t,s_1)\chi(s_1)\xi(s_1) G(s_1,s_2)\chi(s_2)\xi(s_2)\mathrm{d}s_1\mathrm{d}s_2
			=\frac{1}{2}\left(\int_{\mathbb{R}}G(t,s)\chi(s)\xi(s)dx\right)^2\,.
		\end{gather*}
		entails
		\begin{align*}
			\Upsilon_{\delta/2}(G\ast[\chi\xi G\ast\chi\xi])
			=\frac{1}{2}\Upsilon_{\delta/2}\bigg[\bigg(
			\int\limits_{\mathbb{R}} G(t,s)\chi(s)\xi(s)\mathrm{d}s
			\bigg)^2\bigg]\,.
		\end{align*}
		Consistency forces $G(t,t)=1/2$, which is in agreement with the realization of the solutions of Equation \eqref{Eq: SDE - general form} in the sense of Stratonovich \cite{Oksendal}.
		This feature is not a drawback since every SDE under scrutiny admits always a Stratonovich counterpart and viceversa.
	\end{remark}
	
	\section{Proof of the SDE-MSR correspondence}\label{Sec: proof of the SDE-MSR correspondence}
	
	In this section we will present the proof of the SDE-MSR correspondence outlined in Section \ref{Sec: introduction}. We divide the analysis in two main scenarios, namely in Section \ref{Subsec: additive SDE: beta=1} we prove Theorem \ref{Thm: main results} for the additive case, that is setting $\beta=1$ in Equation \eqref{Eq: SDE - general form}, while in Section \ref{Subsec: multiplicative SDE: alpha=0} we consider the multiplicative scenario, \textit{i.e.} $\alpha=0$.
	
	\begin{remark}\label{Rmk: MSR expectation value - graph representation with squared vertices}
		In the following frequently we make an extensive application of the graph rules \ref{Graph: Rules 1}, \ref{Graph: Rules 2} and \ref{Graph: Rules 3} to compute the graph expansion of the correlation functions in the MSR formalism.
		In order to simplify the notation and to make the analysis clearer to a reader, in Section \ref{Subsec: additive SDE: beta=1} we slightly alter the graph rules \ref{Graph: Rules 3} as follows: We denote by
		\begin{itemize}
			\item  \begin{tikzpicture}[every node/.style={sloped,allow upside down}]
				\node[rectangle,draw,fill=white] at (1,0) {$0$};
			\end{tikzpicture}
			any contribution due to $\chi\alpha$, 
			\item 
			\begin{tikzpicture}[every node/.style={sloped,allow upside down}]
				\node[rectangle,draw,fill=white] at (1,0) {$k$};
			\end{tikzpicture}
			any contribution due to $\chi\alpha_k:=\chi\partial_x^k\alpha$.
		\end{itemize}
		Observe that we will adopt the same convention when dealing with $\beta$ and with its derivatives in Section \ref{Subsec: multiplicative SDE: alpha=0}.
		There is no risk of confusion using the same notation since we will be analyzing separately the additive and the multiplicative cases, in which only one among $\alpha$ and $\beta$ contributes effectively to the analysis. 
		
		With this convention the complexity of the polynomial interaction is codified in a single squared vertex which recollect at once all straight external lines.
		In addition we stress that all the graph rules \ref{Graph: Rules 2} still apply, {\it e.g.} 
		\begin{center}
			\begin{tikzpicture}[every node/.style={sloped,allow upside down}]
				\draw[snake] (2,0)--(3,0);
				\node[rectangle,draw,fill=white] at (0,0) {$0$};
				\filldraw (1,0) node {$\cdot_G$};
				\filldraw (2,0) circle (1.5pt) node[left] {$s$};
				\draw (3.5,0) node {$=$};
			\end{tikzpicture}
			\begin{tikzpicture}[every node/.style={sloped,allow upside down}]
				\draw[snake] (1,0)--(2,0);
				\node[rectangle,draw,fill=white] at (0,0) {$0$};
				\filldraw (1,0) circle (1.5pt) node[left] {$s$};
				\draw (2.5,0) node {$+$};
			\end{tikzpicture}
			\begin{tikzpicture}[every node/.style={sloped,allow upside down}]
				\draw (0,0) -- node {\midarrow} (1,0);
				\node[rectangle,draw,fill=white] at (0,0) {$1$};
				\filldraw (1,0) circle (1.5pt) node[right] {$s$};
			\end{tikzpicture}
		\end{center}
	\end{remark}
	
	\subsection{Additive SDE: $\beta=1$} \label{Subsec: additive SDE: beta=1}
	
	In this part of the section we prove Theorem \ref{Thm: main results} for the case of a purely additive SDE, that is,
	\begin{align}\label{Eq: SDE - additive case}
		\dot{x}_\eta(t)=\chi(t)\left[\alpha(x_\eta(t),t)+\sqrt{\sigma}\eta(t)\right]\,,
	\end{align}
	where $\alpha(x(t),t)$ is a polynomial function of $x(t)$ with coefficients smoothly depending on $t$.
	Following the rationale of Section \ref{Sec: algebraic approach to SDE} we consider the integral form of Equation \eqref{Eq: SDE - additive case}, that is Equation \eqref{Eq: SDE fixed point} setting therein $\beta=1$. To it we associate as per Definition \ref{Def: Pxi-functionals} a $\xi$-functional $x_\xi\in\mathcal{P}_\xi[[\chi]]$ defined as per Equation \eqref{Eq: SDE fixed point}:
	\begin{align}\label{Eq: SDE fixed point - additive case}
		x_\xi=x_0
		+G\ast\chi(\alpha(x_\xi)+\sqrt{\sigma}\xi)\,.
	\end{align}
	We recall both that the symbol $[[\chi]]$ entails that the functional is realized as a formal power series with respect to $\chi$ and that, for all $F\in\mathcal{P}_x$ as per Remark \ref{Rem: tildex-independent-functionals}, expectation values $\mathbb{E}_\eta[F(x_\eta)]$ are defined by means of Equation \eqref{Eq: expectation values in the algebraic SDE approach}.
	
	Within the MSR approach as described in Section \ref{Sec: algebraic approach to Martin-Sigga-Rose formalism}, the expectation values of any $x$-polynomial functional $F\in\mathcal{P}_x$, see Remark \ref{Rem: tildex-independent-functionals}, can be computed by means of the map $\llangle\;\rrangle_{\alpha,1,\vartheta_0}$ introduced in Definition \ref{Def: MSR expectation value}:
	\begin{align}
		\llangle F\rrangle_{\alpha,1,\vartheta_0}
		=[e_{\cdot_G}^{\langle V_{\alpha,1,\vartheta_0}\rangle}
		\cdot_GF]\bigg|_{\substack{x=x_0\\\tilde{x}=0}}\,,
		\qquad
		V_{\alpha,1,\vartheta_0}(x,\tilde{x})
		:=\chi\tilde{x}\alpha
		+\frac{\sigma}{2}\chi^2\tilde{x}^2\,.
	\end{align}
	Notice that, since Equation \eqref{Eq: SDE - additive case} is purely additive ($\beta=1$) there is no dependence in $\llangle\;\rrangle_{\alpha,1,\vartheta_0}$ on the choice of the value of $\vartheta_0$ as argued in Remark \ref{Rmk: vartheta0-generalized interacting vertex; multiplicative ambiguity in MSR}.
	
	We state the main result of this section which is a particular case of Theorem \ref{Thm: main results}.
	
	\begin{theorem}\label{Thm: SDE-MSR correspondence - additive case}
		Let $x_\xi\in\mathcal{P}_\xi[[\chi]]$ be defined by Equation \eqref{Eq: SDE fixed point - additive case}. Then for all $F\in\mathcal{P}_x$ it holds true that
		\begin{align}\label{Eq: SDE-MSR correspondence - additive case}
			\Gamma_{\delta/2}[F(x_\xi)]|_{\xi=0}
			=\llangle F\rrangle_{\alpha,1,\vartheta_0}\,.
		\end{align}
	\end{theorem}
	
	\noindent The proof of Theorem \ref{Thm: SDE-MSR correspondence - additive case} is based on the graph structure introduced in Section \ref{Sec: algebraic approach to Martin-Sigga-Rose formalism}. First of all we prove an ancillary lemma:
	
	\begin{lemma}\label{Lem: SDE-MSR expectation values - additive case}
		For all $F\in\mathcal{P}_x$ it holds
		\begin{align}
			\label{Eq: MSR expectation value - additive case}
			\llangle F\rrangle_{\alpha,1,\vartheta_0}
			&=\Gamma_{\frac{\sigma}{2}Q_{\chi^2}}[(e^{\langle\chi,\tilde{x}\alpha\rangle}_{\cdot_G}\cdot_GF)\big|_{\tilde{x}=0}]\big|_{x=x_0}
			\\
			\label{Eq: SDE expectation value - additive case}
			\Gamma_{\delta/2}[F(x_\xi)]|_{\xi=0}
			&=\Gamma_{\frac{\sigma}{2}Q_{\chi^2}}[F_\chi]\big|_{x=x_0}\,,
		\end{align}
		where $F_\chi$ is defined as per Equation \eqref{Eq: Fchi functional on xi-solution}
	\end{lemma}
	\begin{proof}
		Equation \eqref{Eq: MSR expectation value - additive case} is nothing but Equation \eqref{Eq: alphabeta expectation value useful identity} ---\textit{cf.} Proposition \ref{Prop: alphabeta expectation value useful identity} and Remark \ref{Rmk: alphabeta expectation value useful identity - particular cases}.
		Hence we need only to prove the identity in Equation \eqref{Eq: SDE expectation value - additive case}. As a preliminary observation, we highlight that 
		$x_\xi$, defined as per Equation \eqref{Eq: SDE fixed point - additive case}, can be read as a functional of $x_0+G\ast\chi\sqrt{\sigma}\xi$. Consequently, for any $F\in\mathcal{P}_x$,  we can define $F_\chi\in\mathcal{P}_x[[\chi]]$ as 
		\begin{align}\label{Eq: Fchi functional on xi-solution}
			F_\chi(x_0+G\ast\chi\sqrt{\sigma}\xi)=F(x_\xi)
			\qquad\forall x_0\in C^\infty(\mathbb{R})\,,
		\end{align}
		where $x_\xi\in\mathcal{P}_\xi[[\chi]]$ is defined by Equation \eqref{Eq: SDE fixed point - additive case}. Consequently, 
		\begin{align*}
			\frac{\delta F(x_\xi)}{\delta\xi(s)}
			=\int\limits_{\mathbb{R}}\frac{\delta F_\chi}{\delta x(t)}\bigg|_{x=x_0+G\ast\chi\sqrt{\sigma}\xi}
			G(t,s)\chi(s)\sqrt{\sigma}\mathrm{d}t\,,
		\end{align*}
		which implies in turn
		\begin{multline*}
			\Upsilon_{\delta/2}[F(x_\xi)]|_{\xi=0}
			=\frac{1}{2}\int\limits_{\mathbb{R}}\frac{\delta^2F(x_\xi)}{\delta\xi(s)^2}\bigg|_{\xi=0}\mathrm{d}s
			\\
			=\frac{1}{2}\int\limits_{\mathbb{R}^2}\frac{\delta^2 F_\chi}{\delta x(t_1)\delta x(t_2)}\bigg|_{x=x_0}
			G(t_1,s)G(t_2,s)\chi(s)^2\sigma
			\mathrm{d}t_1\mathrm{d}t_2\mathrm{d}s
			=\Upsilon_{\frac{\sigma}{2}Q_{\chi^2}}[F_\chi]|_{x=x_0}\,,
		\end{multline*}
		where we employed Equation \eqref{Eq: Qbeta-operator} for $\beta=1$. This proves the sought identity.
	\end{proof}
	
	\begin{proof}[Proof of Theorem \ref{Thm: SDE-MSR correspondence - additive case}.]
		Lemma \ref{Lem: SDE-MSR expectation values - additive case} implies that Theorem \ref{Thm: SDE-MSR correspondence - additive case} is proved if we show that
		\begin{align}\label{Eq: SDE-MSR correspondence - additive case - reduced thesis}
			F_\chi(x_0)
			=[e^{\langle\chi,\tilde{x}\alpha\rangle}_{\cdot_G}\cdot_GF]\bigg|_{\substack{\tilde{x}=0\\x=x_0}}\,,
		\end{align}
		where $F\in\mathcal{P}_x$ while $F_\chi$ is defined as per Equation \eqref{Eq: Fchi functional on xi-solution}. In turn, it suffices to check whether it holds true at the level of generators of $\mathcal{P}_x$, that is for
		\begin{align*}
			F(x)=x(t_1)\cdots x(t_k)\,.
		\end{align*}
		To begin with, let us consider $F(x)=x(t)$ \textit{i.e.} $F_\chi(x_0)=x_{\xi=0}(t)$.
		To prove Equation \eqref{Eq: SDE-MSR correspondence - additive case - reduced thesis} we shall work with a graph representation as outlined in Section \ref{Sec: algebraic approach to Martin-Sigga-Rose formalism} and \ref{Sec: algebraic approach to SDE}. 
		
		\vskip .3cm
		
		\noindent {\bf SDE --} To begin with we present the graph expansion for the fixed point Equation \eqref{Eq: SDE fixed point - additive case} when $\xi=0$. To this avail, we denote by $x_{\xi=0,n}$ the $n$-th order term of the expansion of $x_{\xi=0}$ in powers of $\chi$. Equation \eqref{Eq: SDE fixed point - additive case} leads to
		\begin{align}\label{Eq: SDE n-th order expansion - additive case}
			x_{\xi=0,n}
			=\begin{dcases*}
				x_0
				\qquad n=0
				\\
				G\ast\chi\alpha_0
				\qquad n=1
				\\
				\sum_{k\geq 0}G\ast\bigg[\chi\alpha_k
				\sum_{\substack{j_1+\ldots+j_{n-1}=k\\
						\sum_\ell\ell j_\ell=n-1}}
				\frac{1}{j_1!\cdots j_{n-1}!}x_{\xi=0,1}^{j_1}\cdots x_{\xi=0,n-1}^{j_{n-1}}\bigg]
				\qquad n\geq 2
			\end{dcases*}\,,		
		\end{align}
		where $\alpha_k(t):=\partial_x^k\alpha(x_0,t)$.
		The ensuing graph expansion of $x_{\xi=0}$ can be obtained using the graph rules \ref{Graph: Rules 1} and \ref{Graph: Rules 2} and Remark \ref{Rmk: MSR expectation value - graph representation with squared vertices}, the first three orders of $x_{\xi=0}(t)$ reading
		\begin{center}
			\begin{tikzpicture}[every node/.style={sloped,allow upside down}]
				\filldraw (0.25,0.5) node {$\chi^0$};
				\draw (0,0) -- (0.5,0);
				\filldraw (0,0) circle (1.5pt) node[left] {$t$};
			\end{tikzpicture}
			$+$
			\begin{tikzpicture}[every node/.style={sloped,allow upside down}]
				\filldraw (0.25,0.5) node {$\chi^1$};
				\draw (0,0) -- node {\midarrow} (1,0);
				\filldraw (0,0) circle (1.5pt) node[left] {$t$};
				\node[rectangle,draw,fill=white] at (1,0) {$0$};
			\end{tikzpicture}
			$+$
			\begin{tikzpicture}[every node/.style={sloped,allow upside down}]
				\filldraw (0.25,0.5) node {$\chi^2$};
				\filldraw (0,0) circle (1.5pt) node[left] {$t$};
				\draw (0,0) -- node {\midarrow} (1,0);
				\draw (1,0) -- node {\midarrow} (2,0);
				\node[rectangle,draw,fill=white] at (1,0) {$1$};
				\node[rectangle,draw,fill=white] at (2,0) {$0$};
			\end{tikzpicture}
			$+$
			\begin{tikzpicture}[every node/.style={sloped,allow upside down}]
				\filldraw (0.25,0.5) node {$\chi^3$};
				\filldraw (0,0) circle (1.5pt) node[left] {$t$};
				\draw (0,0) -- node {\midarrow} (1,0);
				\draw (1,0) -- node {\midarrow} (2,0);
				\draw (2,0) -- node {\midarrow} (3,0);
				\node[rectangle,draw,fill=white] at (1,0) {$1$};
				\node[rectangle,draw,fill=white] at (2,0) {$1$};
				\node[rectangle,draw,fill=white] at (3,0) {$0$};
			\end{tikzpicture}
			$+\frac{1}{2}$
			\begin{tikzpicture}[every node/.style={sloped,allow upside down}]
				\filldraw (0.25,0.5) node {$\chi^3$};
				\filldraw (0,0) circle (1.5pt) node[left] {$t$};
				\draw (0,0) -- node {\midarrow} (1,0);
				\draw (1,0) -- node {\midarrow} (2,1);
				\draw (1,0) -- node {\midarrow} (2,0);
				\node[rectangle,draw,fill=white] at (1,0) {$2$};
				\node[rectangle,draw,fill=white] at (2,0) {$0$};
				\node[rectangle,draw,fill=white] at (2,1) {$0$};
			\end{tikzpicture}
			$+\;\mathcal{O}(\chi^4)$
		\end{center}
		This last expansion can be regarded as grafting a leaf on one of the nodes of the lower order contribution in all possible ways, \textit{cf.} \cite{Oudom_Guin_2008}.
		Denoting by $\mathcal{G}_{\textsc{sde}}^{(n)}$ the space of graphs contributing to the sum in Equation \eqref{Eq: SDE n-th order expansion - additive case}, observe that each element $\gamma\in\mathcal{G}_{\textsc{sde}}^{(n)}$, $n\geq 1$, fulfills the following properties:
		\begin{enumerate}[(I)]
			\item\label{Item: SDE graph representation of additive SDE - number of vertices and edges}
			$\gamma$ is a connected directed graph containing neither loops nor closed oriented paths.
			It is composed by $n$ squared vertices
			\begin{tikzpicture}[every node/.style={sloped,allow upside down}]
				\node[rectangle,draw,fill=white] at (0,0) {$\cdot$};
			\end{tikzpicture}\;,
			a single $t$-vertex
			\begin{tikzpicture}[every node/.style={sloped,allow upside down}]
				\filldraw (0,0) circle (1.5pt) node[left] {$t$};
			\end{tikzpicture}
			and $n$ directed edges
			\begin{tikzpicture}[every node/.style={sloped,allow upside down}]
				\draw (0,0) -- node {\midarrow} (0.5,0);
			\end{tikzpicture}.
			Moreover, the $t$-vertex has a single outgoing arrow pointing to a squared vertex
			\begin{tikzpicture}[every node/.style={sloped,allow upside down}]
				\node[rectangle,draw,fill=white] at (0,0) {$\cdot$};
			\end{tikzpicture}\,.
			
			\item\label{Item: SDE graph representation of additive SDE - properties of squares vertices}
			The number of squared vertices in each $\gamma$ coincides with the order in perturbation theory.
			Furthermore, given any
			\begin{tikzpicture}[every node/.style={sloped,allow upside down}]
				\node[rectangle,draw,fill=white] at (0,0) {$j$};
			\end{tikzpicture},
			$j$ coincides with the number of outgoing edges.
			
			\item\label{Item: SDE graph representation of additive SDE - blooming recursion}
			For any $\gamma_{n+1}\in\mathcal{G}_{\textsc{sde}}^{(n+1)}$ there exists a possibly non-unique graph $\gamma_n\in\mathcal{G}_{\textsc{sde}}^{(n)}$ such that $\gamma_{n+1}$ is obtained from $\gamma_n$ by a "blooming" procedure.
			The latter consists of replacing a single squared vertex
			\begin{tikzpicture}[every node/.style={sloped,allow upside down}]
				\node[rectangle,draw,fill=white] at (0,0) {$\cdot$};
			\end{tikzpicture}
			of $\gamma_n$ with a "bloomed" vertex namely
			\begin{tikzpicture}[every node/.style={sloped,allow upside down}]
				\draw (0,0) -- node {\midarrow} (1.5,0);
				\node[rectangle,draw,fill=white] at (0,0) {$\cdot+1$};
				\node[rectangle,draw,fill=white] at (1.5,0) {$0$};
			\end{tikzpicture}\;.
		\end{enumerate}
		The expansion of $x_{\xi=0}(t)$ subordinated to Equation \eqref{Eq: SDE n-th order expansion - additive case} can be written as the following sum of graphs $\gamma_n\in\mathcal{G}_{\textsc{sde}}^{(n)}$, $n\geq 0$:
		\begin{align}\label{Eq: SDE order expansion - additive case - graph version}
			x_{\xi=0}(t)
			=\sum_{n\geq 0}\sum_{\gamma\in\mathcal{G}_{\textsc{sde}}^{(n)}}
			\frac{1}{|\operatorname{Aut}_{\mathcal{G}_{\textsc{sde}}^{(n)}}(\gamma)|}\gamma\,.
		\end{align}
		Here $\operatorname{Aut}_{\mathcal{G}_{\textsc{sde}}^{(n)}}(\gamma)$ is the group of automorphisms of $\gamma$, that is the collection of all edge-preserving permutations of the squared vertices
		\begin{tikzpicture}[every node/.style={sloped,allow upside down}]
			\node[rectangle,draw,fill=white] at (0,0) {$\cdot$};
		\end{tikzpicture}\,. 
		By edge-preserving we mean that, if two vertices $v,v'$ are connected by a directed edge, then, given $\varsigma\in\operatorname{Aut}_{\mathcal{G}_{\textsc{sde}}^{(n)}}(\gamma)$, $\varsigma(v),\varsigma(v')$ are connected by a directed edge with the same orientation. The factor $|\operatorname{Aut}_{\mathcal{G}_{\textsc{sde}}^{(n)}}(\gamma)|^{-1}$ is necessary to avoid overcounting when summing in Equation \eqref{Eq: SDE order expansion - additive case - graph version}.
		To wit, the graph
		
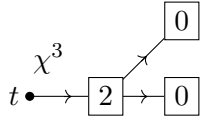
\begin{center}
			\begin{tikzpicture}[every node/.style={sloped,allow upside down}]
				\filldraw (0.25,0.5) node {$\chi^3$};
				\filldraw (0,0) circle (1.5pt) node[left] {$t$};
				\draw (0,0) -- node {\midarrow} (1,0);
				\draw (1,0) -- node {\midarrow} (2,1);
				\draw (1,0) -- node {\midarrow} (2,0);
				\node[rectangle,draw,fill=white] at (1,0) {$2$};
				\node[rectangle,draw,fill=white] at (2,0) {$0$};
				\node[rectangle,draw,fill=white] at (2,1) {$0$};
			\end{tikzpicture}
		\end{center}
		has an automorphism group made by the identity and the permutation of the two
		\begin{tikzpicture}[every node/.style={sloped,allow upside down}]
			\node[rectangle,draw,fill=white] at (0,0) {$0$};
		\end{tikzpicture}
		squared vertices, thus leading to an overall factor $1/2$.
		
		\vskip .3cm	
		
		\noindent{\bf MSR --}
		We discuss the graphical expansion of the expectation value
		\begin{align*}
			\llangle x(t)\rrangle_{\alpha,1,\vartheta_0}\,,
		\end{align*}
		within the MSR formalism described in Section \ref{Sec: algebraic approach to Martin-Sigga-Rose formalism} using the graph rules \ref{Graph: Rules 1}, \ref{Graph: Rules 2} and \ref{Graph: Rules 3}. A direct application of Equation \eqref{Eq: graph series expansion of MSR expectation value - general case} leads to
		\begin{align}\label{Eq: MSR order expansion - additive case - graph version}
			\llangle x(t)\rrangle_{\alpha,1,\vartheta_0}
			=\sum_{n\geq 0}\sum_{\gamma\in\mathcal{G}_{\textsc{msr}}^{(n)}}
			\frac{1}{|\operatorname{Aut}_{\mathcal{G}_{\textsc{msr}}^{(n)}}(\gamma)|}\gamma\,.
		\end{align}
		Here $\mathcal{G}_{\textsc{msr}}^{(n)}:=\mathcal{G}_{\alpha,1}^{(n,x(t))}$ is the set of graphs built as per the graph rules \ref{Graph: Rules 3}.
		In particular, for all $n\geq 1$ :
		\begin{enumerate}[(A)]
			\item\label{Item: MSR graph representation of additive SDE}
			$\gamma$ is a connected directed graph containing neither loops nor closed oriented paths.
			It is realized out of $n$-vertices of the form 
			\begin{tikzpicture}[every node/.style={sloped,allow upside down}]
				\draw[snake] (0,0) -- (1,0);
				\node[rectangle,draw,fill=white] at (0,0) {$0$};
			\end{tikzpicture}
			each representing a factor $\chi\tilde{x}\alpha$ and out of a single $t$-vertex
			\begin{tikzpicture}[every node/.style={sloped,allow upside down}]
				\filldraw (0,0) circle (1.5pt) node[left] {$t$};
				\draw (0,0) -- (0.5,0);
			\end{tikzpicture}
			by applying the graph rules \ref{Graph: Rules 1}-\ref{Graph: Rules 2}.
			We are also exploiting Remark \ref{Rmk: MSR expectation value - graph representation with squared vertices} {\it e.g.} for computing
			\begin{center}
				$\chi\tilde{x}\alpha\cdot_G\tilde{x}=$
				\begin{tikzpicture}[every node/.style={sloped,allow upside down}]
					\draw[snake] (0,0) -- (1,0);
					\node[rectangle,draw,fill=white] at (0,0) {$0$};
				\end{tikzpicture}
				$\cdot_G$
				\begin{tikzpicture}[every node/.style={sloped,allow upside down}]
					\filldraw (0,0) circle (1.5pt);
					\draw[snake] (0,0) -- (1,0);
				\end{tikzpicture}
				$\stackrel{\tilde{x}=0}{=}$
				\begin{tikzpicture}[every node/.style={sloped,allow upside down}]
					\draw[snake] (0,0) -- (1,0.5);
					\draw (0,0) -- node {\midarrow} (1,0);
					\filldraw (1,0) circle (1.5pt);
					\node[rectangle,draw,fill=white] at (0,0) {$1$};
				\end{tikzpicture}
			\end{center}
			
			\item\label{Item: MSR graph authomorphism of additive SDE}
			Recalling Equation \eqref{Eq: graph series expansion of MSR expectation value - general case} and the subsequent discussion, an automorphism $\varsigma\in\operatorname{Aut}_{\mathcal{G}_{\textsc{msr}}^{(n)}}(\gamma)$ of $\gamma\in\mathcal{G}_{\textsc{msr}}^{(n)}$ is an arbitrary permutation of the $n$ interacting vertices $\langle V_{\alpha,1,\vartheta_0}\rangle$ which preserves the number and orientation of the directed edges.
		\end{enumerate}
		It follows from item \ref{Item: MSR graph representation of additive SDE} that $\gamma\in\mathcal{G}_{\textsc{msr}}^{(n)}$ if and only if $\gamma$ fulfills the conditions in item \ref{Item: SDE graph representation of additive SDE - number of vertices and edges}.
		At the same time a graph $\gamma\in\mathcal{G}_{\textsc{sde}}^{(n)}$ abides by the constraints outlined in item \ref{Item: MSR graph representation of additive SDE}: This implies that $\mathcal{G}_{\textsc{msr}}^{(n)}=\mathcal{G}_{\textsc{sde}}^{(n)}$.
		Moreover, for all $\gamma\in\mathcal{G}_{\textsc{msr}}^{(n)}$, the permutation of the vertices $\langle V_{\alpha,1,\vartheta_0}\rangle$ due to $\varsigma\in\operatorname{Aut}_{\mathcal{G}_{\textsc{msr}}^{(n)}}(\gamma)$ can be read as an edge-preserving permutation of the squared vertices
		\begin{tikzpicture}[every node/.style={sloped,allow upside down}]
			\node[rectangle,draw,fill=white] at (0,0) {$\cdot$};
		\end{tikzpicture}.
		This entails that $\operatorname{Aut}_{\mathcal{G}_{\textsc{msr}}^{(n)}}(\gamma)\simeq\operatorname{Aut}_{\mathcal{G}_{\textsc{sde}}^{(n)}}(\gamma)$.
		Thus Equations \eqref{Eq: SDE order expansion - additive case - graph version} and \eqref{Eq: MSR order expansion - additive case - graph version} coincide and therefore
		\begin{align*}
			\llangle x(t)\rrangle_{\alpha,1,\vartheta_0}=\Gamma_{\delta/2}(x_\xi)|_{\xi=0}\,,
		\end{align*}
		proving Equation \eqref{Eq: SDE-MSR correspondence - additive case} for $F=x(t)$.
		
		\bigskip
		
		Let us consider now a generic generator, {\it i.e.} $F(x)=x(t_1)\cdots x(t_k)\in\mathcal{P}_x$ and, in turn, $F_\chi(x_0):=x_{\xi=0}(t_1)\cdots x_{\xi=0}(t_k)$ as per Equation \eqref{Eq: Fchi functional on xi-solution}. In this case we can consider the graphical expansion of $[e^{\langle\chi,\tilde{x}\alpha\rangle}_{\cdot_G}\cdot_Gx(t_1)\cdots x(t_k)]\bigg|_{\substack{\tilde{x}=0\\x=x_0}}$ using the same rationale as in the first part of proof.
		
		As a matter of facts, the thesis is a byproduct of the identity
		\begin{align}\label{Eq: triviality of MSR reduced correlations - additive case}
			[e^{\langle\chi,\tilde{x}\alpha\rangle}_{\cdot_G}\cdot_Gx(t_1)\cdots x(t_k)]\bigg|_{\substack{\tilde{x}=0\\x=x_0}}
			=[e^{\langle\chi,\tilde{x}\alpha\rangle}_{\cdot_G}\cdot_Gx(t_1)]\bigg|_{\substack{\tilde{x}=0\\x=x_0}}
			\cdots[e^{\langle\chi,\tilde{x}\alpha\rangle}_{\cdot_G}\cdot_G x(t_k)]\bigg|_{\substack{\tilde{x}=0\\x=x_0}}\,.
		\end{align}
		To prove it, we observe that, once expanded in graphs, the $n$-th order contribution of
		\begin{align*}
			[e^{\langle\chi,\tilde{x}\alpha\rangle}_{\cdot_G}\cdot_Gx(t_1)\cdots x(t_k)]\bigg|_{\substack{\tilde{x}=0\\x=x_0}}\,,
		\end{align*}
		consists of the sum of all graphs obtained from the $k+n$ graphs
		\begin{center}
			\begin{tikzpicture}[every node/.style={sloped,allow upside down}]
				\draw (0,0) -- (0.5,0);
				\filldraw (0,0) circle(1.5pt) node[left] {$t_1$};
			\end{tikzpicture},
			$\ldots,$
			\begin{tikzpicture}[every node/.style={sloped,allow upside down}]
				\draw (0,0) -- (0.5,0);
				\filldraw (0,0) circle (1.5pt) node[left] {$t_k$};
			\end{tikzpicture}\,,\,
			\begin{tikzpicture}[every node/.style={sloped,allow upside down}]
				\draw[snake] (0,0) -- (1,0);
				\node[rectangle,draw,fill=white] at (0,0) {$0$};
			\end{tikzpicture}
			$,\ldots,$
			\begin{tikzpicture}[every node/.style={sloped,allow upside down}]
				\draw[snake] (0,0) -- (1,0);
				\node[rectangle,draw,fill=white] at (0,0) {$0$};
			\end{tikzpicture}
		\end{center}
		applying in all admissible ways the graph rules \ref{Graph: Rules 1}, \ref{Graph: Rules 2} and Remark \ref{Rmk: MSR expectation value - graph representation with squared vertices}.
		
		Since each snaky edge can connect with at most a single vertex \begin{tikzpicture}[every node/.style={sloped,allow upside down}]
			\draw (0,0) -- (0.5,0);
			\filldraw (0,0) circle(1.5pt) node[left] {$t$};
		\end{tikzpicture},
		or to another vertex
		\begin{tikzpicture}[every node/.style={sloped,allow upside down}]
			\draw[snake] (0,0) -- (1,0);
			\node[rectangle,draw,fill=white] at (0,0) {$\cdot$};
		\end{tikzpicture},
		the graph expansion of $[e^{\langle\chi,\tilde{x}\alpha\rangle}_{\cdot_G}\cdot_Gx(t_1)\cdots x(t_k)]\bigg|_{\substack{\tilde{x}=0\\x=x_0}}$ amounts to the sum of the product of the graph expansions of each individual factor $[e^{\langle\chi,\tilde{x}\alpha\rangle}_{\cdot_G}\cdot_Gx(t_1)]\bigg|_{\substack{\tilde{x}=0\\x=x_0}}, \ldots,[e^{\langle\chi,\tilde{x}\alpha\rangle}_{\cdot_G}\cdot_Gx(t_k)]\bigg|_{\substack{\tilde{x}=0\\x=x_0}}$.
		This proves Equation \eqref{Eq: triviality of MSR reduced correlations - additive case} and thus Equation \eqref{Eq: SDE-MSR correspondence - additive case - reduced thesis} for the case $F(x)=x(t_1)\cdots x(t_k)$.
	\end{proof}
	\begin{remark}\label{Rmk: on B-series}
		The algebraic approach to SDEs adopted in this paper to represent the expectation values $\mathbb{E}_\eta[F(x_\eta)]$ is close in spirit to the numerical schemes for the approximation of the solution to stochastic differential equations.
		In particular the perturbative expansion of the solution $x_\xi$ appearing in Equation \eqref{Eq: SDE n-th order expansion - additive case} coincides with its Butcher series \cite{Butcher-1972}, see also Equation \eqref{Eq: SDE n-th order expansion - multiplicative case} for the multiplicative scenario.
		The latter leads to a graphical representation of the solution $x_\xi$ in terms of forests of rooted, non-planar trees which are in one-to-one correspondence with elementary differentials, modulo automorphisms.
		Among other aspects, like their group structure \cite{Hairer-2006,Hairer-1974}, B-series are of paramount relevance for numerical approximations since a large class of numerical integration methods fall under this series expansion on graphs. In particular, when the map $\Upsilon_{\delta/2}$ is considered, the graphs arising from the expansion of $x_\xi$ coincide with the graphical expression of an exotic aromatic Butcher series \cite{Munthe-Kaas_Verdier_2016,Laurent-2020}. The latter are characterized by the presence of aromas in the form of closed, oriented loops \cite{Bogfjellmo-2019}. In the present setting aromas are suppressed on account of the retardation properties of $G$ with the exception of a single closed loop which appears when considering $\vartheta_0=1/2$, \emph{i.e.}, the Stratonovich prescription.
		In exotic Butcher series the additional structure carried by the noise is codified by the presence of an additional type of vertices, called \emph{grafted nodes}, which, at the level of expectations, amount to a new type of edges, referred to as \emph{lianas}. These account for the contractions into covariances of the stochastic forcing \cite{Laurent-2020}.
		They coincide with the dashed edges drew in Section \ref{Subsec: multiplicative SDE: alpha=0}. Our perturbative construction of expectations can also be interpreted as the weak Taylor expansion mentioned in \cite{Rossler-2004, Rossler-2006}.
		\\
		Interestingly, there has been a strong effort in understanding the algebraic structure involved in exotic aromatic Butcher series \cite{Bogfjellmo-2019}, which naturally accounts for the combinatorics associated to the group of automorphisms, whose dimension is called \emph{symmetry coefficient} in the literature.
		In particular, the comparison between S-series over grafted forests and exotic S-series performed in \cite{Bronasco-2024} unveils the combinatorial nature of the prefactor in Equation \eqref{Eq: SDE multiplicative correlation - quotient graph expansion of Gamma delta map}. We reckon that this thorough understanding of the Hopf algebraic structure of the space of rooted trees involved in our expansion could lead to more straightforward proofs, as well as to novel results. We postpone the study of these aspects to future works. 
	\end{remark}

	\subsection{Multiplicative SDE: $\alpha=0$}
	\label{Subsec: multiplicative SDE: alpha=0}
	
	In this part of the section we prove Theorem \ref{Thm: main results} for the case of a purely multiplicative SDE, that is,
	\begin{align}\label{Eq: SDE - multiplicative case}
		\dot{x}_\eta(t)=\chi(t)\left[\sqrt{\sigma}\eta(t)\beta(x_\eta(t),t)\right]\,,
	\end{align}
	where $\beta(x(t),t)$ is a polynomial in $x(t)$ with coefficients smoothly depending on $t$. The $\xi$-functional $x_\xi\in\mathcal{P}_\xi[[\chi]]$ defined as per Equation \eqref{Eq: SDE fixed point} reads
	\begin{align}\label{Eq: SDE fixed point - multiplicative case}
		x_\xi=x_0
		+G\ast\chi\sqrt{\sigma}\xi\beta\,.
	\end{align}
	According to Section \ref{Sec: algebraic approach to SDE}, the expectation values $\mathbb{E}_\eta[F(x_\eta)]$, $F\in\mathcal{P}_x$, are computed by means of Equation \eqref{Eq: expectation values in the algebraic SDE approach}.
	We stress that, since we consider $\beta$ such that $\beta_1\neq0$, the SDE \eqref{Eq: SDE - multiplicative case} is ambiguous and the discussion of Remark \ref{Rmk: SDE formalism deals with Stratonovich only} applies. In particular the algebraic SDE approach will provide the expectation values for the solution $x_\eta$ in the sense of Stratonovich.
	
	The MSR approach introduced in Section \ref{Sec: algebraic approach to Martin-Sigga-Rose formalism}, allows to compute the expectation values of $F\in\mathcal{P}_x$ in terms of the map $\llangle\;\rrangle_{0,\beta,\vartheta_0}$ ---\textit{cf.} Definition \ref{Def: MSR expectation value}. 
	Since the algebraic SDE approach forces us to work in the Stratonovich framework, see Remark \ref{Rmk: SDE formalism deals with Stratonovich only}, we are led to considering $\vartheta_0=1/2$, namely
	\begin{align}
		\llangle F\rrangle_{0,\beta,1/2}
		=[e_{\cdot_G}^{\langle V_{0,\beta,1/2}\rangle}
		\cdot_GF]\bigg|_{\substack{x=x_0\\\tilde{x}=0}}\,,
		\qquad
		V_{\alpha,1,\vartheta_0}(x,\tilde{x})
		:=\frac{1}{2}\tilde{x}\chi\sigma\beta\beta_1
		+\frac{\sigma^2}{2}\chi^2\tilde{x}^2\beta^2\,.
	\end{align}
	
	\noindent The adaptation of Theorem \ref{Thm: main results} to the case in hand reads
	\begin{theorem}\label{Thm: SDE-MSR correspondence - multiplicative case}
		Let $x_\xi$ be defined by Equation \eqref{Eq: SDE fixed point - multiplicative case}.
		Then for all $F\in\mathcal{P}_x$ it holds
		\begin{align}\label{Eq: SDE-MSR correspondence - multiplicative case}
			\Gamma_{\delta/2}[F(x_\xi)]|_{\xi=0}
			=\llangle F\rrangle_{0,\beta,1/2}\,.
		\end{align}
	\end{theorem}
	
	Similarly to Section \ref{Subsec: additive SDE: beta=1}, the $n$-th order in the perturbative expansion of of $x_\xi$ defined by Equation \eqref{Eq: SDE fixed point - multiplicative case} obeys a recursive formula given by
	\begin{align}\label{Eq: SDE n-th order expansion - multiplicative case}
		x_{\xi,n}
		=\begin{dcases}
			x_0
			&n=0
			\\
			G\ast\chi\sqrt{\sigma}\xi\beta_0
			&n=1
			\\
			\sum_{k\geq 0}G\ast\bigg[\chi\sqrt{\sigma}\xi\beta_k
			\sum_{\substack{j_1+\ldots+j_{n-1}=k\\
					\sum_\ell\ell j_\ell=n-1}}
			\frac{1}{j_1!\cdots j_{n-1}!}x_{\xi,1}^{j_1}\cdots x_{\xi,n-1}^{j_{n-1}}\bigg]
			& n\geq 2
		\end{dcases}\,,		
	\end{align}
	where $\beta_k(t):=\partial_x^k\beta(x_0,t)$.
	Alas, Equation \eqref{Eq: SDE n-th order expansion - multiplicative case} does not suffice to prove Equation \eqref{Eq: SDE-MSR correspondence - multiplicative case} since one needs to compute $\Gamma_{\delta/2}[x_\xi(t_1)\cdots x_{\xi}(t_k)]|_{\xi=0}$, which involves taking $\xi$-derivatives of Equation \eqref{Eq: SDE n-th order expansion - multiplicative case}.
	Hence there is no analogous of Lemma \ref{Lem: SDE-MSR expectation values - additive case} for the case in hand.
	
	\begin{proof}[Proof of Theorem \ref{Thm: SDE-MSR correspondence - multiplicative case}.]
		We prove Equation \eqref{Eq: SDE-MSR correspondence - multiplicative case} at the level of generators: 
		$F=x(t_1)\cdots x(t_k)$, $k\in\mathbb{N}$.
		On account of Definition \ref{Def: xtildex local functionals} this would suffice to prove Equation \eqref{Eq: SDE-MSR correspondence - multiplicative case} for all $F\in\mathcal{P}_x$.
		
		The strategy of the proof is similar to the one of Theorem \ref{Thm: SDE-MSR correspondence - additive case}.
		At first we will provide a graphical expansion for both sides of Equation \eqref{Eq: SDE-MSR correspondence - multiplicative case}, \textit{cf.} Equations \eqref{Eq: SDE multiplicative correlation - quotient graph expansion of Gamma delta map}-\eqref{Eq: MSR order expansion - multiplicative case - graph expansion}.
		Eventually we will prove that the two formulae lead to the same contribution.
		
		\vskip .3cm
		
		\noindent {\bf SDE --} We shall now discuss the expansion in terms of graphs of
		\begin{align*}
			\Gamma_{\delta/2}[x_\xi(t_1)\cdots x_\xi(t_k)]|_{\xi=0}\,.
		\end{align*}
		To this avail, it is convenient to find a more explicit analytical expression.
		On account of Definition \ref{Def: Gamma delta map} it descends
		\begin{align}\label{Eq: SDE multiplicative correlation - series of Gamma delta map}
			\left.\Gamma_{\delta/2}[x_\xi(t_1)\cdots x_\xi(t_k)]\right|_{\xi=0}
			=\sum_{n\geq 0}\frac{1}{2^n n!}
			\int\limits_{\mathbb{R}^n}\prod_{j=1}^n\frac{\delta^2}{\delta\xi(s_j)^2}
			x_\xi(t_1)\cdots x_\xi(t_k)\bigg|_{\xi=0}\mathrm{d}s_1\cdots\mathrm{d}s_n\,.
		\end{align}
		Notice that, Equation \eqref{Eq: SDE fixed point - multiplicative case} entails that each $\xi$-derivatives raises the perturbative order in $\chi$ by $1$, that is, $\dfrac{\delta x_\xi}{\delta\xi}=O(\chi)$. Hence 
		\begin{align}\label{Eq: SDE Gammadelta expansion without odd orders}
			\Gamma_{\delta/2}[x_\xi(t_1)\cdots x_\xi(t_k)]|_{\xi=0}
			=\sum_{\substack{n\geq 0\\p_1+\ldots+p_k=2n}}\frac{1}{2^n n!}
			\int\limits_{\mathbb{R}^n}\prod_{j=1}^n\frac{\delta^2}{\delta\xi(s_j)^2}
			x_{\xi,p_1}(t_1)\cdots x_{\xi,p_k}(t_k)\bigg|_{\xi=0}\mathrm{d}s_1\cdots\mathrm{d}s_n\,,
		\end{align}
		is finite at each perturbative order in $\chi$ and the odd terms are vanishing. To proceed further we need to discuss the action of $2n$ $\xi$-derivatives on the product $x_\xi(t_1)\cdots x_\xi(t_k)$:
		\begin{align*}
			\prod_{j=1}^n\frac{\delta^2}{\delta\xi(s_j)^2}
			x_\xi(t_1)\cdots x_\xi(t_k)\,.
		\end{align*}
		In what follows it will be convenient to distinguish whether a factor $\dfrac{\delta^2}{\delta\xi(s_j)^2}$ in Equation \eqref{Eq: SDE multiplicative correlation - series of Gamma delta map} acts on a single $x_\xi$ or not.
		To this avail, for all $i=1,\ldots,k$ we denote by
		\begin{multline}\label{Eq: M,N sets}
			N_i:=\{j\in\{1,\ldots,n\}\,|\,
			\dfrac{\delta^2}{\delta\xi(s_j)^2}
			\textrm{ acts on $x_\xi(t_i)$ and on $x_\xi(t_\ell)$, $\ell\neq i$}\}\,,
			\\
			M_i:=\{j\in\{1,\ldots,n\}\,|\,
			\dfrac{\delta^2}{\delta\xi(s_j)^2}
			\textrm{ acts only on $x_\xi(t_i)$}\}\,.
		\end{multline}
		Notice that, for all $i=1,\dots,k$, $N_i,M_i$ are such that
		\begin{multline}\label{Eq: N,M sets properties}
			N_i\cap M_j
			=\varnothing
			\quad\forall i,j\,,
			\qquad
			N_i\cap N_h\cap N_\ell
			=\varnothing
			\quad\forall i\neq h\neq\ell\,,
			\\
			\bigcup_{i=1}^kN_i
			=\bigg[\{1,\ldots,n\}
			\setminus\bigcup_{i=1}^kM_i\bigg]
			\sqcup\bigg[\{1,\ldots,n\}
			\setminus\bigcup_{i=1}^kM_i\bigg]\,,
		\end{multline}
		where $\sqcup$ denotes disjoint union while the last equality expresses the fact that, for all $j\in\{1,\ldots,n\}\setminus\bigcup_{i=1}^kM_i$ there are exactly two $i,\ell\in\{1,\ldots,k\}$ such that $j\in N_i\cap N_\ell$.
		In what follows we shall denote by $\tilde{N}:=\{1,\ldots,n\}\setminus\bigcup_{i=1}^kM_i$ and $N_{i\ell}:=N_i\cap N_\ell$, $i\neq \ell$, so that
		\begin{align}\label{Eq: Ntilde as union of Niell}
			\tilde{N}=\bigcup_{i\neq \ell}N_{i\ell}\,.
		\end{align}
		For the sake of simplicity we also set
		\begin{align*}
			\frac{\delta^{|N_i|+2|M_i|}x_\xi(t_i)}{\delta\xi_{N_i}\delta\xi_{M_i}^2}
			:=\frac{\delta^{|N_i|+2|M_i|}x_\xi(t_i)}{\prod_{j\in N_i}\delta\xi(s_j)\prod_{\ell\in M_i}\delta\xi(s_\ell)^2}\,.
		\end{align*}
		Notice that each $j\in M_i$ contributes for two $\xi$-derivatives, \textit{cf.} Equation \eqref{Eq: M,N sets}.
		With these definitions, Equation \eqref{Eq: SDE multiplicative correlation - series of Gamma delta map} reduces to
		\begin{align}\label{Eq: SDE multiplicative correlation - set series of Gamma delta map}
			\Gamma_{\delta/2}[x_\xi(t_1)\cdots x_\xi(t_k)]\bigg|_{\xi=0}
			=\sum_{\substack{n\geq 0\\p_1+\ldots+p_k=2n}}\frac{1}{2^n}
			\sum_{\substack{N_1\ldots,N_k\\M_1,\ldots,M_k}}
			2^{|\tilde{N}|}
			\int\limits_{\mathbb{R}^n}\prod_{i=1}^k\frac{\delta^{|N_i|+2|M_i|}x_{\xi,p_i}(t_i)}{\delta\xi_{N_i}\delta\xi_{M_i}^2}\bigg|_{\xi=0}
			\mathrm{d}s_1\cdots\mathrm{d}s_n\,,
		\end{align}
		where the sum over $N_1,\ldots,N_k,M_1,\ldots,M_k$ runs over the sets defined in Equation \eqref{Eq: M,N sets}.
		
		Notice that, for all $j\in N_{i\ell}$, there are two possible ways to distribute $\dfrac{\delta^2}{\delta\xi(s_j)^2}$ on the product $x_\xi(t_i)x_\xi(t_\ell)$ so that each $x_\xi$-factor is derived only once, thus producing overall the same analytical contribution.
		This occurs a number of times equal to $\sum_{i\neq \ell}|N_{i\ell}|=|\cup_{i\neq\ell}N_{i\ell}|=|\tilde{N}|$, leading to the factor $2^{|\tilde{N}|}$ in Equation \eqref{Eq: SDE multiplicative correlation - set series of Gamma delta map}.
		
		
		On account of the integration over $s_1,\ldots,s_n$ each term in the series \eqref{Eq: SDE multiplicative correlation - set series of Gamma delta map} does not depend on the particular choice of $N_1,\ldots,N_k,M_1,\ldots,M_k$ but only on the cardinality $m_i:=|M_i|$, $n_i:=|N_i|$, $\tilde{n}:=|\tilde{N}|$ and $n_{i\ell}:=|N_{i\ell}|$.
		Hence Equation \eqref{Eq: SDE multiplicative correlation - set series of Gamma delta map} simplifies further as
		\begin{multline}\label{Eq: SDE multiplicative correlation - cardinality set series of Gamma delta map}
			\Gamma_{\delta/2}[x_\xi(t_1)\cdots x_\xi(t_k)]\bigg|_{\xi=0}
			=\sum_{\substack{n\geq 0\\p_1+\ldots+p_k=2n}}
			\sum_{n_{i\ell},m_1,\ldots,m_k}
			\frac{2^{\tilde{n}-n}}{\prod_i m_i!\tilde{n}_{ii}!}
			\\
			\int\limits_{\mathbb{R}^n}\prod_{i=1}^k\frac{\delta^{n_i+2m_i}x_{\xi,p_i}(t_i)}{\delta\xi_{N_i}\delta\xi_{M_i}^2}\bigg|_{\xi=0}
			\mathrm{d}s_1\cdots\mathrm{d}s_n\,,
		\end{multline}
		where the sum over $\{n_{i\ell}\}_{i,\ell=1}^k,\{m_j\}_{j=1}^k$ is over the possible cardinality of the sets defined in Equations \eqref{Eq: M,N sets}, while the sets $N_1,\ldots,N_k$, $M_1,\ldots,M_k$ correspond to an arbitrary choice compatible with the chosen cardinality ---notice that $n_i=\sum_{\ell\neq i}n_{i\ell}$ while we set $\tilde{n}_{ii}:=\sum_{\ell> i}n_{i\ell}$.
		
		The factorials in Equation \eqref{Eq: SDE multiplicative correlation - cardinality set series of Gamma delta map} are due to all possible ways to arrange the elements of $M_1,\ldots, M_k$, $N_1,\ldots,N_k$, namely
		\begin{multline*}
			{n\choose m_1}
			{n-m_1\choose m_2}
			\cdots
			{n-(m_1+\ldots+m_{k-1})\choose
				\underbrace{n-(m_1+\ldots+m_k)}_{\tilde{n}}}
			\\
			\cdot
			{\tilde{n}\choose n_1}
			{\tilde{n}-n_1\choose n_2-n_{12}}
			\cdots
			{\tilde{n}-\sum_{i=1}^{k-1}
				(n_i-\sum_{\ell<i}n_{i\ell})
				\choose
				\tilde{n}-\sum_{i=1}^k
				(n_i-\sum_{\ell<i}n_{i\ell})}
			=\frac{n!}{\prod_{i=1}^km_i!\tilde{n}_{ii}!}\,.
		\end{multline*}
		The next step in the discussion will be to convert Equation \eqref{Eq: SDE multiplicative correlation - cardinality set series of Gamma delta map} into a sum over graphs.
		To this avail, first of all we focus our attention on 
		\begin{align*}
			\frac{\delta^nx_\xi(t)}{\delta\xi(s_1)\cdots\delta\xi(s_n)}\bigg|_{\xi=0}\,.
		\end{align*}
		An expansion in graphs of this contribution is obtained using Equation \eqref{Eq: SDE n-th order expansion - multiplicative case}, in particular for $p\geq 2$
		\begin{multline}\label{Eq: SDE n-th xi derivative - multiplicative case}
			\frac{\delta^n x_{\xi,p}(t)}{\delta\xi(s_1)\cdots\delta\xi(s_n)}\bigg|_{\xi=0}
			=\sum_{k\geq n}
			\sum_{\pi\in\mathfrak{S}_n}
			G(t,s_{\pi(1)})\chi(s_{\pi(1)})\beta_k(s_{\pi(1)})
			\\
			\frac{\delta^{n-1}}{\prod_{j=2}^n\delta\xi(s_{\pi(j)})}
			\sum_{\substack{j_1+\ldots+j_{p-1}=k\\
					\sum_\ell\ell j_\ell=p-1}}
			\frac{1}{j_1!\cdots j_{p-1}!}
			[x_{\xi,1}^{j_1}\cdots x_{\xi,p-1}^{j_{p-1}}](s_{\pi(1)})\,,
		\end{multline}
		where $\mathfrak{S}_n$ denotes the $n$-th permutation group.
		Equation \eqref{Eq: SDE n-th xi derivative - multiplicative case} determines recursively the $n$-th $\xi$-derivative of $x_\xi$ at $\xi=0$.
		This can be turned into a graph expansion by an adaptation of the convention in Remark \ref{Rmk: MSR expectation value - graph representation with squared vertices} used in the proof of Theorem \ref{Thm: SDE-MSR correspondence - additive case}.
		In particular we adopt the following rules:
		\begin{enumerate}[(a)]
			\item
			We assign to $x_{\xi=0}(t)=x_0$ the graph
			\begin{tikzpicture}[every node/.style={sloped,allow upside down}]
				\draw (0,0) -- (0.5,0);
				\filldraw (0,0) circle (1.5pt) node[left] {$t$};
			\end{tikzpicture};
			
			\item\label{Item: SDE nth order xi derivative - edge rules}
			We assign to $G(t,s)=\vartheta(t-s)$ the directed edge
			\begin{tikzpicture}[every node/.style={sloped,allow upside down}]
				\draw (0,0) -- node {\midarrow} (0.5,0);
				\filldraw (0,0) circle (1.5pt) node[left] {$t$};
				\filldraw (0.5,0) circle (1.5pt) node[right] {$s$};
			\end{tikzpicture}
			with the convention that, if $t=s$, then the closed loop
			\begin{tikzpicture}[every node/.style={sloped,allow upside down}]
				\path (0,0) edge [loop right] node {} (0,0);
				\filldraw (0,0) circle (1.5pt) node[left] {$t$};
			\end{tikzpicture}
			is not drawn and the graph is multiplied by a factor $1/2$.
			As a matter of fact, this rule implements the convention $G(t,t)=1/2$, whose necessity is explained in Remark \ref{Rmk: SDE formalism deals with Stratonovich only}. Carrying on the analogy with Butcher series, see Remark \ref{Rmk: on B-series}, these loops are the only aromas allowed by the retarded nature of $G$;
			
			\item
			To each factor $\chi\beta_j$ we assign a vertex
			\begin{tikzpicture}[every node/.style={sloped,allow upside down}]
				\node[rectangle,draw,fill=white] at (0,0) {$j$};
			\end{tikzpicture};
			
			\item\label{Item: SDE nth order xi derivative - squared vertex rules}
			To a factor $\chi^2\beta_{j_1}\beta_{j_2}$ we associate a single double squared vertex
			\begin{tikzpicture}[every node/.style={sloped,allow upside down}]
				\node[rectangle split, rectangle split horizontal, rectangle split parts=2,draw,fill=white] at (0,0) {\nodepart{one} $j_1$ \nodepart{two} $j_2$};
			\end{tikzpicture}.
		\end{enumerate}
		The graphs obtained with this procedure abide by the properties outlined in the Items \eqref{Item: SDE graph representation of additive SDE - number of vertices and edges}-\eqref{Item: SDE graph representation of additive SDE - properties of squares vertices}-\eqref{Item: SDE graph representation of additive SDE - blooming recursion}. As an example the graphs associated to the first 3 $\xi$-derivatives are 
		\begin{center}
			\begin{tikzpicture}[every node/.style={sloped,allow upside down}]
				\filldraw (0.25,0.5) node {$n=1$};
				\draw (0,0) -- node {\midarrow} (1,0);
				\filldraw (0,0) circle (1.5pt) node[left] {$t$};
				\node[rectangle,draw,fill=white] at (1,0) {$0$};
			\end{tikzpicture}
			\,,\,
			\begin{tikzpicture}[every node/.style={sloped,allow upside down}]
				\filldraw (0.25,0.5) node {$n=2$};
				\filldraw (0,0) circle (1.5pt) node[left] {$t$};
				\draw (0,0) -- node {\midarrow} (1,0);
				\draw (1,0) -- node {\midarrow} (2,0);
				\node[rectangle,draw,fill=white] at (1,0) {$1$};
				\node[rectangle,draw,fill=white] at (2,0) {$0$};
			\end{tikzpicture}
			\,,\,
			\begin{tikzpicture}[every node/.style={sloped,allow upside down}]
				\filldraw (0.25,0.5) node {$n=3$};
				\filldraw (0,0) circle (1.5pt) node[left] {$t$};
				\draw (0,0) -- node {\midarrow} (1,0);
				\draw (1,0) -- node {\midarrow} (2,0);
				\draw (2,0) -- node {\midarrow} (3,0);
				\node[rectangle,draw,fill=white] at (1,0) {$1$};
				\node[rectangle,draw,fill=white] at (2,0) {$1$};
				\node[rectangle,draw,fill=white] at (3,0) {$0$};
			\end{tikzpicture}
			$+$
			\begin{tikzpicture}[every node/.style={sloped,allow upside down}]
				\filldraw (0,0) circle (1.5pt) node[left] {$t$};
				\draw (0,0) -- node {\midarrow} (1,0);
				\draw (1,0) -- node {\midarrow} (2,1);
				\draw (1,0) -- node {\midarrow} (2,0);
				\node[rectangle,draw,fill=white] at (1,0) {$2$};
				\node[rectangle,draw,fill=white] at (2,0) {$0$};
				\node[rectangle,draw,fill=white] at (2,1) {$0$};
			\end{tikzpicture}
		\end{center}
		Focusing on the case $n=2$ and making explicit the base point coordinates we find
		\begin{center}
			$\dfrac{\delta^2x_\xi(t)}{\delta\xi(s_1)\delta\xi(s_2)}=$
			\begin{tikzpicture}[every node/.style={sloped,allow upside down}]
				\filldraw (1,0.5) node {$s_1$};
				\filldraw (2,0.5) node {$s_2$};
				\filldraw (0,0) circle (1.5pt) node[left] {$t$};
				\draw (0,0) -- node {\midarrow} (1,0);
				\draw (1,0) -- node {\midarrow} (2,0);
				\node[rectangle,draw,fill=white] at (1,0) {$1$};
				\node[rectangle,draw,fill=white] at (2,0) {$0$};
			\end{tikzpicture}
			$+$
			\begin{tikzpicture}[every node/.style={sloped,allow upside down}]
				\filldraw (1,0.5) node {$s_2$};
				\filldraw (2,0.5) node {$s_1$};
				\filldraw (0,0) circle (1.5pt) node[left] {$t$};
				\draw (0,0) -- node {\midarrow} (1,0);
				\draw (1,0) -- node {\midarrow} (2,0);
				\node[rectangle,draw,fill=white] at (1,0) {$1$};
				\node[rectangle,draw,fill=white] at (2,0) {$0$};
			\end{tikzpicture}
			\,,\quad
			$\dfrac{\delta^2x_\xi(t)}{\delta\xi(s)^2}=$
			\begin{tikzpicture}[every node/.style={sloped,allow upside down}]
				\filldraw (1.5,0.5) node {$s$};
				\filldraw (0,0) circle (1.5pt) node[left] {$t$};
				\draw (0,0) -- node {\midarrow} (1.5,0);
				\node[rectangle split, rectangle split horizontal, rectangle split parts=2,draw,fill=white] at (1.5,0) {\nodepart{one} $1$ \nodepart{two} $0$};
			\end{tikzpicture}
			\,,
		\end{center}
		where in the last graph we considered the convention $\vartheta_0=1/2$. Considering in addition the graph rules \ref{Graph: Rules 1}, the graph expansion of
		\begin{align*}
			\sum_{n_{i\ell},m_1,\ldots,m_k}
			\frac{2^{\tilde{n}-n}}{\prod_i m_i!\tilde{n}_{ii}!}
			\int\limits_{\mathbb{R}^n}\prod_{i=1}^k\frac{\delta^{n_i+2m_i}x_{\xi,p_i}(t_i)}{\delta\xi_{N_i}\delta\xi_{M_i}^2}\bigg|_{\xi=0}
			\mathrm{d}s_1\ldots\mathrm{d}s_n\,,
		\end{align*}
		consists of a sum over graphs $\gamma\in\mathcal{G}_{\textsc{sde}}^{(n,k)}$, $n:=\sum_{i=1}^k (n_i+2m_i)$, abiding by the following properties:
		\begin{enumerate}[(i)]
			\item\label{Item: SDE graph representation of multiplicative SDE - vertices-edges structure}
			Each $\gamma\in\mathcal{G}_{\textsc{sde}}^{(n,k)}$ is an oriented graph which contains neither loops nor closed oriented paths with $n$ doubled-squared vertices
			\begin{tikzpicture}[every node/.style={sloped,allow upside down}]
				\node[rectangle split, rectangle split horizontal, rectangle split parts=2,draw,fill=white] at (4,0) {\nodepart{one} $\cdot$ \nodepart{two} $\cdot$};
			\end{tikzpicture}\;,
			$k$ vertices
			\begin{tikzpicture}[every node/.style={sloped,allow upside down}]
				\filldraw (0,0) circle (1.5pt) node[left]{$t_1$};
			\end{tikzpicture}
			,\,\ldots,
			\begin{tikzpicture}[every node/.style={sloped,allow upside down}]
				\filldraw (0,0) circle (1.5pt) node[left]{$t_k$};
			\end{tikzpicture}
			and $n$ directed edges
			\begin{tikzpicture}[every node/.style={sloped,allow upside down}]
				\draw (0,0) -- node {\midarrow} (0.5,0);
			\end{tikzpicture}\,.
			An example is given by
			
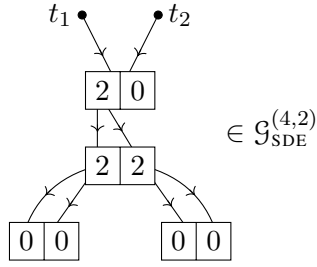
\begin{center}				
				\begin{tikzpicture}[every node/.style={sloped,allow upside down}]
					\filldraw (-0.5,0) circle (1.5pt) node[left] {$t_1$};
					\filldraw (0.5,0) circle (1.5pt) node[right] {$t_2$};
					\draw (-0.5,0) -- node {\midarrow} (0,-1);
					\draw (0.5,0) -- node {\midarrow} (0,-1);
					\draw (-0.3,-1) -- node {\midarrow} (-0.3,-2);
					\draw (-0.3,-1) -- node {\midarrow} (0.3,-2);
					\draw (-0.3,-2) -- node {\midarrow} (-1,-3);
					\draw (-0.3,-2) edge [bend right] node {\midarrow} (-1.3,-3);
					\draw (0.3,-2) -- node {\midarrow} (1,-3);
					\draw (0.3,-2) edge [bend left] node {\midarrow} (1.3,-3);
					\node[rectangle split, rectangle split horizontal, rectangle split parts=2,draw,fill=white] at (0,-1) {\nodepart{one} $2$ \nodepart{two} $0$};
					\node[rectangle split, rectangle split horizontal, rectangle split parts=2,draw,fill=white] at (0,-2) {\nodepart{one} $2$ \nodepart{two} $2$};
					\node[rectangle split, rectangle split horizontal, rectangle split parts=2,draw,fill=white] at (-1,-3) {\nodepart{one} $0$ \nodepart{two} $0$};
					\node[rectangle split, rectangle split horizontal, rectangle split parts=2,draw,fill=white] at (1,-3) {\nodepart{one} $0$ \nodepart{two} $0$};
					\node at (2,-1.5) {$\in\mathcal{G}_{\textsc{sde}}^{(4,2)}$};
				\end{tikzpicture}
			\end{center}
			which is one of the graphs obtained in the contraction
			\begin{align*}
				\frac{\delta^7 x_\xi(t_1)}{\delta\xi(s_1)\delta\xi(s_2)^2\delta\xi(s_3)^2\delta\xi(s_4)^2}
				\frac{\delta x_\xi(t_2)}{\delta\xi(s_1)}\,.
			\end{align*}
			
			\item\label{Item: SDE graph representation of multiplicative SDE - double vertex structure}
			Within $\gamma\in\mathcal{G}_{\textsc{sde}}^{(n,k)}$, each vertex
			\begin{tikzpicture}[every node/.style={sloped,allow upside down}]
				\filldraw (0,0) circle (1.5pt) node {};
			\end{tikzpicture}
			has a single outgoing edge, pointing to a double squared vertex
			\begin{tikzpicture}[every node/.style={sloped,allow upside down}]
				\node[rectangle split, rectangle split horizontal, rectangle split parts=2,draw,fill=white] at (0,0) {\nodepart{one} $\cdot$ \nodepart{two} $\cdot$};
			\end{tikzpicture}\;.
			Each of these has either one or two incoming edges.
			If
			\begin{tikzpicture}[every node/.style={sloped,allow upside down}]
				\node[rectangle split, rectangle split horizontal, rectangle split parts=2,draw,fill=white] at (0,0) {\nodepart{one} $j_1$ \nodepart{two} $j_2$};
			\end{tikzpicture}
			has two incoming edges, then there are $j_1$ outgoing edges from the squared vertex
			\begin{tikzpicture}[every node/.style={sloped,allow upside down}]
				\node[rectangle,draw,fill=white] at (0,0) {$j_1$};
			\end{tikzpicture}
			and $j_2$ outgoing edges from the squared vertex
			\begin{tikzpicture}[every node/.style={sloped,allow upside down}]
				\node[rectangle,draw,fill=white] at (0,0) {$j_2$};
			\end{tikzpicture}\;.
			Instead,
			if
			\begin{tikzpicture}[every node/.style={sloped,allow upside down}]
				\node[rectangle split, rectangle split horizontal, rectangle split parts=2,draw,fill=white] at (0,0) {\nodepart{one} $j_1$ \nodepart{two} $j_2$};
			\end{tikzpicture}
			has a single incoming edge, then
			\begin{tikzpicture}[every node/.style={sloped,allow upside down}]
				\node[rectangle,draw,fill=white] at (0,0) {$j_1$};
			\end{tikzpicture}
			has $\ell_1$ outgoing edges while
			\begin{tikzpicture}[every node/.style={sloped,allow upside down}]
				\node[rectangle,draw,fill=white] at (0,0) {$j_2$};
			\end{tikzpicture}
			has $\ell_2$ outgoing edges where either $(\ell_1,\ell_2)=(j_1,j_2-1)$ or $(\ell_1,\ell_2)=(j_1-1,j_2)$.
			
			Double squared vertices with one incoming edge are generated when two squared vertices which are joined by an edge share the same base point coordinate ---\textit{cf.} Items \eqref{Item: SDE nth order xi derivative - edge rules} and \eqref{Item: SDE nth order xi derivative - squared vertex rules}: An example reads
			\begin{center}
				$\dfrac{\delta^2x_\xi(t)}{\delta\xi(s)^2}=$
				\begin{tikzpicture}[every node/.style={sloped,allow upside down}]
					\filldraw (1.5,0.5) node {$s$};
					\filldraw (0,0) circle (1.5pt) node[left] {$t$};
					\draw (0,0) -- node {\midarrow} (1.5,0);
					\node[rectangle split, rectangle split horizontal, rectangle split parts=2,draw,fill=white] at (1.5,0) {\nodepart{one} $1$ \nodepart{two} $0$};
				\end{tikzpicture}
			\end{center}
			
			\item\label{Item: SDE multiplicative case - incoming edges ambiguity}
			For all double squared vertices
			\begin{tikzpicture}[every node/.style={sloped,allow upside down}]
				\node[rectangle split, rectangle split horizontal, rectangle split parts=2,draw,fill=white] at (0,0) {\nodepart{one} $j_1$ \nodepart{two} $j_2$};
			\end{tikzpicture}
			the incoming edges connect to either the squared vertex
			\begin{tikzpicture}[every node/.style={sloped,allow upside down}]
				\node[rectangle, draw,fill=white] at (0,0) {$j_1$};
			\end{tikzpicture}
			or to
			\begin{tikzpicture}[every node/.style={sloped,allow upside down}]
				\node[rectangle, draw,fill=white] at (0,0) {$j_2$};
			\end{tikzpicture}
			with the constraint that a single squared vertex
			\begin{tikzpicture}[every node/.style={sloped,allow upside down}]
				\node[rectangle, draw,fill=white] at (0,0) {$\cdot$};
			\end{tikzpicture}
			can be joined by at most one incoming edge.
			In what follows it will be relevant to keep track of whether an incoming edge connects to
			\begin{tikzpicture}[every node/.style={sloped,allow upside down}]
				\node[rectangle split, rectangle split horizontal, rectangle split parts=2,draw,fill=white] at (0,0) {\nodepart{one} $j_1$ \nodepart{two} $j_2$};
			\end{tikzpicture}
			through
			\begin{tikzpicture}[every node/.style={sloped,allow upside down}]
				\node[rectangle,draw,fill=white] at (0,0) {$j_1$};
			\end{tikzpicture}
			or through
			\begin{tikzpicture}[every node/.style={sloped,allow upside down}]
				\node[rectangle,draw,fill=white] at (0,0) {$j_2$};
			\end{tikzpicture}.
			Thus the following graphs are considered as different although they produce the same analytical contribution.
			In the following example this is marked by the different positions of the dotted and dashed edges:
			
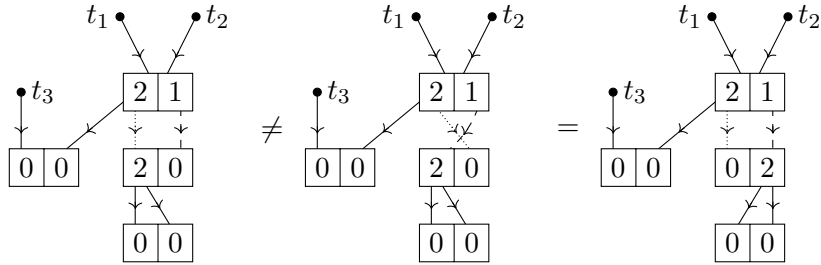
\begin{center}				
				\begin{tikzpicture}[every node/.style={sloped,allow upside down}]
					\filldraw (-0.5,0) circle (1.5pt) node[left] {$t_1$};
					\filldraw (0.5,0) circle (1.5pt) node[right] {$t_2$};
					\filldraw (-1.5-0.3,0) circle (1.5pt) node[right] {$t_3$};
					\draw (-1.5-0.3,0) -- node {\midarrow} (-1.5-0.3,-1);
					\draw (-0.5,0) -- node {\midarrow} (0,-1);
					\draw (0.5,0) -- node {\midarrow} (0,-1);
					\draw[densely dotted] (-0.3,-1) -- node {\midarrow} (-0.3,-2);
					\draw[dashed] (0.3,-1) -- node {\midarrow} (0.3,-2);
					\draw (-0.3,-1) -- node {\midarrow} (-1.5,-2);
					\draw (-1.5-0.3,-1) -- node {\midarrow} (-1.5-0.3,-2);
					\draw (-0.3,-2) -- node {\midarrow} (-0.3,-3);
					\draw (-0.3,-2) -- node {\midarrow} (0.3,-3);
					\node[rectangle split, rectangle split horizontal, rectangle split parts=2,draw,fill=white] at (0,-1) {\nodepart{one} $2$ \nodepart{two} $1$};
					\node[rectangle split, rectangle split horizontal, rectangle split parts=2,draw,fill=white] at (0,-2) {\nodepart{one} $2$ \nodepart{two} $0$};
					\node[rectangle split, rectangle split horizontal, rectangle split parts=2,draw,fill=white] at (-1.5,-2) {\nodepart{one} $0$ \nodepart{two} $0$};
					\node[rectangle split, rectangle split horizontal, rectangle split parts=2,draw,fill=white] at (-2,-1) {\nodepart{one} $0$ \nodepart{two} $2$};
					\node[rectangle split, rectangle split horizontal, rectangle split parts=2,draw,fill=white] at (0,-3) {\nodepart{one} $0$ \nodepart{two} $0$};
					\node at (1.5,-1.5) {$\neq$};
				\end{tikzpicture}
				\begin{tikzpicture}[every node/.style={sloped,allow upside down}]
					\filldraw (-0.5,0) circle (1.5pt) node[left] {$t_1$};
					\filldraw (0.5,0) circle (1.5pt) node[right] {$t_2$};
					\filldraw (-1.5-0.3,0) circle (1.5pt) node[right] {$t_3$};
					\draw (-1.5-0.3,0) -- node {\midarrow} (-1.5-0.3,-1);
					\draw (-0.5,0) -- node {\midarrow} (0,-1);
					\draw (0.5,0) -- node {\midarrow} (0,-1);
					\draw[densely dotted] (-0.4,-1) -- node {\midarrow} (0.4,-2);
					\draw[dashed] (0.4,-0.8) edge [bend left] node {\midarrow} (-0.45,-2);
					\draw (-0.3,-1) -- node {\midarrow} (-1.5,-2);
					\draw (-1.5-0.3,-1) -- node {\midarrow} (-1.5-0.3,-2);
					\draw (-0.3,-2) -- node {\midarrow} (-0.3,-3);
					\draw (-0.3,-2) -- node {\midarrow} (0.3,-3);
					\node[rectangle split, rectangle split horizontal, rectangle split parts=2,draw,fill=white] at (0,-1) {\nodepart{one} $2$ \nodepart{two} $1$};
					\node[rectangle split, rectangle split horizontal, rectangle split parts=2,draw,fill=white] at (0,-2) {\nodepart{one} $2$ \nodepart{two} $0$};
					\node[rectangle split, rectangle split horizontal, rectangle split parts=2,draw,fill=white] at (-1.5,-2) {\nodepart{one} $0$ \nodepart{two} $0$};
					\node[rectangle split, rectangle split horizontal, rectangle split parts=2,draw,fill=white] at (-2,-1) {\nodepart{one} $0$ \nodepart{two} $2$};
					\node[rectangle split, rectangle split horizontal, rectangle split parts=2,draw,fill=white] at (0,-3) {\nodepart{one} $0$ \nodepart{two} $0$};
					\node at (1.5,-1.5) {$=$};
				\end{tikzpicture}
				\begin{tikzpicture}[every node/.style={sloped,allow upside down}]
					\filldraw (-0.5,0) circle (1.5pt) node[left] {$t_1$};
					\filldraw (0.5,0) circle (1.5pt) node[right] {$t_2$};
					\filldraw (-1.5-0.3,0) circle (1.5pt) node[right] {$t_3$};
					\draw (-1.5-0.3,0) -- node {\midarrow} (-1.5-0.3,-1);
					\draw (-0.5,0) -- node {\midarrow} (0,-1);
					\draw (0.5,0) -- node {\midarrow} (0,-1);
					\draw[densely dotted] (-0.3,-1) -- node {\midarrow} (-0.3,-2);
					\draw[dashed] (0.3,-1) -- node {\midarrow} (0.3,-2);
					\draw (-0.3,-1) -- node {\midarrow} (-1.5,-2);
					\draw (-1.5-0.3,-1) -- node {\midarrow} (-1.5-0.3,-2);
					\draw (0.3,-2) -- node {\midarrow} (-0.3,-3);
					\draw (0.3,-2) -- node {\midarrow} (0.3,-3);
					\node[rectangle split, rectangle split horizontal, rectangle split parts=2,draw,fill=white] at (0,-1) {\nodepart{one} $2$ \nodepart{two} $1$};
					\node[rectangle split, rectangle split horizontal, rectangle split parts=2,draw,fill=white] at (0,-2) {\nodepart{one} $0$ \nodepart{two} $2$};
					\node[rectangle split, rectangle split horizontal, rectangle split parts=2,draw,fill=white] at (-1.5,-2) {\nodepart{one} $0$ \nodepart{two} $0$};
					\node[rectangle split, rectangle split horizontal, rectangle split parts=2,draw,fill=white] at (-2,-1) {\nodepart{one} $0$ \nodepart{two} $2$};
					\node[rectangle split, rectangle split horizontal, rectangle split parts=2,draw,fill=white] at (0,-3) {\nodepart{one} $0$ \nodepart{two} $0$};
				\end{tikzpicture}
			\end{center}
			Keeping track of this difference allows to recover the graph expansion of the $\xi$-derivatives from which $\gamma$ is generated.
			For example
			
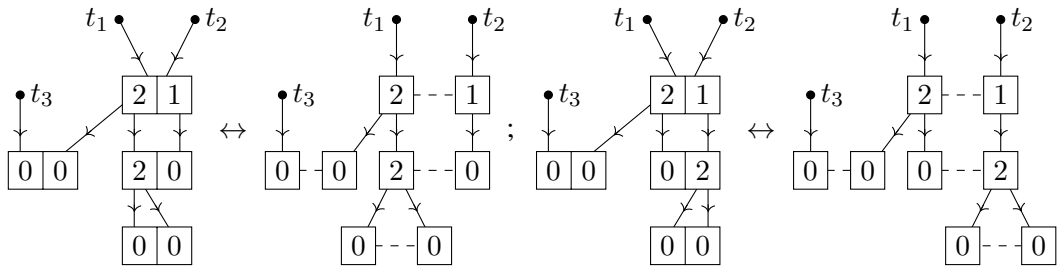
\begin{center}		
				\begin{tikzpicture}[every node/.style={sloped,allow upside down}]
					\filldraw (-0.5,0) circle (1.5pt) node[left] {$t_1$};
					\filldraw (0.5,0) circle (1.5pt) node[right] {$t_2$};
					\filldraw (-2,0) circle (1.5pt) node[right] {$t_3$};
					\draw[dashed] (-2,-1) edge [bend right] (-2-0.6,-2);
					\draw (-2,0) -- node {\midarrow} (-2,-1);
					\draw (-0.5,0) -- node {\midarrow} (-0.5,-1);
					\draw (0.5,0) -- node {\midarrow} (0.5,-1);
					\draw[dashed] (-0.5,-1) -- (0.5,-1);
					\draw (-0.5,-1) -- node {\midarrow} (-0.5,-2);
					\draw[dashed] (-0.5,-2) -- (0.5,-2);
					\draw[dashed] (-2,-2) -- (-1.25,-2);
					\draw (-2,-1) -- node {\midarrow} (-2,-2);
					\draw (-0.5,-1) -- node {\midarrow} (-1.25,-2);
					\draw (0.5,-1) -- node {\midarrow} (0.5,-2);
					\draw[dashed] (-1,-3) -- (0,-3);
					\draw (-0.5,-2) -- node {\midarrow} (-1,-3);
					\draw (-0.5,-2) -- node {\midarrow} (0,-3);
					\node[rectangle,draw,fill=white] at (-2,-1) {$2$};
					\node[rectangle,draw,fill=white] at (-2-0.6,-2) {$0$};
					\node[rectangle,draw,fill=white] at (-0.5,-1) {$2$};
					\node[rectangle,draw,fill=white] at (0.5,-1) {$1$};
					\node[rectangle,draw,fill=white] at (-0.5,-2) {$2$};
					\node[rectangle,draw,fill=white] at (0.5,-2) {$0$};
					\node[rectangle,draw,fill=white] at (-2,-2) {$0$};
					\node[rectangle,draw,fill=white] at (-1.25,-2) {$0$};
					\node[rectangle,draw,fill=white] at (0,-3) {$0$};
					\node[rectangle,draw,fill=white] at (-1,-3) {$0$};
					\node at (1,-1.5) {$\leftrightarrow$};
				\end{tikzpicture}	
				\begin{tikzpicture}[every node/.style={sloped,allow upside down}]
					\filldraw (-0.5,0) circle (1.5pt) node[left] {$t_1$};
					\filldraw (0.5,0) circle (1.5pt) node[right] {$t_2$};
					\filldraw (-1.25-0.3,0) circle (1.5pt) node[right] {$t_3$};
					\draw (-1.25-0.3,0) -- node {\midarrow} (-1.25-0.3,-1);
					\draw (-0.5,0) -- node {\midarrow} (0,-1);
					\draw (0.5,0) -- node {\midarrow} (0,-1);
					\draw(-0.3,-1) -- node {\midarrow} (-0.3,-2);
					\draw (0.3,-1) -- node {\midarrow} (0.3,-2);
					\draw (-0.3,-1) -- node {\midarrow} (-1.25,-2);
					\draw (-1.25-0.3,-1) -- node {\midarrow} (-1.25-0.3,-2);
					\draw (-0.3,-2) -- node {\midarrow} (-0.3,-3);
					\draw (-0.3,-2) -- node {\midarrow} (0.3,-3);
					\node[rectangle split, rectangle split horizontal, rectangle split parts=2,draw,fill=white] at (0,-1) {\nodepart{one} $2$ \nodepart{two} $1$};
					\node[rectangle split, rectangle split horizontal, rectangle split parts=2,draw,fill=white] at (0,-2) {\nodepart{one} $2$ \nodepart{two} $0$};
					\node[rectangle split, rectangle split horizontal, rectangle split parts=2,draw,fill=white] at (-1.25,-2) {\nodepart{one} $0$ \nodepart{two} $0$};
					\node[rectangle split, rectangle split horizontal, rectangle split parts=2,draw,fill=white] at (-1.75,-1) {\nodepart{one} $0$ \nodepart{two} $2$};
					\node[rectangle split, rectangle split horizontal, rectangle split parts=2,draw,fill=white] at (0,-3) {\nodepart{one} $0$ \nodepart{two} $0$};
					\draw (1,-1.5) node {;};
				\end{tikzpicture}
				\begin{tikzpicture}[every node/.style={sloped,allow upside down}]
					\filldraw (-0.5,0) circle (1.5pt) node[left] {$t_1$};
					\filldraw (0.5,0) circle (1.5pt) node[right] {$t_2$};
					\filldraw (-1.25-0.3,0) circle (1.5pt) node[right] {$t_3$};
					\draw (-1.25-0.3,0) -- node {\midarrow} (-1.25-0.3,-1);
					\draw (-0.5,0) -- node {\midarrow} (0,-1);
					\draw (0.5,0) -- node {\midarrow} (0,-1);
					\draw(-0.3,-1) -- node {\midarrow} (-0.3,-2);
					\draw (0.3,-1) -- node {\midarrow} (0.3,-2);
					\draw (-0.3,-1) -- node {\midarrow} (-1.25,-2);
					\draw (-1.25-0.3,-1) -- node {\midarrow} (-1.25-0.3,-2);
					\draw (0.3,-2) -- node {\midarrow} (-0.3,-3);
					\draw (0.3,-2) -- node {\midarrow} (0.3,-3);
					\node[rectangle split, rectangle split horizontal, rectangle split parts=2,draw,fill=white] at (0,-1) {\nodepart{one} $2$ \nodepart{two} $1$};
					\node[rectangle split, rectangle split horizontal, rectangle split parts=2,draw,fill=white] at (0,-2) {\nodepart{one} $0$ \nodepart{two} $2$};
					\node[rectangle split, rectangle split horizontal, rectangle split parts=2,draw,fill=white] at (-1.25,-2) {\nodepart{one} $0$ \nodepart{two} $0$};
					\node[rectangle split, rectangle split horizontal, rectangle split parts=2,draw,fill=white] at (-1.75,-1) {\nodepart{one} $0$ \nodepart{two} $2$};
					\node[rectangle split, rectangle split horizontal, rectangle split parts=2,draw,fill=white] at (0,-3) {\nodepart{one} $0$ \nodepart{two} $0$};
					\node at (1,-1.5) {$\leftrightarrow$};
				\end{tikzpicture}
				\begin{tikzpicture}[every node/.style={sloped,allow upside down}]
					\filldraw (-0.5,0) circle (1.5pt) node[left] {$t_1$};
					\filldraw (0.5,0) circle (1.5pt) node[right] {$t_2$};
					\filldraw (-2,0) circle (1.5pt) node[right] {$t_3$};
					\draw[dashed] (-2,-1) edge [bend right] (-2-0.6,-2);
					\draw (-2,0) -- node {\midarrow} (-2,-1);
					\draw (-0.5,0) -- node {\midarrow} (-0.5,-1);
					\draw (0.5,0) -- node {\midarrow} (0.5,-1);
					\draw[dashed] (-0.5,-1) -- (0.5,-1);
					\draw (-0.5,-1) -- node {\midarrow} (-0.5,-2);
					\draw[dashed] (-0.5,-2) -- (0.5,-2);
					\draw[dashed] (-2,-2) -- (-1.25,-2);
					\draw (-2,-1) -- node {\midarrow} (-2,-2);
					\draw (-0.5,-1) -- node {\midarrow} (-1.25,-2);
					\draw (0.5,-1) -- node {\midarrow} (0.5,-2);
					\draw[dashed] (1,-3) -- (0,-3);
					\draw (0.5,-2) -- node {\midarrow} (1,-3);
					\draw (0.5,-2) -- node {\midarrow} (0,-3);
					\node[rectangle,draw,fill=white] at (-2,-1) {$2$};
					\node[rectangle,draw,fill=white] at (-2-0.6,-2) {$0$};
					\node[rectangle,draw,fill=white] at (-0.5,-1) {$2$};
					\node[rectangle,draw,fill=white] at (0.5,-1) {$1$};
					\node[rectangle,draw,fill=white] at (-0.5,-2) {$0$};
					\node[rectangle,draw,fill=white] at (0.5,-2) {$2$};
					\node[rectangle,draw,fill=white] at (-2,-2) {$0$};
					\node[rectangle,draw,fill=white] at (-1.25,-2) {$0$};
					\node[rectangle,draw,fill=white] at (0,-3) {$0$};
					\node[rectangle,draw,fill=white] at (1,-3) {$0$};
				\end{tikzpicture}
			\end{center}
			contributes to the terms
			\begin{multline*}
				\frac{\delta^3 x_\xi(t_3)}{\delta\xi(s_1)\delta\xi(s_2)^2}
				\frac{\delta^5x_\xi(t_1)}{\delta\xi(s_1)\delta\xi(s_3)\delta\xi(s_4)\delta\xi(s_5)^2}
				\frac{\delta^2x_\xi(t_2)}{\delta\xi(s_3)\delta\xi(s_4)}\,,
				\\
				\frac{\delta^3 x_\xi(t_3)}{\delta\xi(s_1)\delta\xi(s_2)^2}
				\frac{\delta^3x_\xi(t_1)}{\delta\xi(s_1)\delta\xi(s_3)\delta\xi(s_4)}
				\frac{\delta^4x_\xi(t_2)}{\delta\xi(s_3)\delta\xi(s_4)\delta\xi(s_5)^2}\,.
			\end{multline*}
			In the graph above the dashed lines underline the squared vertices which are joined together.
			
			\item
			Per definition, $\varsigma\in\operatorname{Aut}_{\mathcal{G}_{\textsc{sde}}^{(n,k)}}(\gamma)$, $\gamma\in\mathcal{G}_{\textsc{sde}}^{(n,k)}$, is an arbitrary permutation of the double squared vertices
			\begin{tikzpicture}[every node/.style={sloped,allow upside down}]
				\node[rectangle split, rectangle split horizontal, rectangle split parts=2,draw,fill=white] at (4,0) {\nodepart{one} $\cdot$ \nodepart{two} $\cdot$};
			\end{tikzpicture}
			which preserves all edge connections and orientations;
			
			\item
			For fixed $n_{j,\ell},m_1,\ldots,m_k$, $\gamma\in\mathcal{G}_{\textsc{sde}}^{(n,k)}$, on account of the integration over $s_1,\ldots,s_n$, $n=\sum_{i=1}^k(m_i+n_{ii})$, appears exactly
			\begin{align*}
				\frac{1}{2^{\tilde{n}}}\sum_{i=1}^k\frac{m_i!\tilde{n}_{ii}!}{|\operatorname{Aut}_{\mathcal{G}_{\textsc{sde}}^{(n,k)}}(\gamma)|}\,,
			\end{align*}
			times in the expansion in Equation \eqref{Eq: SDE multiplicative correlation - cardinality set series of Gamma delta map}. The factor $2^{-\tilde{n}}$ accounts for a redundancy of the graph $\gamma\in\mathcal{G}_{\textsc{sde}}^{(n,k)}$ which we will make explicit by introducing a suitable equivalence relation on $\mathcal{G}_{\textsc{sde}}^{(n,k)}$ ---\textit{cf.} Equation \eqref{Eq: MSR order expansion - multiplicative case - quotient graph expansion}.
		\end{enumerate}
		Overall, the graph expansion of Equation \eqref{Eq: SDE multiplicative correlation - cardinality set series of Gamma delta map} amounts to
		\begin{align}\label{Eq: SDE multiplicative correlation - graph expansion of Gamma delta map}
			\Gamma_{\delta/2}[x_\xi(t_1)\cdots x_\xi(t_k)]\bigg|_{\xi=0}
			=\sum_{n\geq 0}
			\sum_{\gamma\in\mathcal{G}_{\textsc{sde}}^{(n,k)}}
			\frac{2^{-n}}{|\operatorname{Aut}_{\mathcal{G}_{\textsc{sde}}^{(n,k)}}(\gamma)|}\gamma\,.
		\end{align}
		At this stage, we may further simplify Equation \eqref{Eq: SDE multiplicative correlation - graph expansion of Gamma delta map} by observing that two graphs $\gamma,\gamma'\in\mathcal{G}_{\textsc{sde}}^{(n,k)}$ may lead to the same analytical contribution, \textit{e.g.}
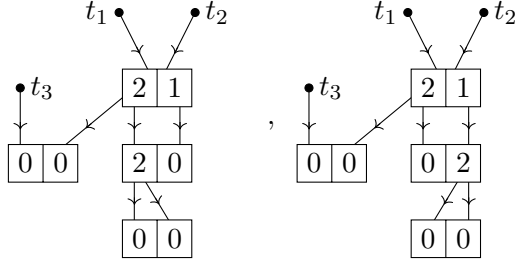
		\begin{center}
			\begin{tikzpicture}[every node/.style={sloped,allow upside down}]
				\filldraw (-0.5,0) circle (1.5pt) node[left] {$t_1$};
				\filldraw (0.5,0) circle (1.5pt) node[right] {$t_2$};
				\filldraw (-1.25-0.3,0) circle (1.5pt) node[right] {$t_3$};
				\draw (-1.25-0.3,0) -- node {\midarrow} (-1.25-0.3,-1);
				\draw (-0.5,0) -- node {\midarrow} (0,-1);
				\draw (0.5,0) -- node {\midarrow} (0,-1);
				\draw(-0.3,-1) -- node {\midarrow} (-0.3,-2);
				\draw (0.3,-1) -- node {\midarrow} (0.3,-2);
				\draw (-0.3,-1) -- node {\midarrow} (-1.25,-2);
				\draw (-1.25-0.3,-1) -- node {\midarrow} (-1.25-0.3,-2);
				\draw (-0.3,-2) -- node {\midarrow} (-0.3,-3);
				\draw (-0.3,-2) -- node {\midarrow} (0.3,-3);
				\node[rectangle split, rectangle split horizontal, rectangle split parts=2,draw,fill=white] at (0,-1) {\nodepart{one} $2$ \nodepart{two} $1$};
				\node[rectangle split, rectangle split horizontal, rectangle split parts=2,draw,fill=white] at (0,-2) {\nodepart{one} $2$ \nodepart{two} $0$};
				\node[rectangle split, rectangle split horizontal, rectangle split parts=2,draw,fill=white] at (-1.25,-2) {\nodepart{one} $0$ \nodepart{two} $0$};
				\node[rectangle split, rectangle split horizontal, rectangle split parts=2,draw,fill=white] at (-1.75,-1) {\nodepart{one} $0$ \nodepart{two} $2$};
				\node[rectangle split, rectangle split horizontal, rectangle split parts=2,draw,fill=white] at (0,-3) {\nodepart{one} $0$ \nodepart{two} $0$};
				\draw (1,-1.5) node {=};
			\end{tikzpicture}
			\begin{tikzpicture}[every node/.style={sloped,allow upside down}]
				\filldraw (-0.5,0) circle (1.5pt) node[left] {$t_1$};
				\filldraw (0.5,0) circle (1.5pt) node[right] {$t_2$};
				\filldraw (-1.25-0.3,0) circle (1.5pt) node[right] {$t_3$};
				\draw (-1.25-0.3,0) -- node {\midarrow} (-1.25-0.3,-1);
				\draw (-0.5,0) -- node {\midarrow} (0,-1);
				\draw (0.5,0) -- node {\midarrow} (0,-1);
				\draw(-0.3,-1) -- node {\midarrow} (-0.3,-2);
				\draw (0.3,-1) -- node {\midarrow} (0.3,-2);
				\draw (-0.3,-1) -- node {\midarrow} (-1.25,-2);
				\draw (-1.25-0.3,-1) -- node {\midarrow} (-1.25-0.3,-2);
				\draw (0.3,-2) -- node {\midarrow} (-0.3,-3);
				\draw (0.3,-2) -- node {\midarrow} (0.3,-3);
				\node[rectangle split, rectangle split horizontal, rectangle split parts=2,draw,fill=white] at (0,-1) {\nodepart{one} $2$ \nodepart{two} $1$};
				\node[rectangle split, rectangle split horizontal, rectangle split parts=2,draw,fill=white] at (0,-2) {\nodepart{one} $0$ \nodepart{two} $2$};
				\node[rectangle split, rectangle split horizontal, rectangle split parts=2,draw,fill=white] at (-1.25,-2) {\nodepart{one} $0$ \nodepart{two} $0$};
				\node[rectangle split, rectangle split horizontal, rectangle split parts=2,draw,fill=white] at (-1.75,-1) {\nodepart{one} $0$ \nodepart{two} $2$};
				\node[rectangle split, rectangle split horizontal, rectangle split parts=2,draw,fill=white] at (0,-3) {\nodepart{one} $0$ \nodepart{two} $0$};
			\end{tikzpicture}
		\end{center}
		are different graphs in $\mathcal{G}_{\textsc{sde}}^{(4,3)}$ but leads to the same analytical contribution.
		This multiplicity extends the one already encountered when discussing the factor $2^{\tilde{n}}$ within Equation \eqref{Eq: SDE multiplicative correlation - set series of Gamma delta map}.
		Indeed, for any $j\in N_{i\ell}$ the same analytical contribution is produced by two identical graphs due to the action of $\dfrac{\delta^2}{\delta\xi(s_j)^2}$ on $x_\xi(t_1)x_\xi(t_\ell)$ in such a way that each $x_\xi$-factor is derived only once. This suggests to introduce an equivalence relation between elements of $\mathcal{G}_{\textsc{sde}}^{(n,k)}$ by declaring $\gamma\sim\gamma'$ if their analytical contributions coincide.
		
		Denoting by $[\mathcal{G}]_{\textsc{sde}}^{(n,k)}:=\mathcal{G}_{\textsc{sde}}^{(n,k)}/\sim$ the resulting quotient space, we observe that any equivalence class $[\gamma]\in[\mathcal{G}]_{\textsc{sde}}^{(n,k)}$, $\gamma\in\mathcal{G}_{\textsc{sde}}^{(n,k)}$, has $2^{c(\gamma)}$ representatives.
		Here $c(\gamma)=|C(\gamma)|$ is the cardinality of the set of squared vertices
		\begin{tikzpicture}[every node/.style={sloped,allow upside down}]
			\node[rectangle split, rectangle split horizontal, rectangle split parts=2,draw,fill=white] at (0,0) {\nodepart{one} $\cdot$ \nodepart{two} $\cdot$};
		\end{tikzpicture}
		with two incoming edges emanated from two different vertices, that is either a pair of distinct squared vertices
		\begin{tikzpicture}[every node/.style={sloped,allow upside down}]
			\node[rectangle,draw,fill=white] at (0,0) {$j_1$};
		\end{tikzpicture}\,,
		\begin{tikzpicture}[every node/.style={sloped,allow upside down}]
			\node[rectangle,draw,fill=white] at (0,0) {$j_2$};
		\end{tikzpicture}
		or a pair of vertices
		\begin{tikzpicture}[every node/.style={sloped,allow upside down}]
			\filldraw (0,0) circle (1.5pt) node {};
		\end{tikzpicture}, or a squared vertex together with a single vertex \begin{tikzpicture}[every node/.style={sloped,allow upside down}]
			\filldraw (0,0) circle (1.5pt) node {};
		\end{tikzpicture}.
		As a matter of fact, for all
		\begin{tikzpicture}[every node/.style={sloped,allow upside down}]
			\node[rectangle split, rectangle split horizontal, rectangle split parts=2,draw,fill=white] at (0,0) {\nodepart{one} $\cdot$ \nodepart{two} $\cdot$};
		\end{tikzpicture}
		$\in C(\gamma)$
		we may flip the sources of the incoming edges producing a different graph with the same analytical contribution.
		For example, in the following graph $\gamma\in\mathcal{G}_{\textsc{sde}}^{(4,2)}$ the dashed double squared vertices
		\begin{tikzpicture}[every node/.style={sloped,allow upside down}]
			\node[rectangle split, rectangle split horizontal, rectangle split parts=2,draw,fill=white,dashed] at (0,0) {\nodepart{one} $\cdot$ \nodepart{two} $\cdot$};
		\end{tikzpicture}
		denote elements lying in $C(\gamma)$:
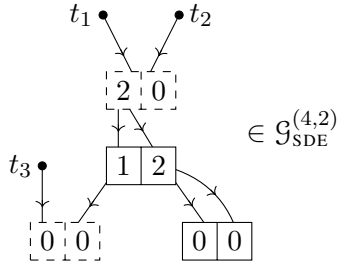
		\begin{center}				
			\begin{tikzpicture}[every node/.style={sloped,allow upside down}]
				\filldraw (-0.5,0) circle (1.5pt) node[left] {$t_1$};
				\filldraw (0.5,0) circle (1.5pt) node[right] {$t_2$};
				\filldraw (-1.3,-2) circle (1.5pt) node[left] {$t_3$};
				\draw (-0.5,0) -- node {\midarrow} (0,-1);
				\draw (0.5,0) -- node {\midarrow} (0,-1);
				\draw (-0.3,-1) -- node {\midarrow} (-0.3,-2);
				\draw (-0.3,-1) -- node {\midarrow} (0.3,-2);
				\draw (-0.3,-2) -- node {\midarrow} (-1,-3);
				\draw (-1.3,-2) -- node {\midarrow} (-1.3,-3);
				\draw (0.3,-2) -- node {\midarrow} (1,-3);
				\draw (0.3,-2) edge [bend left] node {\midarrow} (1.3,-3);
				\node[dashed,rectangle split, rectangle split horizontal, rectangle split parts=2,draw,fill=white] at (0,-1) {\nodepart{one} $2$ \nodepart{two} $0$};
				\node[rectangle split, rectangle split horizontal, rectangle split parts=2,draw,fill=white] at (0,-2) {\nodepart{one} $1$ \nodepart{two} $2$};
				\node[dashed,rectangle split, rectangle split horizontal, rectangle split parts=2,draw,fill=white] at (-1,-3) {\nodepart{one} $0$ \nodepart{two} $0$};
				\node[rectangle split, rectangle split horizontal, rectangle split parts=2,draw,fill=white] at (1,-3) {\nodepart{one} $0$ \nodepart{two} $0$};
				\node at (2,-1.5) {$\in\mathcal{G}_{\textsc{sde}}^{(4,2)}$};
			\end{tikzpicture}
		\end{center}
		Notice that $\operatorname{Aut}_{\mathcal{G}_{\textsc{sde}}^{(n,k)}}(\gamma)=\operatorname{Aut}_{\mathcal{G}_{\textsc{sde}}^{(n,k)}}(\gamma')$ if $\gamma\sim\gamma'$.
		This leads to a simplified version of Equation \eqref{Eq: SDE multiplicative correlation - graph expansion of Gamma delta map} in terms of equivalent classes of graphs:		
		\begin{align}\label{Eq: SDE multiplicative correlation - quotient graph expansion of Gamma delta map}
			\Gamma_{\delta/2}[x_\xi(t_1)\cdots x_\xi(t_k)]\bigg|_{\xi=0}
			=\sum_{n\geq 0}
			\sum_{\gamma\in[\mathcal{G}]_{\textsc{sde}}^{(n,k)}}
			\frac{2^{c(\gamma)-n}}{|\operatorname{Aut}_{\mathcal{G}_{\textsc{sde}}^{(n,k)}}(\gamma)|}[\gamma]\,.
		\end{align}
		\begin{remark}
		By identifying $|\operatorname{Aut}_{\mathcal{G}_{\textsc{sde}}^{(n,k)}}(\gamma)|$ with the symmetry coefficient of $\gamma$ \cite{Hairer-2006}, the factor $2^{c(\gamma)-n}$ coincides with $e(\gamma)$ in the notation adopted in \cite{Hairer-2010}, where $e$ are the coefficients of the S-series of the exact solution \cite{Bronasco-2024}, seen as elements of the algebraic dual of the space of exotic rooted trees.
		\end{remark}
		
		\noindent {\bf MSR --}
		We discuss the expansion in terms of graphs of the expectation value
		\begin{align*}
			\llangle x(t_1)\cdots x(t_k)\rrangle_{0,\beta,1/2}\,,
		\end{align*}
		using the graph rules \ref{Graph: Rules 1}, \ref{Graph: Rules 2} and \ref{Graph: Rules 3} together with Remark \ref{Rmk: MSR expectation value - graph representation with squared vertices}, as well as the MSR formalism introduced in Section \ref{Sec: algebraic approach to Martin-Sigga-Rose formalism} .
		For the multiplicative case Equation \eqref{Eq: graph series expansion of MSR expectation value - general case} reduces to
		\begin{align}\label{Eq: MSR order expansion - multiplicative case - graph expansion}
			\llangle x(t_1)\cdots x(t_k)\rrangle_{0,\beta,1/2}
			=\sum_{n\geq 0}\sum_{\gamma_{0,\beta}^{(n,k)}\in\mathcal{G}_{0,\beta}^{(n,k)}}
			\frac{1}{|\operatorname{Aut}_{\mathcal{G}_{0,\beta}^{(n,k)}}(\gamma_{0,\beta}^{(n,k)})|}\gamma_{0,\beta}^{(n,k)}\,,
		\end{align}
		where $\mathcal{G}_{0,\beta}^{(n,k)}:=\mathcal{G}_{0,\beta}^{(n,x(t_1)\cdots x(t_k))}$ is the set of graphs built as per Remark \ref{Graph: Rules 3}. Observe that $\gamma_{0,\beta}^{(n,k)}\in\mathcal{G}_{0,\beta}^{(n,k)}$ provides a contribution of order $2n$. This is consistent with Equation \eqref{Eq: SDE Gammadelta expansion without odd orders} which contains only even orders in $\chi$.
		
		In order to compare Equation \eqref{Eq: MSR order expansion - multiplicative case - graph expansion} with Equation \eqref{Eq: SDE multiplicative correlation - quotient graph expansion of Gamma delta map} we specialize $\mathcal{G}_{0,\beta}^{(n,k)}$ to the case in hand. More precisely, we want to describe any $\gamma_{0,\beta}^{(n,k)}\in\mathcal{G}_{0,\beta}^{(n,k)}$ in terms of the vertices
		\begin{center}
			\begin{tikzpicture}[every node/.style={sloped,allow upside down}]
				\filldraw (0,0) circle (1.5pt) node[left] {$t_1$};
				\filldraw (1,0) node[below] {$,\ldots,$};
				\filldraw (2,0) circle (1.5pt) node[right] {$t_k$};
			\end{tikzpicture}
			\quad
			\begin{tikzpicture}[every node/.style={sloped,allow upside down}]
				\draw (-1,0) node {,};
				\node[rectangle split, rectangle split horizontal, rectangle split parts=2,draw,fill=white] at (0,0) {\nodepart{one} $\cdot$ \nodepart{two} $\cdot$};
			\end{tikzpicture}
		\end{center}
		As a matter of fact, according to graph rules \ref{Graph: Rules 3} any $\gamma_{0,\beta}^{(n,k)}\in\mathcal{G}_{0,\beta}^{(n,k)}$
		is obtained from $k$ graphs
		\begin{tikzpicture}[every node/.style={sloped,allow upside down}]
			\filldraw (0,0) circle (1.5pt) node[left] {$t_1$};
			\draw (0,0) -- (0.5,0);
		\end{tikzpicture}
		$,\ldots,$
		\begin{tikzpicture}[every node/.style={sloped,allow upside down}]
			\filldraw (0,0) circle (1.5pt) node[left] {$t_k$};
			\draw (0,0) -- (0.5,0);
		\end{tikzpicture}
		and $n$ interacting vertices $V_{0,\beta,1/2}$
		\begin{center}
			\begin{tikzpicture}[every node/.style={sloped,allow upside down}]
				\draw (-0.7,0) node {$\frac{1}{2}$};
				\draw[snake] (0,0) -- (-0.5,1);
				\draw[snake] (0,0) -- (0.5,1);
				\node[rectangle split, rectangle split horizontal, rectangle split parts=2,draw,fill=white] at (0,0) {\nodepart{one} $0$ \nodepart{two} $0$};
			\end{tikzpicture}
			\begin{tikzpicture}[every node/.style={sloped,allow upside down}]
				\draw (-0.8,0) node {$+\frac{1}{2}$};
				\draw[snake] (0,0) -- (0,1);
				\node[rectangle split, rectangle split horizontal, rectangle split parts=2,draw,fill=white] at (0,0) {\nodepart{one} $0$ \nodepart{two} $1$};
				\draw (1,0) node[right] {$=\frac{1}{2}\chi^2\tilde{x}^2\beta^2+\frac{1}{2}\chi^2\tilde{x}\beta\beta_1.$};
			\end{tikzpicture}
		\end{center}
		by applying Remark \ref{Rmk: MSR expectation value - graph representation with squared vertices} as well as the graph rules \ref{Graph: Rules 2}. In particular observe that all snaky edges
		\begin{tikzpicture}[every node/.style={sloped,allow upside down}]
			\draw[snake] (0,0) -- (1,0);
		\end{tikzpicture}
		joined to a straight external line
		\begin{tikzpicture}[every node/.style={sloped,allow upside down}]
			\draw (0,0) node {};
			\draw (0,0) -- (0.5,0);
		\end{tikzpicture}
		generate a directed edge
		\begin{tikzpicture}[every node/.style={sloped,allow upside down}]
			\draw (0,0) edge node {\midarrow} (0.5,0);
		\end{tikzpicture}. This information is implicitly codified in the double squared vertices
		\begin{tikzpicture}[every node/.style={sloped,allow upside down}]
			\node[rectangle split, rectangle split horizontal, rectangle split parts=2,draw,fill=white] at (0,0) {\nodepart{one} $\cdot$ \nodepart{two} $\cdot$};
		\end{tikzpicture}\,, see Remark \ref{Rmk: MSR expectation value - graph representation with squared vertices}.
		
		It follows that any $\gamma_{0,\beta}^{(n,k)}\in\mathcal{G}_{0,\beta}^{(n,k)}$ is an oriented graph which contains neither loops nor closed, oriented paths. In addition it possesses $k$ vertices
		\begin{tikzpicture}[every node/.style={sloped,allow upside down}]
			\filldraw (0,0) circle(1.5pt) node {};
		\end{tikzpicture}
		and $n$ squared vertices
		\begin{tikzpicture}[every node/.style={sloped,allow upside down}]
			\node[rectangle split, rectangle split horizontal, rectangle split parts=2,draw,fill=white] at (0,0) {\nodepart{one} $\cdot$ \nodepart{two} $\cdot$};
		\end{tikzpicture}\,.
		Moreover, $\gamma_{0,\beta}^{(n,k)}$ abides by the properties introduced in Items \eqref{Item: SDE graph representation of multiplicative SDE - vertices-edges structure} and \eqref{Item: SDE graph representation of multiplicative SDE - double vertex structure}.
		Additionally, $\gamma_{0,\beta}^{(n,k)}$ is insensible to the target of the outgoing edges of a doubles squared vertex
		\begin{tikzpicture}[every node/.style={sloped,allow upside down}]
			\node[rectangle split, rectangle split horizontal, rectangle split parts=2,draw,fill=white] at (0,0) {\nodepart{one} $\cdot$ \nodepart{two} $\cdot$};
		\end{tikzpicture}\,,
		as long as they are consistent with Item \eqref{Item: SDE graph representation of multiplicative SDE - double vertex structure}.
		This implies that to any $\gamma_{0,\beta}^{(n,k)}\in\mathcal{G}_{0,\beta}^{(n,k)}$ we may assign an equivalent class $[\gamma]\in[\mathcal{G}]_{\textsc{sde}}^{(n,k)}$, furthermore, the map
		\begin{align*}
			\mathcal{G}_{0,\beta}^{(n,k)}\ni\gamma_{0,\beta}^{(n,k)}
			\mapsto[\gamma]\in[\mathcal{G}]_{\textsc{sde}}^{(n,k)}\,,
		\end{align*}
		is bijective.
		In addition it holds that  $\operatorname{Aut}_{\mathcal{G}_{0,\beta}^{(n,k)}}(\gamma_{0,\beta}^{(n,k)})=\operatorname{Aut}_{\mathcal{G}_{\textsc{sde}}^{(n,k)}}(\gamma)$, because to each permutation $\varsigma\in\operatorname{Aut}_{\mathcal{G}_{\textsc{sde}}^{(n,k)}}(\gamma)$ of squared vertices
		\begin{tikzpicture}[every node/.style={sloped,allow upside down}]
			\node[rectangle split, rectangle split horizontal, rectangle split parts=2,draw,fill=white] at (0,0) {\nodepart{one} $\cdot$ \nodepart{two} $\cdot$};
		\end{tikzpicture}
		it corresponds a counterpart $\varsigma_{0,\beta}^{(n,k)}\in\operatorname{Aut}(\gamma_{0,\beta}^{(n,k)})$ acting on the interacting vertices $V_{0,\beta,1/2}$.
		
		This shows that Equation \eqref{Eq: MSR order expansion - multiplicative case - graph expansion} can be expressed as a sum over $[\gamma]\in\mathcal{G}_{\textsc{sde}}^{(n,k)}$.
		However, we need also to account for the factors $1/2$ appearing in the interacting vertices
		\begin{center}
			\begin{tikzpicture}[every node/.style={sloped,allow upside down}]
				\draw (-0.7,0) node {$\frac{1}{2}$};
				\draw[snake] (0,0) -- (-0.5,1);
				\draw[snake] (0,0) -- (0.5,1);
				\node[rectangle split, rectangle split horizontal, rectangle split parts=2,draw,fill=white] at (0,0) {\nodepart{one} $0$ \nodepart{two} $0$};
			\end{tikzpicture}
			\begin{tikzpicture}[every node/.style={sloped,allow upside down}]
				\draw (-0.8,0) node {$+\frac{1}{2}$};
				\draw[snake] (0,0) -- (0,1);
				\node[rectangle split, rectangle split horizontal, rectangle split parts=2,draw,fill=white] at (0,0) {\nodepart{one} $0$ \nodepart{two} $1$};
			\end{tikzpicture}
			\,,
		\end{center}
		which are therefore part of $\gamma_{0,\beta}^{(n,k)}\in\mathcal{G}_{0,\beta}^{(n,k)}$. This entails that, when we consider Equation \eqref{Eq: MSR order expansion - multiplicative case - graph expansion} and we convert the sum over $\gamma_{0,\beta}^{(n,k)}\in\mathcal{G}_{0,\beta}^{(n,k)}$ in one over the equivalence classes $[\gamma]\in\mathcal{G}_{\textsc{sde}}^{(n,k)}$ we have to insert a factor $2^{-n}$. This accounts for the factor $1/2$ due to each double squared vertex
		\begin{tikzpicture}[every node/.style={sloped,allow upside down}]
			\node[rectangle split, rectangle split horizontal, rectangle split parts=2,draw,fill=white] at (0,0) {\nodepart{one} $\cdot$ \nodepart{two} $\cdot$};
		\end{tikzpicture}\,.
		Moreover, the graph rules \ref{Graph: Rules 2} and Remark \ref{Rmk: MSR expectation value - graph representation with squared vertices} entail the presence of additional multiplicative factors $2$, {\it e.g.} when considering term such as
		\begin{center}
			\begin{tikzpicture}[every node/.style={sloped,allow upside down}]
				\node[rectangle split, rectangle split horizontal, rectangle split parts=2,draw,fill=white] at (0,0) {\nodepart{one} $j_1$ \nodepart{two} $j_2$};
				\draw (1,0) node {$\cdot_G$};
			\end{tikzpicture}
			\begin{tikzpicture}[every node/.style={sloped,allow upside down}]
				\draw (-0.6,0) node {$\frac{1}{2}$};
				\draw[snake] (0,0) -- (-0.5,1);
				\draw[snake] (0,0) -- (0.5,1);
				\node[rectangle split, rectangle split horizontal, rectangle split parts=2,draw,fill=white] at (0,0) {\nodepart{one} $0$ \nodepart{two} $0$};
				\draw (1,0) node {$\stackrel{\tilde{x}=0}{=}$};
			\end{tikzpicture}
			\begin{tikzpicture}[every node/.style={sloped,allow upside down}]
				\draw (-0.3,1) -- node {\midarrow} (0,0);
				\draw (0.3,1) -- node {\midarrow} (0,0);
				\node[rectangle split, rectangle split horizontal, rectangle split parts=2,draw,fill=white] at (0,0) {\nodepart{one} $0$ \nodepart{two} $0$};
				\node[rectangle split, rectangle split horizontal, rectangle split parts=2,draw,fill=white] at (0,1) {\nodepart{one} $j_1+1$ \nodepart{two} $j_2+1$};
				\draw (1,0) node {$+$};
			\end{tikzpicture}
			\begin{tikzpicture}[every node/.style={sloped,allow upside down}]
				\draw (-0.6,0) node {$\frac{1}{2}$};
				\draw (-0.3,1) edge [bend left] node {\midarrow} (0.3,0);
				\draw (-0.3,1) -- node {\midarrow} (0,0);
				\node[rectangle split, rectangle split horizontal, rectangle split parts=2,draw,fill=white] at (0,0) {\nodepart{one} $0$ \nodepart{two} $0$};
				\node[rectangle split, rectangle split horizontal, rectangle split parts=2,draw,fill=white] at (0,1) {\nodepart{one} $j_1+2$ \nodepart{two} $j_1$};
				\draw (1,0) node {$+$};
			\end{tikzpicture}
			\begin{tikzpicture}[every node/.style={sloped,allow upside down}]
				\draw (-0.6,0) node {$\frac{1}{2}$};
				\draw (0.3,1) edge [bend left] node {\midarrow} (0,0);
				\draw (0.3,1) -- node {\midarrow} (-0.3,0);
				\node[rectangle split, rectangle split horizontal, rectangle split parts=2,draw,fill=white] at (0,0) {\nodepart{one} $0$ \nodepart{two} $0$};
				\node[rectangle split, rectangle split horizontal, rectangle split parts=2,draw,fill=white] at (0,1) {\nodepart{one} $j_1$ \nodepart{two} $j_2+2$};
			\end{tikzpicture}
		\end{center}
		The factor $1/2$ is cancelled in the first graph on the right-hand side on account of the two possible ways of connecting the snaky external lines
		\begin{tikzpicture}[every node/.style={sloped,allow upside down}]
			\draw[snake] (0,0) -- (1,0);
		\end{tikzpicture}
		with the double squared vertex
		\begin{tikzpicture}[every node/.style={sloped,allow upside down}]
			\node[rectangle split, rectangle split horizontal, rectangle split parts=2,draw,fill=white] at (0,0) {\nodepart{one} $j_1$ \nodepart{two} $j_2$};
		\end{tikzpicture}
		so that both squared vertices
		\begin{tikzpicture}[every node/.style={sloped,allow upside down}]
			\node[rectangle,draw,fill=white] at (0,0) {$j_1$};
		\end{tikzpicture}\,,
		\begin{tikzpicture}[every node/.style={sloped,allow upside down}]
			\node[rectangle,draw,fill=white] at (0,0) {$j_2$};
		\end{tikzpicture}
		are increased by $1$.
		On the contrary, if we consider terms of the form
		\begin{center}
			\begin{tikzpicture}[every node/.style={sloped,allow upside down}]
				\node[rectangle split, rectangle split horizontal, rectangle split parts=2,draw,fill=white] at (0,0) {\nodepart{one} $j_1$ \nodepart{two} $j_2$};
				\draw (1,0) node {$\cdot_G$};
			\end{tikzpicture}
			\begin{tikzpicture}[every node/.style={sloped,allow upside down}]
				\draw (-0.6,0) node {$\frac{1}{2}$};
				\draw[snake] (0,0) -- (0,1);
				\node[rectangle split, rectangle split horizontal, rectangle split parts=2,draw,fill=white] at (0,0) {\nodepart{one} $0$ \nodepart{two} $1$};
				\draw (1,0) node {$\stackrel{\tilde{x}=0}{=}$};
			\end{tikzpicture}
			\begin{tikzpicture}[every node/.style={sloped,allow upside down}]
				\draw (-0.6,0) node {$\frac{1}{2}$};
				\draw (-0.3,1) -- node {\midarrow} (0,0);
				\node[rectangle split, rectangle split horizontal, rectangle split parts=2,draw,fill=white] at (0,0) {\nodepart{one} $0$ \nodepart{two} $1$};
				\node[rectangle split, rectangle split horizontal, rectangle split parts=2,draw,fill=white] at (0,1) {\nodepart{one} $j_1+1$ \nodepart{two} $j_2$};
				\draw (1,0) node {$+$};
			\end{tikzpicture}
			\begin{tikzpicture}[every node/.style={sloped,allow upside down}]
				\draw (-0.6,0) node {$\frac{1}{2}$};
				\draw (0.3,1) -- node {\midarrow} (0,0);
				\node[rectangle split, rectangle split horizontal, rectangle split parts=2,draw,fill=white] at (0,0) {\nodepart{one} $0$ \nodepart{two} $1$};
				\node[rectangle split, rectangle split horizontal, rectangle split parts=2,draw,fill=white] at (0,1) {\nodepart{one} $j_1$ \nodepart{two} $j_2+1$};
			\end{tikzpicture}
		\end{center}
		the factor $1/2$ is not canceled because there is a single snaky external line
		\begin{tikzpicture}[every node/.style={sloped,allow upside down}]
			\draw[snake] (0,0) -- (1,0);
		\end{tikzpicture}\,. These examples suggest that each double squared vertex
		\begin{tikzpicture}[every node/.style={sloped,allow upside down}]
			\node[rectangle split, rectangle split horizontal, rectangle split parts=2,draw,fill=white] at (0,1) {\nodepart{one} $\cdot$ \nodepart{two} $\cdot$};
		\end{tikzpicture}
		contributes to a factor $2^a$, $a\in\mathbb{Z}$, where $a$ is computed as follows:
		\begin{enumerate}
			\item\label{Item: MSR double vertex power rules - a double vertex counts 1/2}
			For each double squared vertex
			\begin{tikzpicture}[every node/.style={sloped,allow upside down}]
				\node[rectangle split, rectangle split horizontal, rectangle split parts=2,draw,fill=white] at (0,1) {\nodepart{one} $\cdot$ \nodepart{two} $\cdot$};
			\end{tikzpicture}
			there is a factor $1/2$.
			This is due to the explicit form of the interacting vertex $V_{0,\beta,1/2}$, \textit{cf.} Equation \eqref{Eq: MSR interacting potential - vartheta0 generic case}, and it contributes to an overall factor $2^{-n}$ for all $[\gamma]\in[\mathcal{G}]_{\textsc{sde}}^{(n,k)}$.
			
			\item\label{Item: MSR double vertex power rules - two incoming edges from different squared vertices count 2}
			Let assume that
			\begin{tikzpicture}[every node/.style={sloped,allow upside down}]
				\node[rectangle split, rectangle split horizontal, rectangle split parts=2,draw,fill=white] at (0,1) {\nodepart{one} $\cdot$ \nodepart{two} $\cdot$};
			\end{tikzpicture}
			has two incoming edges starting from two different vertices, \textit{e.g.} 
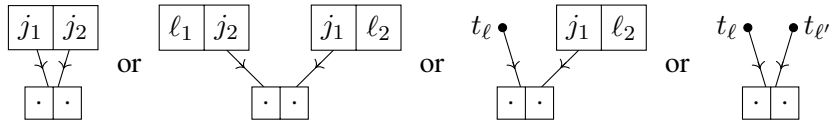
			\begin{center}
				\begin{tikzpicture}[every node/.style={sloped,allow upside down}]
					\draw (0.3,0) -- node {\midarrow} (0,-1);
					\draw (-0.3,0) -- node {\midarrow} (0,-1);
					\node[rectangle split, rectangle split horizontal, rectangle split parts=2,draw,fill=white] at (0,0) {\nodepart{one} $j_1$ \nodepart{two} $j_2$};
					\node[rectangle split, rectangle split horizontal, rectangle split parts=2,draw,fill=white] at (0,-1) {\nodepart{one} $\cdot$ \nodepart{two} $\cdot$};
					\draw (1,-0.5) node {or};
				\end{tikzpicture}
				\begin{tikzpicture}[every node/.style={sloped,allow upside down}]
					\draw (1,0) -- node {\midarrow} (0,-1);
					\draw (-1,0) -- node {\midarrow} (0,-1);
					\node[rectangle split, rectangle split horizontal, rectangle split parts=2,draw,fill=white] at (-1,0) {\nodepart{one} $\ell_1$ \nodepart{two} $j_2$};
					\node[rectangle split, rectangle split horizontal, rectangle split parts=2,draw,fill=white] at (1,0) {\nodepart{one} $j_1$ \nodepart{two} $\ell_2$};
					\node[rectangle split, rectangle split horizontal, rectangle split parts=2,draw,fill=white] at (0,-1) {\nodepart{one} $\cdot$ \nodepart{two} $\cdot$};
					\draw (2,-0.5) node {or};
				\end{tikzpicture}
				\begin{tikzpicture}[every node/.style={sloped,allow upside down}]
					\draw (1,0) -- node {\midarrow} (0,-1);
					\draw (-0.3,0) -- node {\midarrow} (0,-1);
					\filldraw (-0.3,0) circle (1.5pt) node[left] {$t_\ell$};
					\node[rectangle split, rectangle split horizontal, rectangle split parts=2,draw,fill=white] at (1,0) {\nodepart{one} $j_1$ \nodepart{two} $\ell_2$};
					\node[rectangle split, rectangle split horizontal, rectangle split parts=2,draw,fill=white] at (0,-1) {\nodepart{one} $\cdot$ \nodepart{two} $\cdot$};
					\draw (2,-0.5) node {or};
				\end{tikzpicture}
				\begin{tikzpicture}[every node/.style={sloped,allow upside down}]
					\draw (0.3,0) -- node {\midarrow} (0,-1);
					\draw (-0.3,0) -- node {\midarrow} (0,-1);
					\filldraw (-0.3,0) circle (1.5pt) node[left] {$t_\ell$};
					\filldraw (0.3,0) circle (1.5pt) node[right] {$t_{\ell'}$};
					\node[rectangle split, rectangle split horizontal, rectangle split parts=2,draw,fill=white] at (0,-1) {\nodepart{one} $\cdot$ \nodepart{two} $\cdot$};
				\end{tikzpicture}
			\end{center}
			Then one needs to include a factor $2$ which takes into account the two possible ways of connecting the snaky external lines
			\begin{tikzpicture}[every node/.style={sloped,allow upside down}]
				\draw[snake] (0,0) -- (1,0);
			\end{tikzpicture}
			of
			\begin{tikzpicture}[every node/.style={sloped,allow upside down}]
				\draw[snake] (0,0) -- (-0.5,1);
				\draw[snake] (0,0) -- (0.5,1);
				\node[rectangle split, rectangle split horizontal, rectangle split parts=2,draw,fill=white] at (0,0) {\nodepart{one} $\cdot$ \nodepart{two} $\cdot$};
			\end{tikzpicture}
			to each squared vertex
			\begin{tikzpicture}[every node/.style={sloped,allow upside down}]
				\node[rectangle,draw,fill=white] at (0,0) {$j_1$};
			\end{tikzpicture}\,,
			\begin{tikzpicture}[every node/.style={sloped,allow upside down}]
				\node[rectangle,draw,fill=white] at (0,0) {$j_2$};
			\end{tikzpicture}
			(or
			\begin{tikzpicture}[every node/.style={sloped,allow upside down}]
				\node[rectangle,draw,fill=white] at (0,0) {$j_1$};
			\end{tikzpicture}\,,
			\begin{tikzpicture}[every node/.style={sloped,allow upside down}]
				\filldraw circle (1.5pt) node[left] {$t_\ell$};
			\end{tikzpicture})\,.
			This leads to an overall factor $2^{c(\gamma)}$, where $c(\gamma)=|C(\gamma)|$ is the cardinality of the set of double squared vertices
			\begin{tikzpicture}[every node/.style={sloped,allow upside down}]
				\node[rectangle split, rectangle split horizontal, rectangle split parts=2,draw,fill=white] at (0,1) {\nodepart{one} $\cdot$ \nodepart{two} $\cdot$};
			\end{tikzpicture}
			with two incoming edges starting from two different vertices. 
		\end{enumerate}

		\noindent All in all
		\begin{align}\label{Eq: MSR order expansion - multiplicative case - quotient graph expansion}
			\llangle x(t_1)\cdots x(t_k)\rrangle_{0,\beta,1/2}
			=\sum_{n\geq 0}\sum_{\gamma\in[\mathcal{G}]_{\textsc{sde}}^{(n,k)}}
			\frac{2^{c(\gamma)-n}}{|\operatorname{Aut}_{\mathcal{G}_{\textsc{sde}}^{(n,k)}}(\gamma)|}[\gamma]\,,
		\end{align}
		which is nothing but Equation \eqref{Eq: SDE multiplicative correlation - quotient graph expansion of Gamma delta map}.
	\end{proof}

\appendix

\section{Proofs of Propositions \ref{Prop: vanishing expectation if tildex is everywhere}-\ref{Prop: alphabeta expectation value useful identity}}\label{Sec: appendix with proofs}

\begin{proof}[Proof of Proposition \ref{Prop: vanishing expectation if tildex is everywhere}]
	We provide an informal proof based on the graph rules \ref{Graph: Rules 1}-\ref{Graph: Rules 2}.
	Eventually an equivalent analytic proof is also described.
	The value $\Gamma_{G}(F_1\cdots F_n)|_{\tilde{x}=0}$ can be computed by a graph expansion as explained in the graph rules \ref{Graph: Rules 1}-\ref{Graph: Rules 2}.
	In particular the graph $\gamma_0$ associated to $\Gamma_{G}(F_1\cdots F_n)|_{\tilde{x}=0}$ is obtained from the graph $\gamma$ associated with $\Gamma_{G}(F_1\cdots F_n)$.
	We observe that, since an evaluation at $\tilde{x}=0$ is considered, $\gamma_0$ is the sum of the subgraphs in $\gamma$ with no snaky external lines.
	We recall that $\gamma$ is a directed graph with no directed loops nor closed directed paths. Moreover, since $F_1,\ldots, F_n\in\mathcal{P}_{x\tilde{x}}^{\textsc{loc}}$, $\gamma$ has exactly $n$ vertices, each associated with one of the $F$-functional.
	To prove that $\Gamma_{G}(F_1\cdots F_n)|_{\tilde{x}=0}$ vanishes we observe that, on account of the assumption $F_j(x,0)=0$, each vertex of $\gamma$ has at least one snaky external line.
	Only those graphs where these lines have been removed provide a non-vanishing contribution to $\gamma_0$.
	According to the graph rules \ref{Graph: Rules 2}, every time we remove a snaky external line from a vertex $v$ we have to insert a directed edge pointing to $v$ and starting from another vertex $v'$.
	Since this procedure has to be repeated at least once for all vertices, the resulting graph will contain at least one closed directed path, hence yielding a contradiction. Thus, the resulting contribution vanishes.
	
	An analytic proof of the identity $\Gamma_G(F_1\cdots F_n)|_{\tilde{x}=0}=0$ can be obtained by direct inspection.
	Indeed,
	\begin{multline*}
		\Gamma_G(F_1\cdots F_n)\big|_{\tilde{x}=0}
		=\sum_{m\geq 0}\frac{1}{m!}
		\int\limits_{\mathbb{R}^{2m}}\prod_{j=1}^m G(t_j,s_j)
		\\
		\frac{\delta^m}{\delta x(t_1)\cdots\delta x(t_m)}
		\frac{\delta^m}{\delta \tilde{x}(s_1)\cdots\delta \tilde{x}(s_m)}
		F_1\cdots F_n\big|_{\tilde{x}=0}
		\mathrm{d}t_1\cdots\mathrm{d}t_m\mathrm{d}s_1\cdots\mathrm{d}s_m\,.
	\end{multline*}
	On account of the assumption $F_j(x,0)=0$, $j\in\{0,\ldots,n\}$, the contributions for $m<n$ vanish.
	Moreover, if $m\geq n$, each $F_j$ has to be $\tilde{x}$-derived at least once.
	Expanding the functional derivatives it descends
	\begin{multline*}
		\prod_{j=1}^m G(t_j,s_j)
		\frac{\delta^m}{\delta x(t_1)\cdots\delta x(t_n)}
		\frac{\delta^m}{\delta \tilde{x}(s_1)\cdots\delta \tilde{x}(s_n)}
		F_1\cdots F_n\big|_{\tilde{x}=0}
		\\
		=\prod_{j=1}^m G(t_j,s_j)
		\sum_{\substack{L_1,\ldots,L_n\\K_1,\ldots,K_n}}\frac{m!^2}{\prod_{j=1}^n\ell_j!k_j!}
		\frac{\delta^{\ell_1+k_1} F_1}{\delta x(t_{L_1})\delta\tilde{x}(s_{K_1})}
		\cdots\frac{\delta^{\ell_n+k_n} F_n}{\delta x(t_{L_n})\delta\tilde{x}(s_{K_n})}\bigg|_{\tilde{x}=0}\,,
	\end{multline*}
	where the sum is over the partitions of $\{1,\ldots,m\}$ into $L_1,\ldots,L_n$ disjoint, possibly empty, subsets while $K_1,\ldots,K_n$ run over the partitions of $\{1,\ldots,m\}$ into non-empty disjoint subsets.
	In the above formula we set $\ell_j:=|L_j|$ and $k_j:=|K_j|$, moreover, we denoted
	\begin{align*}
		\frac{\delta^{\ell_h+k_h} F_h}{\delta x(t_{L_h})\delta\tilde{x}(s_{K_h})}
		:=\frac{\delta^{\ell_h+k_h} F_h}{\prod_{i\in L_h}\delta x(t_i)\prod_{j\in K_h}\delta\tilde{x}(s_j)}\,.
	\end{align*}	
	Since $F_h\in\mathcal{P}_{x\tilde{x}}^{\textsc{loc}}$ for all $h\in\{1,\ldots,n\}$, it follows that
	\begin{align*}
		\frac{\delta^{\ell_h+k_h} F_h}{\prod_{i\in L_h}\delta x(t_i)\prod_{j\in K_h}\delta\tilde{x}(s_j)}\,,
	\end{align*}
	is a distribution supported on $t_i=s_j$ for all $i\in L_h$ and $j\in K_h$.
	Moreover, the product $\prod_{j=1}^m G(t_j,s_j)$ is supported on $t_j>s_j$ for all $j\in\{1,\ldots,n\}$.
	On account of the convention $G(t,t)=\vartheta_0=0$ and of the support properties of the functional derivatives of all $F_j$, it follows that there exists a sequence $t'_1,\ldots,t'_p$, $p\in\mathbb{N}$, such that $G(t_1',t_2')\cdots G(t_{p-1}',t_p')G(t_p',t_1')$ appears in $\prod_{j=1}^m G(t_j,s_j)$.
	On account of the support properties of $G$, the overall contribution vanishes.
\end{proof}

\begin{proof}[Proof of Proposition \ref{Prop: alphabeta expectation value useful identity}]
	The proof is divided in two separate parts.	In the first one we consider a simpler setting which allows to present succinctly and exhaustively all the computational steps which are necessary to prove the sought statement. The general scenario follows suit, yet with formulae which are structurally identical, though they need to account for longer and more cumbersome expressions. 
	
	\vskip .2cm	
	
	\noindent{\em Part 1--} We start considering $F\in\mathcal{P}_{x}$, see Remark \ref{Rem: tildex-independent-functionals} and setting $\alpha=0=\vartheta_0$. Combining Equation \ref{Eq: MSR expectation value} with \ref{Eq: series expansion of MSR expectation value - general case}, a direct computation yields
	\begin{multline}\label{Eq: intermediate}
		\llangle F\rrangle_{0,\beta,0}
		=\Gamma_{G}(Fe^{\frac{\sigma}{2}\langle\chi^2,\tilde{x}^2\beta^2\rangle})\big|_{\tilde{x}=0}
		\\
		=\sum_{n=0}^\infty\frac{1}{n!}\int\limits_{\mathbb{R}^{2n}}
		\prod_{j=1}^nG(t_j,s_j)
		\frac{\delta^n}{\delta x(t_1)\ldots\delta x(t_n)}\bigg[
		F\frac{\delta^n e^{\frac{\sigma}{2}\langle\chi^2,\tilde{x}^2\beta^2\rangle}}{\delta\tilde{x}(s_1)\ldots\delta\tilde{x}(s_n)}\bigg|_{\tilde{x}=0}
		\bigg]
		\\\mathrm{d}t_1\ldots\mathrm{d}t_n\mathrm{d}s_1\ldots\mathrm{d}s_n\, ,
	\end{multline}
	where in the second equality we work at the level of integral kernels and we have applied Definition \ref{Def: Gamma G map}. Since $e^{\frac{\sigma}{2}\langle\chi^2,\tilde{x}^2\beta^2\rangle}$ is even in $\tilde{x}$ and since we are evaluating at $\tilde{x}=0$, the non-vanishing contributions occur only for $n=2m$, $m\in\mathbb{N}\cup\{0\}$.
	In particular
	\begin{align*}
		\frac{\delta^{2m} e^{\frac{\sigma}{2}\langle\chi^2,\tilde{x}^2\beta^2\rangle}}{\delta\tilde{x}(s_1)\ldots\delta\tilde{x}(s_{2m})}\bigg|_{\tilde{x}=0}
		=\sum_{\substack{\textrm{pair partition}\\\wp \textrm{ of }\{1,\ldots, 2m\}}}
		\prod_{(i,j)\in\wp}\sigma\chi^2(s_j)\beta^2(x(s_j),s_j)\delta(s_i-s_j)\,.
	\end{align*}
	Replacing this identity in Equation \eqref{Eq: intermediate} and integrating with respect to the variables $s_1,\ldots, s_{2m}$ yields
	\begin{multline*}
		\Gamma_G(Fe^{\frac{\sigma}{2}\langle\chi^2,\tilde{x}^2\beta^2\rangle})\big|_{\tilde{x}=0}
		=\sum_{m=0}^\infty\frac{1}{(2m)!}\int\limits_{\mathbb{R}^{2m}}
		\frac{\delta^{2m}}{\delta x(t_1)\ldots\delta x(t_{2m})}\bigg[
		F\\\sum_{\substack{\textrm{pair partition}\\\wp \textrm{ of }\{1,\ldots, 2m\}}}\prod_{(i,j)\in\wp}\sigma Q_{\chi^2\beta^2}(t_i,t_j)
		\bigg]
		\mathrm{d}t_1\ldots\mathrm{d}t_{2m}\,,
	\end{multline*}
	where $Q_{\chi^2\beta^2}$ is defined in Equation \eqref{Eq: Qbeta-operator}.
	Observe in particular that
	\begin{align*}
		\frac{\delta}{\delta x(t_1)}Q_{\chi^2\beta^2}(t_2,t_3)
		=G(t_2,t_1)G(t_3,t_1)\chi(t_1)^2\beta_1(x(t_1),t_1)
		=0\qquad\textrm{ if }t_1\geq \min\{t_2,t_3\}\,.
	\end{align*}
	It follows that, for every element of $\wp$, we may order the $x$-functional derivatives and the $Q$-factors according to the ordering of the pairs $(i,j)$, where $(i,j)\leq (i',j')$ if and only if $\min\{t_i,t_j\}\leq\min\{t_i',t_j'\}$.
	Up to a relabel of the variables $t_1,\ldots,t_{2m}$, the support properties of the Heaviside function entail that 
	\begin{multline*}
		\int\limits_{\mathbb{R}^{2m}}
		\frac{\delta^{2m}}{\delta x(t_1)\ldots\delta x(t_{2m})}\bigg[
		F\prod_{(i,j)\in\wp}\sigma Q_{\chi^2\beta^2}(t_i,t_j)
		\bigg]
		\mathrm{d}t_1\ldots\mathrm{d}t_{2m}
		\\=m!\int_{\tau_1\geq\ldots\geq\tau_m}\bigg[\prod_{j=1}^m\sigma Q_{\chi^2\beta^2}(t_j,t_{j+m})
		\frac{\delta^2}{\delta x(t_j)\delta x(t_{j+m})}\bigg]F
		\mathrm{d}t_1\ldots\mathrm{d}t_{2m}
		\\=:\textrm{T-}\int\bigg[\prod_{j=1}^m\sigma Q_{\chi^2\beta^2}(t_j,t_{j+m})
		\frac{\delta^2}{\delta x(t_j)\delta x(t_{j+m})}\bigg]F
		\mathrm{d}t_1\ldots\mathrm{d}t_{2m}\,,
	\end{multline*}
	being $\tau_j:=\min\{t_j,t_{j+m}\}$ whereas $\textrm{T-}\int$ indicates a time-ordered integration.
	Thus, every pair partition $\wp$ leads to the same final contribution.
	Since the number of pair partitions of $t_1,\ldots t_{2m}$ is $(2m-1)!!$ we end up with
	\begin{align*}
		\Gamma_G(Fe^{\frac{\sigma}{2}\langle\chi^2,\tilde{x}^2\beta^2\rangle})\big|_{\tilde{x}=0}
		&=\sum_{m=0}^\infty\frac{1}{m!}\textrm{T-}\int
		\prod_{j=1}^m\bigg[\frac{\sigma}{2}Q_{\chi^2\beta^2}(t_j,t_{j+m})\frac{\delta^2}{\delta x(t_j)\delta x(t_{j+m})}\bigg]F
		\mathrm{d}t_1\ldots\mathrm{d}t_{2m}
		\\&=T_{\frac{\sigma}{2}Q_{\chi^2\beta^2}}(F)\,.
	\end{align*}
	
	\vskip .2cm	
	
	\noindent{\em Part 2--} 
	We consider the generic case where no constraint has been imposed on the values of $\alpha,\beta,\vartheta_0$.
	Let $F\in\mathcal{P}_x$ and let us set $F_{\alpha,\vartheta_0}:=Fe^{\langle\chi,\tilde{x}\alpha\rangle+\langle\chi^2,\sigma\vartheta_0\tilde{x}\beta\beta_1\rangle}$.
	Then, applying once more Equation \eqref{Eq: MSR expectation value} and \eqref{Eq: series expansion of MSR expectation value - general case}, it descends
	\begin{multline*}
		\llangle F\rrangle_{\alpha,\beta,\vartheta_0}
		=\Gamma_G\left[Fe^{\langle\chi,\tilde{x}\alpha\rangle
			+\langle\chi^2,\sigma\vartheta_0\tilde{x}\beta\beta_1\rangle
			+\frac{\sigma}{2}\langle\chi^2,\tilde{x}^2\beta^2\rangle}\right]\bigg|_{\tilde{x}=0}
		\\=\sum_{n=0}^\infty\frac{1}{n!}
		\sum_{a+b=n}\frac{n!}{a!b!}\int\limits_{\mathbb{R}^{2n}}
		\prod_{j=1}^nG(t_j,s_j)
		\frac{\delta^n}{\delta x(t_1)\ldots\delta x(t_n)}\bigg[
		\frac{\delta^a F_{\alpha,\vartheta_0}}
		{\delta\tilde{x}(s_1)\ldots\delta\tilde{x}(s_a)}\bigg|_{\tilde{x}=0}
		\\\frac{\delta^b e^{\frac{\sigma}{2}\langle\chi^2,\tilde{x}^2\beta^2\rangle}}
		{\delta\tilde{x}(s_{a+1})\ldots\delta\tilde{x}(s_{a+b})}\bigg|_{\tilde{x}=0}
		\bigg]\mathrm{d}t_1\ldots\mathrm{d}t_n\mathrm{d}s_1\ldots\mathrm{d}s_n\,.
	\end{multline*}
	For the same reasons outlined in the first part of this proof, we can set $b=2m$, $m\in\mathbb{N}\cup\{0\}$, and repeating the same computations, mutatis mutandis, it descends that
	\begin{multline*}
		\llangle F\rrangle_{\alpha,\beta,\vartheta_0}
		=\sum_{a+2m=0}^\infty\frac{1}{a!m!}\int
		\prod_{j=1}^aG(t_j,s_j)
		\frac{\delta^a}{\delta x(t_1)\ldots\delta x(t_a)}\bigg[
		\\\textrm{T-}\int\bigg(
		\prod_{\ell=1}^m\frac{\sigma}{2}Q_{\chi^2\beta^2}(t_{\ell+a},t_{\ell+a+m})\frac{\delta^2}{\delta x(t_{\ell+a})\delta x(t_{\ell+a+m})}
		\bigg)
		\frac{\delta^a F_{\alpha,\vartheta_0}}
		{\delta\tilde{x}(s_1)\ldots\delta\tilde{x}(s_a)}\bigg|_{\tilde{x}=0}
		\bigg]
		\\\mathrm{d}t_1\ldots\mathrm{d}t_n\mathrm{d}s_1\ldots\mathrm{d}s_a
		\\=\Gamma_G\left[T_{\frac{\sigma}{2}Q_{\chi^2\beta^2}}\left(Fe^{\langle\chi,\tilde{x}\alpha\rangle+\langle\chi^2,\sigma\vartheta_0\tilde{x}\beta\beta_1\rangle}\right)\right]\bigg|_{\tilde{x}=0}\,.
	\end{multline*}
\end{proof}

\paragraph{Acknowledgements.}
A.B. is supported partly by a PhD fellowship of the University of Pavia and partly by the GNFM-Indam Progetto Giovani {\em Feynman propagator for Dirac fields: a microlocal analytic approach} CUP\_E53C22001930001, whose support is gratefully acknowledged.
He is grateful to the Department of Mathematics of the University of Trento for the kind hospitality during the realization of part of this project.
N.D. acknowledge the support of the GNFM-INdAM Progetto Giovani \textit{Non-linear sigma models and the Lorentzian Wetterich equation},  CUP\_E53C22001930001. C.D. is grateful for the support of the GNFN-INdAM.

\paragraph{Data availability statement.}
Data sharing is not applicable to this article as no new data were created or analysed in this study.

\paragraph{Conflict of interest statement.}
The authors certify that they have no affiliations with or involvement in any
organization or entity with any financial interest or non-financial interest in
the subject matter discussed in this manuscript.

\end{document}